\documentclass[reqno,11pt]{amsart}
\usepackage{multind}
\usepackage{graphicx}
\usepackage{amscd}
\usepackage{slashed}
\usepackage{amssymb}
\usepackage{ushort}
\usepackage{mathtools} 
\usepackage[mathscr]{eucal}
\numberwithin{equation}{section}
\allowdisplaybreaks[1]

\title[Perturbative QFT in the Framework of the Fermionic Projector]
{Perturbative Quantum Field Theory in the Framework of the Fermionic Projector}

\author[F.\ Finster]{Felix Finster \\ \\ October 2013}
\address{Fakult\"at f\"ur Mathematik \\ Universit\"at Regensburg \\ D-93040 Regensburg \\ Germany}
\email{finster@ur.de}

\newtheorem{Def}{Definition}[section]
\newtheorem{Thm}[Def]{Theorem}
\newtheorem{Prp}[Def]{Proposition}
\newtheorem{Lemma}[Def]{Lemma}

\newcommand{\Thanks}{\vspace*{.5em} \noindent \thanks}
\newcommand{\beq}{\begin{equation}}
\newcommand{\eeq}{\end{equation}}
\newcommand{\Proof}{\begin{proof}}
\newcommand{\QED}{\end{proof} \noindent}

\newcommand{\la}{\langle}
\newcommand{\ra}{\rangle}
\newcommand{\bra}{\mathopen{<}}
\newcommand{\ket}{\mathclose{>}}
\newcommand{\Sl}{\mathopen{\prec}}
\newcommand{\Sr}{\mathclose{\succ}}

\newcommand{\C}{\mathbb{C}}
\newcommand{\R}{\mathbb{R}}
\newcommand{\1}{\mbox{\rm 1 \hspace{-1.05 em} 1}}
\newcommand{\Z}{\mathbb{Z}}

\newcommand{\Pdd}{\mbox{$\partial$ \hspace{-1.2 em} $/$}}
\newcommand{\slsh}{\mbox{ \hspace{-1.13 em} $/$}}

\renewcommand{\H}{\mathscr{H}}
\newcommand{\J}{\mathscr{J}}
\newcommand{\U}{{\rm{U}}}
\newcommand{\SU}{{\rm{SU}}}

\newcommand{\bep}{\begin{pmatrix}}
\newcommand{\enp}{\end{pmatrix}}

\renewcommand{\O}{\mathscr{O}}

\newcommand{\Fock}{{\mathcal{F}}}

\newcommand{\B}{{\mathscr{B}}}
\renewcommand{\O}{{\mathscr{O}}}
\renewcommand{\L}{{\mathcal{L}}}
\newcommand{\Sact}{{\mathcal{S}}}

\newcommand{\T}{{\mathcal{T}}}

\newcommand{\I}{{\mathscr{I}}}

\newcommand{\np}{n_{\mathrm{p}}}
\newcommand{\na}{n_{\mathrm{a}}}
\newcommand{\Uran}{{\mathscr{U}}}
\newcommand{\Wran}{{\mathscr{W}}}
\newcommand{\Lphase}{L_\text{phase}}
\newcommand{\Lmix}{{L_\text{mix}}}
\newcommand{\as}{{\mathfrak{a}}}
\newcommand{\bs}{{\mathfrak{b}}}
\newcommand{\Bu}{{\underbracket[0.65pt][0pt]{\hat{\B}}}}
\newcommand{\Bd}{{\text{\d{$\hat{\mathscr{B}}$}}}}
\newcommand{\Hint}{H_\text{\rm{int}}}
\newcommand{\Heff}{H_\text{\rm{eff}}}
\newcommand{\ansyb}{\textsf{AnSyB}}
\newcommand{\ansybs}{\textsf{AnSyB}s}

\DeclareMathOperator{\Tr}{Tr}

\DeclareMathOperator{\supp}{supp}
\DeclareMathOperator{\Texp}{Texp}

\DeclareMathOperator{\sign}{sign}

\newcommand{\hM}{\hat{\mathcal{M}}}

\begin{document}
\maketitle

\begin{abstract}
We give a microscopic derivation of perturbative quantum field theory, taking
causal fermion systems and the framework of the fermionic projector as the starting point.
The resulting quantum field theory agrees with standard quantum field theory on the
tree level and reproduces all bosonic loop diagrams. The fermion loops are described in a different
formalism in which no ultraviolet divergences occur.
\end{abstract}

\tableofcontents

\section{Introduction}
In the standard interpretation of quantum mechanics, particles are point-like,
and the absolute square~$|\psi(t, \vec{x})|^2$ of the wave function gives the probability density
for the particle to be observed at the position~$\vec{x}$.
The necessity for the statistical interpretation of the wave function can be understood
if one couples the wave function to a classical field.
In order to work in a simple concrete example, we consider one Schr\"odinger
wave function~$\psi$ coupled to a Coulomb potential,
\beq \label{schmax}
i \hbar \partial_t \psi = \Big( -\frac{\hbar^2}{2m}\, \Delta + V \Big) \psi\:,\qquad
-\Delta V(t,\vec{x}) = e^2\, \big| \psi(t, \vec{x}) \big|^2 \:,
\eeq
where for the charge density entering the Coulomb equation we simply take the expectation
value of the wave function. The coupled system~\eqref{schmax} has the shortcoming that the 
electric potential mediates an interaction of the Schr\"odinger wave function
with itself. For example, in the static situation, the wave function at position~$\vec{x}$
has a charge density~$|\psi(\vec{x})|^2$, which feels the electrostatic repulsion of the
charge density~$|\psi(\vec{y})|^2$ at another position~$\vec{y}$.
As already observed by Schr\"odinger~\cite{schroedinger2}, this ``self-repulsion'' of the
wave function would give corrections to the atomic spectra which are not in agreement
with experimental data. Schr\"odinger concluded that coupling the quantum mechanical equations
to the classical field equations is not the correct physical concept.
In the standard statistical interpretation of quantum mechanics, this
problem is bypassed by giving up the Schr\"odinger wave function as the fundamental
physical object. Instead, one imposes that~$|\psi(\vec{x})|^2$ only gives the probability for a point-like
particle to be at the position~$\vec{x}$. Consequently, the classical field equations
(like the Coulomb or Maxwell equations) are to be coupled to the
point charge, not to the continuous charge distribution as given by the probability density.
The interaction of a particle with itself can be avoided by imposing that the field generated by
a point particle should not couple to the same particle, but only to all other particles.

In relativistic quantum field theory, the self-interaction is described differently.
First, one introduces the free fermionic and bosonic field operators acting on a Fock space
(``second quantization''). Then the interaction is described perturbatively in a formal
power expansion in the coupling constant. The self-interaction is treated order by order in perturbation
theory by renormalizing the divergent loop diagrams.
In this formalism, the physical system is described by a quantum state~$\Psi$ of the Fock space.
This quantum state again has as a probabilistic interpretation, albeit not for the
individual particles, but only for the system as a whole.

The fermionic projector approach is a framework for the formulation of
relativistic quantum theories. A central object is the {\em{fermionic projector}}, which describes the
ensemble of all fermionic wave functions, including states of negative energy
in a configuration which is usually referred to as the Dirac sea (see the survey article~\cite{srev}).
This ensemble of wave functions characterizes the physical system completely, which means
in particular that it encodes the causal structure, the metric of space-time and the bosonic fields.
The point of view of encoding all space-time structures in the wave functions becomes
clearest in the abstract formulation as a {\em{causal fermion system}},
in which the interaction is described by the {\em{causal action principle}} (see~\cite{rrev}
and the references therein). The fermions are quantized in the sense that we use a
many-particle description which includes anti-particles and pair creation.
However, the fermions are not described by a state in the fermionic Fock space.
We consider the ensemble of fermionic wave functions as the basic physical object.
The ``particle character'' of the fermions, however, should arise as a consequence of
the interaction as described by the causal action principle (see the survey article~\cite{dice2010}).
Moreover, in the so-called {\em{continuum limit}}, one obtains an interaction via classical bosonic fields
(see~\cite{sector, lepton} or the survey article~\cite{srev}).
This raises the basic question of how to resolve the problem of the classical self-interaction
of the system~\eqref{schmax}. Also, how does one get quantized bosonic fields?
Is it possible to rewrite the interaction in terms of interacting quantum fields on bosonic and fermionic
Fock spaces? Can one derive a perturbation expansion in terms of Feynman diagrams?
In the present paper, we shall address and give affirmative answers to these questions.

Before entering the discussion of our methods, we mention an approach by Barut,
who gave a detailed discussion of the problem of the coupled Dirac-Maxwell system
\beq \label{dirmax}
(i \Pdd + \slashed{A} - m)\, \psi = 0 \:,\qquad 
\partial_{jk} A^k - \Box A_j = e^2 \,\overline{\psi} \gamma_j \psi \:,
\eeq
and pointed towards possible alternative solutions~\cite{barut}
(from now on we work in natural units~$\hbar=c=1$).
In particular, he takes an attempt to revive Schr\"odinger's concept of regarding the wave function
as the fundamental physical object. To this end, he transforms the system~\eqref{dirmax} to Fourier
space and selects certain combinations of Fourier modes which enter the nonlinear coupling. In our
notation, this construction amounts to replacing~\eqref{dirmax} by
\beq \label{Barut}
(i \Pdd + \slashed{A} - m)\, \psi = 0 \:,\qquad 
A_j = -i \pi e^2\, K_0 \big( \overline{\psi} \gamma_j \psi \big) \:,
\eeq
where~$K_0$ is an integral operator involving the difference of the advanced and retarded Green's function,
\beq \label{K0def}
(K_0 \,J)(x) := \frac{1}{2 \pi i} \int \big( S_0^\vee - S_0^\wedge \big)(x,y)\: J(y) \:d^4y\:.
\eeq
For the connection to Wheeler-Feynman quantum electrodynamics we refer to~\cite[Section~8]{deckert}.
It is remarkable that quantum effects like the Lamb shift can be derived from
this purely classical system (see~\cite{barut+kraus, barut+huele}).
The drawback is that an ad-hoc procedure is used to modify the
Dirac-Maxwell equations~\eqref{dirmax} (note that, since~$K_0$ involves the difference of two
Green's functions, $A$ is a solution of the homogeneous Maxwell equations).
In particular, the above-mentioned ``self-repulsion'' of
the wave function is taken out by hand. Moreover, the agreement with quantum field theory 
seems to be restricted to one-loop corrections.

Our methods for going beyond the Dirac-Maxwell equations~\eqref{dirmax}
make essential use of the concept that all space-time structures are encoded in the ensemble of wave functions.
Namely, this concept makes it possible to regard space-time~$M$ simply as as a point set,
on which the wave functions are defined.
Decomposing space-time points into disjoint subsets
and choosing the wave functions on each subset differently,
we can arrange different space-time structures on the subsets.
Intuitively speaking, space-time becomes a ``mixture'' of many different space-times
which be endowed with different causal structures, different bosonic fields, and so on.
The decomposition of space-time should be fine-grained in the sense that every
macroscopic region of space-time intersects many of the subsets.
Under this assumption, the effective macroscopic dynamics can be described
by ``taking averages'' over the subsystems.
In order clarify this concept of {\em{microscopic mixing}}, we point out that
microscopic mixing is not an ad-hoc procedure to cure the problem of the self-interaction,
but it is in fact a consequence of the causal action principle.
Namely, the causal action diverges (or, if an ultraviolet regularization is present,
becomes very large) if particle or anti-particle states are introduced
into the system, and these divergences can be removed by the microscopic mixing
procedure (for details see Sections~\ref{secwhy} and~\ref{secanti} below).

The method of microscopic mixing was first introduced in~\cite{entangle},
where a mixing of all the particle states including the states of the Dirac sea was considered.
The resulting so-called {\em{decoherent space-time regions}} have an independent dynamics
and do not interact with each other. This concept makes it possible to describe entangled fermionic states
and quantized bosonic fields in the framework of the fermionic projector.
In the present paper, we consider a more general mechanism of microscopic mixing, where
we allow for a microscopic mixing of only a few of the states. More specifically, all the
wave functions of the particles of the system take part in the microscopic mixing,
but most of the states of the Dirac sea are not affected by microscopic mixing.
After this so-called {\em{microscopic mixing of the wave functions}}, the subsystems are not
completely decoherent and still interact with each other.
But as a consequence of microscopic mixing, the effective many-particle wave function
will be totally antisymmetric. As a particular consequence of this anti-symmetrization,
the ``self-repulsion'' of a wave function mentioned above is no longer present.
Instead, an electron feels the electrostatic repulsion only of all the other electronic wave
functions. In this way, the problem of the self-repulsion of the coupled systems~\eqref{schmax}
or~\eqref{dirmax} disappears.

Another ingredient used in our construction is a stochastic bosonic background field
which may depend on the subsystem and thus gives rise to ``correlations'' between the subsystems.
Such correlations give rise to an effect which we refer to as {\em{synchronization}}.
Working with a stochastic field has some similarity with the approaches to explain quantum
effects by adding a stochastic term to the classical equations (see for example
Nelson's stochastic mechanics~\cite{nelson} or~\cite{pena-cetto, khrennikov}). However, in contrast to these approaches, we do not modify the classical equations but only superimpose the
macroscopic field by microscopic fluctuations which are solutions of the homogeneous field equations.
Also, the physical picture is different. In our context, the stochastic background field
can be understood as giving an effective description of microscopic fluctuations.
It can be arbitrarily weak and is thus natural to assume.

Combining these methods, we succeed in rewriting the dynamics in the language of bosonic and
fermionic Fock spaces. In a certain limiting case (the so-called {\em{instantaneous recombination}} in
a {\em{background-synchronized}} system), we obtain complete agreement with the
standard formulation of perturbative quantum field theory, with the only exception of the fermion loops,
which are described in a different mathematical formalism.
In our formulation, the contributions of the fermion loops are all ultraviolet finite.
This can be understood by the fact that the divergent parts of the fermionic loop diagrams
drop out of the Euler-Lagrange equation corresponding to the causal action principle
(as explained in the review paper~\cite{srev}).
Since the connection to perturbative quantum field theory is obtained only in a specific
limiting case, there is the hope that without taking this limiting, we have an extended
theory in which some of the problems of quantum field theory are avoided.
In particular, the divergent bosonic loops appear only in the
limiting case of an instantaneous recombination.

A further potential advantage of our procedure is that the ``quantization'' of the fields
reduces to introducing the microscopic mixing. This procedure is canonical and
seems to apply similarly to any interaction by gauge fields and/or a gravitational field.

The paper is organized as follows.
In Section~\ref{secclass} we review the framework of the fermionic projector and explain
the description in the continuum limit, where the Dirac wave functions interact via classical
bosonic fields. In Section~\ref{secmicmix} we motivate and introduce the concept of microscopic mixing
of the wave functions. Section~\ref{secstoch} is devoted to the stochastic bosonic background field.
In Section~\ref{secasb} we analyze how to take ``averages'' over subsystems. We shall see that
the small-scale fluctuations give rise to destructive interference, except for classes of
anti-symmetrized Feynman diagrams referred to as {\em{anti-symmetrized synchronal blocks}} (\ansybs).
In Section~\ref{secasbdyn} we analyze the dynamics of one \ansyb, whereas Section~\ref{secrecombine}
is devoted to the interaction of several \ansybs.
In Section~\ref{secfock} we rewrite the dynamics in the Fock space formalism.
The connection to the standard formulation of perturbative quantum field theory
is made precise in Theorem~\ref{thmeffective}.
Finally, in Section~\ref{secinterpret} we interpret our results and give an outlook on possible
directions of future research. A more technical issue involved in taking ``averages'' over 
subsystems is worked out in Appendix~\ref{appyoung}.

\section{The Fermionic Projector Coupled to a Classical Bosonic Field} \label{secclass}
\subsection{The Vacuum}
We first introduce the relevant objects in the vacuum. For notational simplicity, we consider
only one type of particles of mass~$m$,
\[ (i \Pdd - m) \Psi = 0 \:, \]
but all our constructions generalize immediately to systems
involving different particles (as introduced in~\cite[\S5.1]{PFP} or~\cite[Section~3]{sector},
\cite[Section~1]{lepton}).
Solving the Dirac equation with plane waves, one obtains a natural splitting of the
solution space into solutions of positive and negative frequency.
The fermionic projector is defined as an operator which maps onto the solutions of
negative frequency. In formulas, the kernel of the fermionic projector is given by
\[ P^\text{vac}(x,y) = \frac{1}{2} \,\big(p_m -k_m \big)(x,y) \,, \label{sea-pk} \]
where
\begin{align}
p_m(x,y)&=\int\frac{d^4q}{(2\pi)^4}\:(\slashed{q}+m)\:\delta(q^2-m^2)\:e^{-iq(x-y)} \label{pmdef} \\
k_m(x,y)&=\int\frac{d^4q}{(2\pi)^4}\:(\slashed{q}+m)\:\delta(q^2-m^2)\:\epsilon(q^0)\:e^{-iq(x-y)} \label{kmdef}
\end{align}
(where~$\epsilon(\tau)$ is the step function taking the values~$1$ if~$\tau>0$ and~$-1$ otherwise).
We also consider~$P^\text{vac}$ as an integral kernel of a corresponding operator~$P^\text{vac}$
(defined for example on the smooth wave functions with compact support).
The image of~$P^\text{vac}$ consists of all negative-frequency solutions of the Dirac equation.
In order to describe fermionic matter, we build in wave functions of particles and anti-particles
by setting
\beq \label{P0def}
P^{(0)}(x,y) = P^\text{vac}(x,y) -\sum_{k=1}^{\np} \Psi_k(x) \overline{\Psi_k(y)}
+\sum_{l=1}^{\na} \Phi_l(x) \overline{\Phi_l(y)}\:.
\eeq
Here~$\Psi_1, \ldots, \Psi_{\np}$ and~$\Phi_1, \ldots, \Phi_{\na}$ are the wave functions
of the particles and anti-particles, respectively, orthonormalized with respect to the
probability scalar product (for details see~\cite[\S2.6]{PFP} or~\cite{rrev}),
\beq \label{normalize}
\int_{\R^3} (\overline{\Psi_k} \gamma^0 \Psi_l)(t, \vec{x}) \, d\vec{x} = \frac{1}{2 \pi}\, \delta_{kl}
= \int_{\R^3} (\overline{\Phi_k} \gamma^0 \Phi_l)(t, \vec{x}) \, d\vec{x} \:.
\eeq
The fermionic projector~$P^{(0)}$ satisfies the free Dirac equation
\[ (i \Pdd - m) \,P^{(0)}(x,y) = 0 \:. \]

\subsection{The Fermions in an External Field}
We next consider the Dirac equation in an external field
\[ (i \Pdd + \B - m) \tilde{\Psi} = 0 \:, \]
where~$\B$ is a multiplication operator, which may depend on time
but is smooth and has suitable decay properties at infinity.
Even in the time-dependent situation there is
a canonical decomposition of the solution space into two subspaces.
Moreover, the fermionic projector can be introduced as an operator whose image
coincides with one of these subspaces (namely the subspace which in the static situation
reduces to the solutions of negative frequency).
These facts were first proven in an expansion in powers of~$\B$
(see~\cite{sea, grotz} or~\cite[\S2.2]{PFP}).
More recently, this construction was carried out non-perturbatively
(see~\cite{finite, hadamard}). 
Here we shall always restrict attention to the perturbative treatment.
Then the fermionic projector~$P$ in the presence of the interaction
is introduced most conveniently using the unitary perturbation flow by
\beq \label{Uflow}
P = U P^{(0)} U^* \:.
\eeq
The operator~$U$ has an an operator product expansion
(see~\cite[Section~5]{grotz}; explicit formulas and a discussion
of the normalization are worked out in~\cite{norm}). Using~\eqref{P0def}, we obtain
\beq \label{Pdef}
P = P^\text{sea} - \sum_{k=1}^{\np} \tilde{\Psi}_k(x) \overline{\tilde{\Psi}_k(y)}
+ \sum_{l=1}^{\na} \tilde{\Phi}_l(x) \overline{\tilde{\Phi}_l(y)}\:,
\eeq
where
\beq \label{PseaU}
P^\text{sea} = U P^\text{vac} U^*
\eeq
and
\beq \label{psitilde}
\tilde{\Psi}_k := U \Psi_k \qquad \text{and} \qquad \tilde{\Phi}_l := U \Psi_l \:.
\eeq
The operator expansion for~$U$ defines~$P^\text{sea}$ perturbatively in terms of an
expansion of the form
\beq \label{Pexform}
P^\text{sea} = \sum_{k=0}^\infty \sum_{\alpha=0}^{\alpha_{\max}(k)} c_\alpha\;
C_{1,\alpha} \, \B\,C_{2,\alpha} \,\B\, \cdots \,\B\, C_{k+1, \alpha} \:,
\eeq
where the factors~$C_{l,\alpha}$ are the Green's functions~$s_m$ or fundamental solutions~$p_m$,
$k_m$ of the free Dirac equation, and the~$c_\alpha$ are combinatorial factors.
Here the Green's function~$s_m$ is the inverse of the Dirac operator,
\[ s_m(q) = \frac{1}{2} \: \lim_{\varepsilon \searrow 0} \sum_\pm
\frac{\slashed{q}+m}{q^2-m^2 \pm i \varepsilon q_0} \:, \]
where the pole is treated as a principal part.
Since~$U$ maps solutions of the free Dirac equation to solutions
in the external field, all the objects in~\eqref{Pdef} satisfy the Dirac equation,
\beq \label{DirB}
(i \Pdd + \B - m) P(x,y) = 0 \:,\qquad
(i \Pdd + \B - m) \tilde{\Psi}_k = 0 = (i \Pdd + \B - m) \tilde{\Phi}_l \:.
\eeq

\begin{figure} %
\begin{picture}(0,0)%
\includegraphics{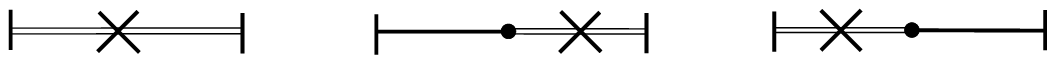}%
\end{picture}%
\setlength{\unitlength}{1533sp}%
\begingroup\makeatletter\ifx\SetFigFont\undefined%
\gdef\SetFigFont#1#2#3#4#5{%
  \reset@font\fontsize{#1}{#2pt}%
  \fontfamily{#3}\fontseries{#4}\fontshape{#5}%
  \selectfont}%
\fi\endgroup%
\begin{picture}(14318,1726)(-2490,1272)
\put(338,2611){\makebox(0,0)[lb]{\smash{{\SetFigFont{11}{13.2}{\rmdefault}{\mddefault}{\updefault}$y$}}}}
\put(1163,2064){\makebox(0,0)[lb]{\smash{{\SetFigFont{11}{13.2}{\rmdefault}{\mddefault}{\updefault}$-$}}}}
\put(6070,2075){\makebox(0,0)[lb]{\smash{{\SetFigFont{11}{13.2}{\rmdefault}{\mddefault}{\updefault}$-$}}}}
\put(10988,2087){\makebox(0,0)[lb]{\smash{{\SetFigFont{11}{13.2}{\rmdefault}{\mddefault}{\updefault}$+$}}}}
\put(11813,2091){\makebox(0,0)[lb]{\smash{{\SetFigFont{11}{13.2}{\rmdefault}{\mddefault}{\updefault}$\cdots$}}}}
\put(-1905,1427){\makebox(0,0)[lb]{\smash{{\SetFigFont{11}{13.2}{\rmdefault}{\mddefault}{\updefault}$P^\text{vac}(x,y)$}}}}
\put(-1185,2582){\makebox(0,0)[lb]{\smash{{\SetFigFont{11}{13.2}{\rmdefault}{\mddefault}{\updefault}$-$}}}}
\put(7728,2582){\makebox(0,0)[lb]{\smash{{\SetFigFont{11}{13.2}{\rmdefault}{\mddefault}{\updefault}$-$}}}}
\put(4499,2582){\makebox(0,0)[lb]{\smash{{\SetFigFont{11}{13.2}{\rmdefault}{\mddefault}{\updefault}$-$}}}}
\put(2028,2567){\makebox(0,0)[lb]{\smash{{\SetFigFont{11}{13.2}{\rmdefault}{\mddefault}{\updefault}$x$}}}}
\put(5322,2589){\makebox(0,0)[lb]{\smash{{\SetFigFont{11}{13.2}{\rmdefault}{\mddefault}{\updefault}$y$}}}}
\put(10233,2607){\makebox(0,0)[lb]{\smash{{\SetFigFont{11}{13.2}{\rmdefault}{\mddefault}{\updefault}$y$}}}}
\put(6953,2582){\makebox(0,0)[lb]{\smash{{\SetFigFont{11}{13.2}{\rmdefault}{\mddefault}{\updefault}$x$}}}}
\put(-2475,2624){\makebox(0,0)[lb]{\smash{{\SetFigFont{11}{13.2}{\rmdefault}{\mddefault}{\updefault}$x$}}}}
\end{picture}%
\caption{A few diagrams of the causal perturbation expansion of the Dirac sea.}
\label{figcausal}
\end{figure}

For what follows, it is very useful to represent the contributions to the causal perturbation
graphically. To this end, we denote every factor~$s_m$ by a straight line. The factors~$p_m$ and~$k_m$,
on the other hand, are depicted by a double line. For distinction, we sometimes add a symbol~$p$
or~$k$, or else we add symbols~``$+$'' for~$(p_m+k_m)/2$ (``positive frequency'')
and~``$-$'' for~$(p_m-k_m)/2$ (``negative frequency''). Every factor~$\B$ is depicted by a point.
Before the first factor and after the last factor in~\eqref{Pexform}, we put
for clarity a delimiter~$|$.
Moreover, we clarify the position of the factor~$P^{(0)}$
in the representation~\eqref{PseaU} by adding to the corresponding line a mark~$\times$.
The delimiters~$|$ and the marks~$\times$ can be viewed as a symbolizing
the ``ket-bra''-notation of a projector $| \Psi \ket \bra \Psi |$.
Finally, we add position and momentum variables when needed.
As an example, Figure~\ref{figcausal} gives a representation of the first terms of the perturbation
expansion of~$P^\text{sea}(x,y)$,
\[ P^\text{sea} = P^\text{vac} - s_m \,\B\: \frac{p_m-k_m}{2}
- \frac{p_m-k_m}{2} \:\B\,  s_m + \cdots \:. \]
In order to distinguish the contribution by~$P^\text{vac}$ in~\eqref{P0def}
from the contribution by the particle and anti-particle wave functions, we sometimes
emphasize the latter contributions by additional symbols~$\Psi_k$, $\Phi_l$
and~$\overline{\Psi_k}$, $\overline{\Phi_l}$.
Note that the resulting diagrams can be viewed as Feynman tree diagrams of a specific form.
Moreover, we remark that the factors~$p_m$ and~$k_m$ are always on-shell,
whereas the Green's functions~$s_m$ have off-shell contributions.
We finally point out that, due to current conservation, the probability
scalar product in~\eqref{normalize} is time independent even in the presence of the
interaction. The interacting wave functions~\eqref{psitilde} should
still be orthonormalized according to~\eqref{normalize},
\[ \int_{\R^3} (\overline{\tilde{\Psi}_k} \gamma^0 \tilde{\Psi}_l)(t, \vec{x}) \, d\vec{x} = \frac{1}{2 \pi}\,\delta_{kl}
= \int_{\R^3} (\overline{\tilde{\Phi}_k} \gamma^0 \tilde{\Phi}_l)(t, \vec{x}) \, d\vec{x} \:. \]

\subsection{Coupling to the Classical Bosonic Field Equations} \label{seccoup}
As worked out in~\cite{sector, lepton}, taking the continuum limit of the causal action principle gives rise to
classical bosonic field equations. For notational simplicity, we here restrict attention to
one abelian bosonic field and write the field equations symbolically as
\beq \label{field}
j^k[\B] - M^2 A^k[\B] = \lambda \,J^k\:,
\eeq
where~$J^k$ is the Dirac current, $A[\B]$ the bosonic potential,
$j^k[\B] = \partial^k_{\;l} A^l - \Box A^k$ the corresponding bosonic current,
$M$ the bosonic mass, and~$\lambda$ the coupling constant
(the generalization to several bosonic fields as considered in~\cite{lepton, quark} 
and to non-abelian gauge fields is straightforward).
In the vectorial case, the Dirac current takes the form
\beq \label{Jdef}
J^i(x) = \sum_{k=1}^{\np} \overline{\tilde{\Psi}_k(x)} \gamma^i \tilde{\Psi}_k(x) 
- \sum_{l=1}^{\na} \overline{\tilde{\Phi}_l(x)} \gamma^i \tilde{\Phi}_l(x) \:,
\eeq
where~$\tilde{\Psi}_k$ and~$\tilde{\Phi}_l$ are the particle and anti-particle wave functions
in~\eqref{Pdef}, respectively (the formulas for chiral or axial currents are analogous).
As shown in~\cite[Section~8]{sector}, the field equations~\eqref{field} come with
several correction terms, including corrections which correspond to the
vacuum polarization (see~\cite[\S8.2]{sector}). In order to take these corrections into
account, it is useful to write the field equations~\eqref{field} in the symbolic form
\beq \label{field2}
j^k[\B] - M^2 A^k[\B] = -\lambda \Tr_{\C^4} \!\big( \gamma^k P(x,x) \big)
- \text{(singular contributions)}\:.
\eeq
Here the ``singular contributions'' denote contributions to the fermionic projector which
are singular on the light cone but drop out of the Euler-Lagrange equations corresponding to the causal action
principle. Moreover, these singular contributions include the corrections to the field
equations worked out in~\cite{sector} which have the form of convolution terms (see~\cite[eq.~(1.1)]{sector}).
For what follows, the specific form of the singular contributions will not be used.
They can be identified with the counter terms needed in QFT in order to
regularize the Dirac current and to make the fermionic loop diagrams finite.
However, it is a major advantage of our approach that these counter terms are not
introduced ad-hoc, but come out of the analysis of the continuum limit
(for details see~\cite{srev}).

\begin{figure} %
\begin{picture}(0,0)%
\includegraphics{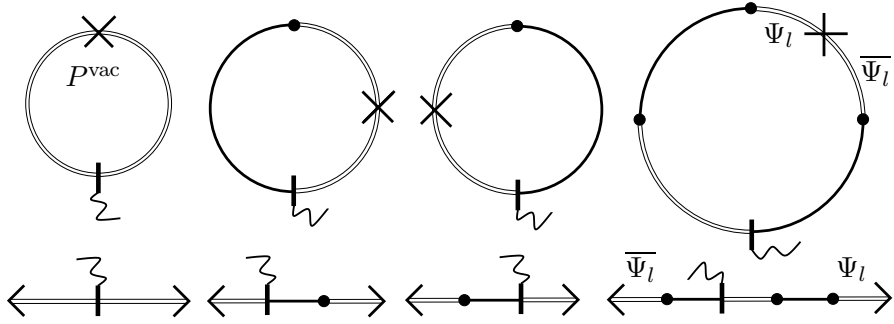}%
\end{picture}%
\setlength{\unitlength}{1533sp}%
\begingroup\makeatletter\ifx\SetFigFont\undefined%
\gdef\SetFigFont#1#2#3#4#5{%
  \reset@font\fontsize{#1}{#2pt}%
  \fontfamily{#3}\fontseries{#4}\fontshape{#5}%
  \selectfont}%
\fi\endgroup%
\begin{picture}(14339,5143)(-5094,-1314)
\put(-4137,2472){\makebox(0,0)[lb]{\smash{{\SetFigFont{11}{13.2}{\rmdefault}{\mddefault}{\updefault}$P^\text{vac}$}}}}
\put(8325,-605){\makebox(0,0)[lb]{\smash{{\SetFigFont{11}{13.2}{\rmdefault}{\mddefault}{\updefault}$\Psi_l$}}}}
\put(4914,-610){\makebox(0,0)[lb]{\smash{{\SetFigFont{11}{13.2}{\rmdefault}{\mddefault}{\updefault}$\overline{\Psi_l}$}}}}
\put(7149,3179){\makebox(0,0)[lb]{\smash{{\SetFigFont{11}{13.2}{\rmdefault}{\mddefault}{\updefault}$\Psi_l$}}}}
\put(8711,2553){\makebox(0,0)[lb]{\smash{{\SetFigFont{11}{13.2}{\rmdefault}{\mddefault}{\updefault}$\overline{\Psi_l}$}}}}
\end{picture}%
\caption{Circular diagrams and corresponding unfolded diagrams.}
\label{figcircular}
\end{figure}

In order to depict the field equations graphically, we close the contributions to the
fermionic projector to a circle and and draw the bosonic field by
a wiggly line (see the left of Figure~\ref{figcircular}).
Note that two delimiters~$|$ at the very left and right of the fermionic projector come together to form
one thick delimiter. We refer to the resulting diagram as a {\em{circular diagram}}.
Sometimes, it is more convenient to ``unfold'' the circular diagram at the line
involving the~$\times$ (see the right of Figure~\ref{figcircular}).
This so-called {\em{unfolded diagram}} has the advantage that, similar as in a scattering process,
the states at the very left and very right are the free states contained in the
fermionic projector~\eqref{P0def} before introducing the interaction.

\begin{figure} %
\begin{picture}(0,0)%
\includegraphics{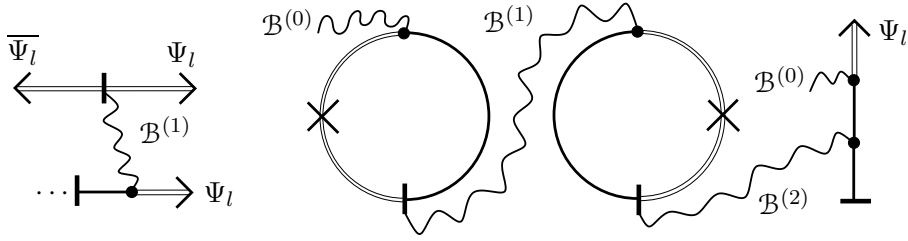}%
\end{picture}%
\setlength{\unitlength}{1533sp}%
\begingroup\makeatletter\ifx\SetFigFont\undefined%
\gdef\SetFigFont#1#2#3#4#5{%
  \reset@font\fontsize{#1}{#2pt}%
  \fontfamily{#3}\fontseries{#4}\fontshape{#5}%
  \selectfont}%
\fi\endgroup%
\begin{picture}(14108,3795)(-5189,-3396)
\put(-3036,-1885){\makebox(0,0)[lb]{\smash{{\SetFigFont{11}{13.2}{\rmdefault}{\mddefault}{\updefault}$\B^{(1)}$}}}}
\put(-1995,-2840){\makebox(0,0)[lb]{\smash{{\SetFigFont{11}{13.2}{\rmdefault}{\mddefault}{\updefault}$\Psi_l$}}}}
\put(2523,-94){\makebox(0,0)[lb]{\smash{{\SetFigFont{11}{13.2}{\rmdefault}{\mddefault}{\updefault}$\B^{(1)}$}}}}
\put(-4719,-2811){\makebox(0,0)[lb]{\smash{{\SetFigFont{11}{13.2}{\rmdefault}{\mddefault}{\updefault}$\cdots$}}}}
\put(-5174,-550){\makebox(0,0)[lb]{\smash{{\SetFigFont{11}{13.2}{\rmdefault}{\mddefault}{\updefault}$\overline{\Psi_l}$}}}}
\put(-2610,-583){\makebox(0,0)[lb]{\smash{{\SetFigFont{11}{13.2}{\rmdefault}{\mddefault}{\updefault}$\Psi_l$}}}}
\put(6968,-3050){\makebox(0,0)[lb]{\smash{{\SetFigFont{11}{13.2}{\rmdefault}{\mddefault}{\updefault}$\B^{(2)}$}}}}
\put(6904,-1071){\makebox(0,0)[lb]{\smash{{\SetFigFont{11}{13.2}{\rmdefault}{\mddefault}{\updefault}$\B^{(0)}$}}}}
\put(8904,-220){\makebox(0,0)[lb]{\smash{{\SetFigFont{11}{13.2}{\rmdefault}{\mddefault}{\updefault}$\Psi_l$}}}}
\put(-1061,-150){\makebox(0,0)[lb]{\smash{{\SetFigFont{11}{13.2}{\rmdefault}{\mddefault}{\updefault}$\B^{(0)}$}}}}
\end{picture}%
\caption{Diagrams of the perturbation expansion in a classical bosonic field.}
\label{fignomix}
\end{figure}

These diagrams are also useful for depicting the perturbation expansion of
the coupled system of partial differential equations~\eqref{DirB} and~\eqref{field}. To this end,
we employ the ansatz
\beq \label{Bpower}
\B = \sum_{p=0}^\infty \lambda^p\, \B^{(p)} \:,
\eeq
and expand the resulting fermionic projector~\eqref{Uflow} in powers of~$\lambda$,
\[ P[\B] = \sum_{p=0}^\infty \lambda^p\, P^{(p)}\:. \]
Then the Dirac equations~\eqref{DirB} are satisfied by construction, whereas the field
equations~\eqref{field} become
\beq \label{jiter}
\hspace*{1.5cm} \bigg\{ \hspace*{-1.8cm}
\begin{split}
j_k[B^{(0)}] - M^2\, A_k[\B^{(0)}] &= 0 \\
j_k[\B^{(p+1)}] - M^2\, A_k[\B^{(p+1)}] &= -\Tr_{\C^4} \!\big( \gamma_k P^{(p)}(x,x) \big) - \text{(s.c.)} \:,
\end{split}
\eeq
where~(s.c.) again denotes the singular contributions as well as the convolution terms.
This system can be solved iteratively with propagator methods. More precisely, one
fixes the gauge of the bosonic field (we do not enter the details because
the procedure is standard and depends on the specific form of the bosonic fields under consideration).
Then one multiplies the field equation~\eqref{jiter} by a corresponding bosonic Green's function~$S_0$
and solves for~$\B^{(p+1)}$, i.e.\ symbolically
\[ \B^{(p+1)} = S_0 \Big( -\Tr_{\C^4} \!\big( \gamma_k P^{(p)}(x,x) \big) + \text{(s.c.)} \Big) \]
(the choice of Green's function will be specified in Section~\ref{secgreen} below).
Generally speaking, the resulting perturbation expansion
involves fermionic loop diagrams, but no bosonic loops.
Using the methods and results in~\cite{sector}, one sees that all the diagrams of this expansion are finite.
Figure~\ref{fignomix} shows a few examples for diagrams.
For an electromagnetic interaction, the diagram on the left corresponds to the analytic expressions
\[ \iint_{M \times M} d^4y\: d^4z\: s_m(x,y) \gamma_i \Psi_l(y)\; S_0^{ij}(y,z)\;
\overline{\Psi}_l(z) \gamma_j \Psi_l(z) \]
whereas the diagram to the right corresponds to
\begin{align*}
&\int d^4y \int d^4z \int d^4u \int d^4v \int d^4\zeta \;
s_m(x,y) \gamma_i s_m(y,z)  \slashed{A}^{(0)}(z) \Psi_l(z)\; S_0^{ij}(y, u) \\
& \;\;\times \Tr \Big( P^\text{vac}(v,u) \gamma_j s_m(u,v) \gamma_k \Big)\; S_0^{kl}(v,w) \;
\Tr \Big( P^\text{vac}(w, \zeta) \slashed{A}^{(0)}(\zeta) s_m(\zeta,w) \gamma_l \Big)
\end{align*}
(where for ease in notation we left out the singular counter terms (s.c.)).

The above expansion differs from the usual perturbation expansion of QFT in two major points:
First, the bosonic loop diagrams are missing. Second, the bosonic field generated by
a wave function~$\Psi_l$ has a back-reaction to the same wave function~$\Psi_l$
(see the diagram on the left hand side of Figure~\ref{fignomix}).
In the next sections, we explain how to overcome these shortcomings.

\section{Microscopic Mixing of the Wave Functions} \label{secmicmix}
\subsection{Why Microscopic Mixing?} \label{secwhy}
The state stability analysis in~\cite[\S5.6]{PFP} and~\cite{reg} makes precise why
a configuration of vacuum Dirac seas is a stable minimizer of the causal action principle.
We now analyze how the particle and anti-particle states in~\eqref{P0def} change the
action. This analysis will reveal that~\eqref{P0def} is not the correct ansatz for
the non-interacting fermionic projector involving matter, making it necessary to
introduce the so-called {\em{microscopic mixing of the wave functions}}.

In preparation, we recall a few methods and results of the state stability analysis
(for details see~\cite{reg}).
In the state stability analysis, one considers the class of homogeneous fermionic projectors,
making it possible to work with a Fourier representation of the form
\[ P^\varepsilon(x,y) = \int \frac{d^4k}{(2 \pi)^4}\: \: \hat{P}^\varepsilon(k)\: e^{-ik(x-y)}\:, \]
where~$\varepsilon$ denotes the length scale of an ultraviolet regularization.
The causal action principle involves a double integral over space-time.
Due to homogeneity, one of the space-time integrals of the causal action gives an infinite constant.
Leaving out this integral, the resulting functional
\[ \Sact = \int \L[A^\varepsilon_{xy}]\: d^4 y \qquad \text{with} \qquad
A^\varepsilon_{xy} = P^\varepsilon(x,y)\, P^\varepsilon(y,x) \]
can be understood as the action per unit volume of Minkowski space.
The first variation of the action can be written as
\beq \label{dS}
\delta \Sact(k) = \Tr \big( \hat{Q}^\varepsilon(k)\, \delta P^\varepsilon(k) \big) ,
\eeq
where the operator~$\hat{Q}^\varepsilon$ is a convolution in momentum space,
\beq \label{MP}
\hat{Q}^\varepsilon(q) = \frac{1}{2}\: (\hM^\varepsilon * \hat{P}^\varepsilon)(q)
= \frac{1}{2} \int \frac{d^4p}{(2 \pi)^4}\: \hM^\varepsilon(p)\:
\hat{P}^\varepsilon(q-p)
\eeq
(and~$\hM^\varepsilon$ is the gradient of the Lagrangian transformed to momentum space).
In~\cite[\S5.6]{PFP} and~\cite{reg} it is shown that by working specific regularizations
(which, technically speaking, have the property of a distributional ${\mathcal{M}} P$-product),
one can arrange that the convolution integral~\eqref{MP} is well-defined and finite
if~$q$ lies in the lower mass cone. More specifically (for details see~\cite[Theorem~2.3 (1)--(3)]{reg}),
\beq \label{Qfinite}
m^5 \lesssim \| \lim_{\varepsilon \searrow 0} \hat{Q}^\varepsilon(q) \|
< \infty \qquad \text{for} \qquad q \in {\mathcal{C}}^\wedge := \{ q \:|\: q^2>0 \text{ and } q^0<0 \}\:,
\eeq
where~$\| . \|$ denotes any norm on the $4 \times 4$-matrices.
We remark that this technical result is the basis for the detailed state stability analysis in~\cite{vacstab},
where it is studied if and how the minima of the eigenvalues of~$\hat{Q}^\varepsilon(q)$
can be arranged to lie precisely on the mass shells of the occupied states of the system.
For the following arguments, however, we only use that~$\hat{Q}^\varepsilon(q)$ is finite inside the
lower mass shell.
Moreover, we need the result proven in~\cite[Theorem~2.3 (4)]{reg} that~$\hat{Q}^\varepsilon(q)$
is infinite for~$q$ in the upper mass shell. More specifically, we know from~\cite[Theorem~5.8]{vacstab}
that
\[ \hM(k) \sim \frac{k\slsh}{k^2} + \O(k^{-2}) \:. \]
Using this result in~\eqref{MP}, we obtain the scaling
\beq \label{Qinfinite}
\| \hat{Q}^\varepsilon(q) \| \sim m^3 \varepsilon^{-2} \qquad \text{for} \qquad q \in {\mathcal{C}}^\vee
:= \{ q \:|\: q^2>0 \text{ and } q^0>0 \}\: .
\eeq

Applying the above formulas to the fermionic projector~\eqref{P0def} gives the following results.
First, one should keep in mind that the particle and anti-particle states can be treated as
first order variations (for details see~\cite[\S5.6]{PFP}).
The anti-particle states~$\Phi_l$ are supported on the lower mass shell. Thus, according to~\eqref{Qfinite}
and~\eqref{dS}, they give a finite contribution to~$\delta \Sact$.
More precisely, working as in~\cite[\S5.6]{PFP} with discrete states in a three-dimensional
box of volume~$V$, we obtain
\beq \label{Sfinite}
(\delta \Sact)[\Phi_l] \simeq c\, m^5
\eeq
with~$c = m /(|k^0|\, V)$.
The particle state~$\Psi_k$, however, are supported on the upper mass shell.
Thus, according to~\eqref{Qinfinite}, they give an infinite positive contribution to~$\delta \Sact$,
\beq \label{Sdiv}
(\delta \Sact)[\Psi_l] \simeq c\, m^3\, \varepsilon^{-2} \:.
\eeq
This contribution diverges as~$\varepsilon \searrow 0$, showing that the
configuration~\eqref{P0def} involving particles and anti-particles is not a minimizer of our action principle.

\subsection{Microscopic Mixing in an Explicit Example} \label{secmixex}
We now explain in the simplest possible example how microscopic mixing of the wave functions
can be used to scale down the undesirable contribution to the action~\eqref{Sdiv}.
We consider a system involving only one particle described by the wave function~$\Psi$.
Moreover, we assume that the Dirac sea~$P^\text{vac}$ is built up of a finite number
of wave functions~$\psi_1, \ldots, \psi_N$.
This corresponds to an ultraviolet regularization which we denote symbolically
by a superscript~$\varepsilon$ (where as in~\cite{PFP, sector} $\varepsilon$ stands for the
regularization length). Then the ansatz~\eqref{P0def} simplifies to
\[ P^\varepsilon(x,y) = -\sum_{n=1}^N \psi_n(x) \,\overline{\psi_n(y)} - \Psi(x) \,\overline{\Psi(y)} \]
(this ansatz can be realized for example by considering the system in finite $3$-volume
with an ultraviolet regularization by a momentum cutoff).
Setting~$\psi_0 = \Psi$, we can write this formula in the more compact form
\[ P^\varepsilon(x,y) = -\sum_{n=0}^N \psi_n(x) \,\overline{\psi_n(y)} \:. \]
We now subdivide Minkowski space~$M$ into sets~$M_1, \ldots, M_\Lmix$, meaning that
\[ M = M_1 \cup \cdots \cup M_\Lmix\qquad \text{and} \qquad M_a \cap M_b = \varnothing \quad
\text{if $a \neq b$}\:. \]
We assume that the number of subsystems is small compared to the total number of particles
(including the sea states),
\beq \label{Lbound}
\Lmix \ll f := N+1\:.
\eeq
In the $a^\text{th}$ subsystem, we want to exchange the particle wave function~$\Psi$ with
the~$a^\text{th}$ sea state. Thus we let~$\sigma_a$ be the operator
which permutes~$0$ with~$a$,
\[ \sigma_a(n) = \left\{ \begin{array}{ll}
a \quad & \text{if~$n=0$} \\
0 & \text{if~$n=a$} \\
n & \text{otherwise} \end{array} \right.\qquad \text{where} \qquad a \in \{1, \ldots, \Lmix\},\:
n \in \{1, \ldots, N\} \]
We set
\begin{align}
P^\varepsilon(x,y) &= \sum_{a,b=1}^\Lmix \chi_{M_a}(x)\,P^{a,b}(x,y)\, \chi_{M_b}(y) \label{Pmic1} \:,
\intertext{where~$\chi_M$ denotes the characteristic function and}
P^{a,b}(x,y) &= -\sum_{n=0}^N \psi_{\sigma_a(n)}(x) \,\overline{\psi_{\sigma_b(n)}(y)} \:. \label{Pmic2}
\end{align}

Let us discuss this ansatz. We first point out that considering~$P^\varepsilon(x,y)$ as the integral
kernel of an operator~$P^\varepsilon$, the image of this operator is spanned by the $N+1$ vectors
\[ \psi_n(x) = \sum_{a=1}^\Lmix \chi_{M_a}(x)\, \psi_{\sigma_a(n)}(x)\:,\qquad n=0,\ldots, N\:. \]
In particular, microscopic mixing does not change the rank of~$P^\varepsilon$.
If the space-time points~$x$ and~$y$ are in the same subsystem,
we can reorder the $n$-summands to obtain the fermionic projector without microscopic
mixing,
\[ P^\varepsilon(x,y) = -\sum_{n=1}^N \psi_n(x) \,\overline{\psi_n(y)} - 
\psi_0(x) \,\overline{\psi_0(y)} \:. \]
However, if the space-time points are in different subsystems, then microscopic mixing
changes the fermionic projector to
\[ P^\varepsilon(x,y) = - \!\!\!\!\!\sum_{n \neq \{0,a,b\} }^N \psi_n(x) \,\overline{\psi_n(y)} 
\:-\: \psi_0(x) \,\overline{\psi_a(y)}
- \psi_b(x) \,\overline{\psi_0(y)} - \psi_a(x) \,\overline{\psi_b(y)}\:, \]
valid if~$x \in M_a$ and~$y \in M_b$ with~$a \neq b$.
In order to analyze how microscopic mixing effects the action, we decompose~$P(x,y)$
similar to~\eqref{P0def} as
\beq \label{Pdec}
P^\varepsilon(x,y) = P^\text{vac}(x,y) + \delta P^\varepsilon(x,y)\:,
\eeq
where
\begin{gather}
\delta P^\varepsilon(x,y) = \sum_{a,b=1}^\Lmix \chi_{M_a}(x) \:\delta P^{a,b}(x,y)\: \chi_{M_b}(y) \label{delP} \\
\intertext{and}
\delta P^{a,b}(x,y) = \left\{ \begin{array}{cl}
-\psi_0(x) \,\overline{\psi_0(y)} & \quad \text{if~$a=b$} \\[0.6em]
\begin{minipage}{6cm}
$-\psi_0(x) \,\overline{\psi_a(y)} - \psi_b(x) \,\overline{\psi_0(y)} $ \\[0.3em]
$- \psi_a(x) \,\overline{\psi_b(y)} + \sum_{n=a,b} \psi_n(x) \,\overline{\psi_n(y)}$
\end{minipage}
&\quad \text{if~$a \neq b$\:.} \end{array} \right.  \label{delPab}
\end{gather}

Let us evaluate how the perturbation~$\delta P^\varepsilon$ in~\eqref{Pdec}
affects the action. To this end, we must evaluate~\eqref{delP}
and~\eqref{delPab} in the formula for~$\delta \Sact$ \eqref{dS}.
To this end, we assume for simplicity that all our wave functions are plane-wave solutions.
In~\eqref{delP}, the wave functions~$\psi_a$ are multiplied by
characteristic functions~$\chi_{M_b}$, and we need to compute their Fourier transform
to momentum space.
In order to evaluate this Fourier transform, we need to specify the sets~$M_a$.
Similar as explained in~\cite[Section~4.1]{entangle}, we assume that the
sets~$M_a$ are {\em{fine-grained on the regularization scale}} in the following sense.
First, every macroscopic region of space-time should intersect all the sets~$M_a$. Moreover,
we assume for simplicity that the sets~$M_a$ are distributed uniformly in space-time, meaning
that when we integrate any macroscopic function~$f$ over one of the sets~$M_a$, this integral can be
approximated by a constant times the Lebesgue integral,
\beq \label{weights}
\int_{M_a} f(x)\, d^4x = c_a \int_M f(x)\, d^4x + \text{(higher orders in~$\varepsilon/\ell_\text{macro}$)}\:.
\eeq
Here~$\varepsilon$ is the regularization length, and~$\ell_\text{macro}$ denotes the macroscopic
length scale on which~$f$ varies. The constants~$c_a>0$ can be understood as the relative densities of
the sets~$M_a$. Since the~$M_a$ form a partition, we know that
\[ \sum_{a=1}^\Lmix c_a = 1\:. \]

Using that the sets~$M_a$ are fine-grained on the regularization scale, we can compute
the Fourier transforms of the sets~$\chi_{M_b} \psi_a$ by
\[ \widehat{ \chi_{M_b} \psi_a }(k) = c_b\, \hat{\psi}_a(k) +
\text{(higher orders in~$\varepsilon/\ell_\text{macro}$)}\:. \]
Here we assumed that the wave functions~$\psi_0, \ldots, \psi_\Lmix$ vary only on the
macroscopic scale. In other words, the energy (= frequency) of these wave functions should be much
smaller than the Planck energy. In order to ensure that this condition can be satisfied,
we need the assumption~\eqref{Lbound}.
Hence the characteristic functions in~\eqref{delP} can be treated in~\eqref{dS}
by factors~$c_a$ and~$c_b$,
\beq \label{delSab}
\delta \Sact(k) = \sum_{a,b=1}^\Lmix c_a c_b \Tr \big( \hat{Q}^\varepsilon(k)\, \delta P^{a,b}(k) \big) +
\text{(higher orders in~$\varepsilon/\ell_\text{macro}$)}\:.
\eeq
This can be computed further using~\eqref{delPab}. If~$a=b$, microscopic mixing has no effect,
so that~\eqref{Sdiv} again applies.
In the case~$a \neq b$, the wave functions~$\psi_n$ with~$n=a,b$ are on the lower mass
shell, giving a finite contribution~\eqref{Sfinite}.
All the other contributions in~\eqref{delPab} are of the form~$\psi_c(x) \overline{\psi_d}(y)$
with~$c \neq d$. They drop out of~\eqref{delSab}, because the two involved wave functions have
different momenta. We conclude that
\[ \delta \Sact(k) \simeq c m^3\, \varepsilon^{-2} \sum_{a=1}^\Lmix c_a^2  + c m^5 \sum_{a \neq b} c_a \,c_b
+ \text{(higher orders in~$\varepsilon/\ell_\text{macro}$)}\:. \]
In order to make this contribution as small as possible, we choose~$c_a \sim 1/\Lmix$.
Then
\beq \label{gain}
\delta \Sact(k) \simeq \frac{c m^3}{\Lmix \,\varepsilon^2} + c m^5 
+ \text{(higher orders in~$\varepsilon/\ell_\text{macro}$)}\:.
\eeq
We conclude that when choosing many subsystems, the microscopic mixing~\eqref{Pmic1} and~\eqref{Pmic2}
indeed makes the divergent contribution to the action~\eqref{Sdiv} smaller.

Before discussing the scalings, we point to another mechanism which
will be important later on. So far, we had to assume that the number of subsystems was much
smaller than the number of particles~\eqref{Lbound}.
In view of~\eqref{gain}, it would be desirable to further increase the number of subsystems.
This can indeed be arranged by inserting suitable {\em{phase factors}} into~\eqref{Pmic2}.
In order to explain the idea, we choose phases~$\varphi_1, \ldots, \varphi_{\Lphase}$.
We label the subsystems by~$M_{a \alpha}$ with~$a \in \{1, \ldots, \Lmix\}$ (again subject to the
condition~\eqref{Lbound}) and~$\alpha \in \{1, \ldots, \Lphase \}$.
We modify~\eqref{Pmic1} and~\eqref{Pmic2} to
\beq \label{Paalpha}
\begin{split}
P^\varepsilon(x,y) &= \sum_{a,b=1}^\Lmix \sum_{\alpha, \beta=1}^{\Lphase}
\chi_{M_{a \alpha}}(x)\,P^{a \alpha,b \beta}(x,y)\, \chi_{M_{b \beta}}(y) \\
P^{a \alpha,b \beta}(x,y) &= -\sum_{n=0}^N \psi^{(\alpha)}_{\sigma_a(n)}(x)
\,\overline{\psi^{(\beta)}_{\sigma_b(n)}(y)} \:,
\end{split}
\eeq
where the index~$(\alpha)$ denotes that the particle wave function~$\Psi$ is
multiplied by the phase factor~$e^{i \varphi_\alpha}$, i.e.
\[ \psi^{(\alpha)}_0 = e^{i \varphi_\alpha} \Psi \qquad \text{and} \qquad
\psi^{(\alpha)}_n = \psi_n \quad \text{for~$n=1,\ldots, f$} \:. \]
This microscopic mixing again leaves the rank of~$P^\varepsilon$ unchanged.
Moreover, one should keep in mind that the phase factors only modify~$\psi_0$.
As a consequence, the decomposition~\eqref{Pdec} remains valid
if~\eqref{delP} and~\eqref{delPab} are modified to
\begin{align*}
\delta P^\varepsilon(x,y) &= \sum_{a,b=1}^\Lmix
\sum_{\alpha, \beta=1}^{\Lphase} \chi_{M_{a \alpha}}(x) \:\delta P^{a \alpha,b \beta}(x,y)\:
\chi_{M_{b \beta}}(y) \\
\delta P^{a \alpha,b \beta}(x,y) &= \left\{ \begin{array}{cl}
-e^{i (\varphi_\alpha - \varphi_\beta)}\: \psi_0(x) \,\overline{\psi_0(y)} & \quad \text{if~$a=b$} \\[0.6em]
\begin{minipage}{6.5cm}
$-e^{i \varphi_\alpha} \psi_0(x) \,\overline{\psi_a(y)} -
e^{-i \varphi_\beta} \psi_b(x) \,\overline{\psi_0(y)} $ \\[0.3em]
$- \psi_a(x) \,\overline{\psi_b(y)} + \sum_{n=a,b} \psi_n(x) \,\overline{\psi_n(y)}$
\end{minipage}
&\quad \text{if~$a \neq b$\:.} \end{array} \right.
\end{align*}
In order to get into the position to compute the sums over the subsystems, we assume that
the phases~$\varphi_\alpha$ are randomly distributed. Then
\[ \sum_{\alpha=1}^{\Lphase} e^{i \varphi_\alpha} \eqsim \frac{1}{\sqrt{\Lphase}} \:. \]
This improves the scaling in~\eqref{gain} to
\beq \label{gain2}
\delta \Sact(k) \eqsim \frac{c m^3}{\Lmix \,\Lphase\, \varepsilon^2} + c m^5 
+ \text{(higher orders in~$\varepsilon/\ell_\text{macro}$)}\:.
\eeq

Let us consider whether the above constructions really make it possible to
remove the divergence of the contribution~\eqref{Sdiv} to the action.
For the cancellations of the terms with~$(a \alpha) \neq (b \beta)$, it
is essential that the wave function~$\chi_{M_{a \alpha}} \Psi$ restricted to
the subsystem~$M_{a \alpha}$ is orthogonal to all the sea states.
Suppose that we consider a discrete space-time with a finite number of
space-time points $\# M < \infty$ (like for example a finite space-time lattice).
Then the maximal number of orthogonal states scales like the number of space-time points.
This gives the following upper bound for the number of subsystems,
\beq \label{LLM}
\Lmix \,\Lphase \lesssim \#M \:.
\eeq
If this scaling is respected, by suitably modifying the wave functions on the
regularization scale we can arrange that the wave functions~$\chi_{M_{a \alpha}} \Psi$ are indeed
orthogonal to all the sea states, implying that the error term in~\eqref{gain2}
vanishes identically. With this in mind, we may disregard this error term in what follows.
In order to determine how~$\#M$ scales in the regularization length~$\varepsilon$,
it is easiest to consider the example of a finite lattice with lattice spacing~$\varepsilon$
(in this situation, variations on the lower mass shell
stay bounded in the limit~$\varepsilon \searrow 0$ according to~\eqref{Sfinite}).
Keeping the total volume of space-time fixed, we get the scaling
\beq \label{Meps}
\# M \simeq \varepsilon^{-4}\:.
\eeq
Using~\eqref{Meps} with~\eqref{LLM} in~\eqref{gain2}, we conclude that
the mechanism of microscopic mixing makes it possible to arrange that~$\delta \Sact$
stays finite in the limit~$\varepsilon \searrow 0$.
More specifically, the number of subsystems can vary in the range
\beq \label{Lrange}
\varepsilon^{-2} \lesssim \Lmix \,\Lphase \lesssim \varepsilon^{-4}\:.
\eeq
In what follows, we will treat~$\Lmix$ and~$\Lphase$ as parameters describing
the unknown microscopic behavior of space-time.
They should comply with the scalings~\eqref{Lbound} and~\eqref{Lrange}
but will remain undetermined otherwise.

\subsection{The General Construction for Free Fields} \label{secfree}
We now work out the mechanism of microscopic mixing systematically for systems
involving particles and anti-particles. Our starting point is the fermionic projector~\eqref{P0def}.
Introducing an ultraviolet regularization involving a finite number of sea states~$N$,
we write~$P^\text{vac}$ as
\[ P^\text{vac}(x,y) = -\sum_{n=1}^N \psi_n(x) \,\overline{\psi_n(y)} \:. \]
Considering on the solutions of the Dirac equation the usual scalar product
\begin{equation} \label{print}
(\psi | \phi) = \int_{t=\text{const}} \Sl \psi(t,\vec{x}) \,|\, \gamma^0 \phi(t,\vec{x}) \Sr \:d^3x\:,
\end{equation}
we obtain a Hilbert space of dimension~$N$, which we denote by~${\mathscr{S}}_0$
(referred to as the {\em{sea space}}; the subscript~$0$ clarifies that we here consider solutions of the
free Dirac equation). Similarly, the vectors in the image of the fermionic projector~\eqref{P0def}
span the Hilbert space~${\mathscr{H}}_0$.
It can be regarded as being composed of all the occupied states of the physical system.
Since the anti-particle states have been removed, its dimension
is given by~$f := \dim \H_0 = N + \np - \na$.

We denote the spinor space by~$(S \simeq \C^4, \Sl .|. \Sr)$.
Introducing the {\em{fermion matrix}} by
\[ \Psi(x) \::\: {\mathscr{H}}_0 \rightarrow \C^4 \:,\qquad
\Psi(x)\, \phi = \phi(x) \:, \]
we can write~\eqref{P0def} as
\[ P^\varepsilon(x,y) = \Psi(x) \Psi(y)^* \:, \]
where the star denotes the adjoint (where on the spinors we clearly take the
adjoint with respect to the spin scalar product, i.e.~$\psi(x)^* = \psi(x)^\dagger \gamma^0$).

We now introduce the microscopic mixing in generalization of~\eqref{Paalpha} by
\begin{align}
P^\varepsilon(x,y) &= \sum_{\as, \bs \in {\mathfrak{M}}} \chi_{M_\as}(x)
\: P^{\as, \bs}(x,y) \:\chi_{M_\bs}(y) 
\label{Pchar} \\
P^{\as, \bs}(x,y) &= -\Psi(x) \, V_\as  \,V_\bs^* \, \Psi(y)^*\:, \label{Pmixansatz}
\end{align}
where~$V_\as \in \U({\mathscr{H}}_0)$ are unitary operators on~$\H_0$, and
${\mathfrak{M}}$ is an index set for the subsystems
(for notational convenience, we combined the latin and greek indices in~\eqref{Paalpha}
to one gothic letter; i.e.~$\as=(a \alpha)$ and~$\bs=(b \beta)$).
In order to specify the operators~$V_\as$, we choose a subspace~$\I_0 \subset \H_0$
(referred to as the {\em{mixing space}}).
A natural choice is to take~$\I_0 = \bra \Psi_1, \ldots, \Psi_{\np} \ket$ as the
span of the particle states. More generally, we could choose~$\I_0$ as a subspace of~$\H_0$
which contains the particle states with~$\dim \I_0 \ll \dim \H_0$. The freedom in choosing~$\I_0$
will be discussed in detail in Section~\ref{secanti}.
Here we merely assume that the dimension of~$\I_0$ is much smaller than that of~$\H_0$,
\[  n := \dim \I_0 \ll \dim \H_0 \:. \]
We denote the orthogonal complement of~$\I_0$ by~${\mathscr{N}}_0$, giving rise to the
direct sum decomposition
\[ \H_0 = \I_0 \oplus {\mathscr{N}}_0\:. \]
The form of the unitary operators~$V_\as$ can be derived from the following guiding principles:
First, we want that all the particle states are mixed with the sea states,
meaning that the subspace~$\I_0$ should be mapped to~${\mathscr{N}}_0$.
Second, microscopic mixing should not change the singularity structure of the distribution~$P(x,y)$
on the light cone. This means that~$V_\as$ should leave as many sea states as possible
unchanged. These assumptions are made precise in the next lemma. In preparation,
we choose a subspace~$\J_0 \subset {\mathscr{N}}_0$
which has the same dimension~$n$ as~$\I_0$.
Setting~${\mathscr{K}}_0 = \J_0^\perp \subset {\mathscr{N}}_0$, we obtain the direct
sum decomposition
\beq \label{dirsum2}
{\mathscr{T}}_0 = \I_0 \oplus \J_0 \oplus {\mathscr{K}}_0 \:.
\eeq

\begin{Lemma}
Suppose that a unitary operator~$V \in \U({\mathscr{T}}_0)$ has the following properties:
\begin{itemize}
\item[(i)] $V$ maps~$\I_0$ to an orthogonal subspace, i.e.
\[ (\phi | U \phi) = 0 \qquad \text{for all~$\phi \in \I_0$}\:. \]
\item[(ii)] There is a subspace~${\mathscr{L}}_0 \subset {\mathscr{K}}_0$
of dimension~$N-2 \na - \np$ on which~$V$ is trivial,
\[ V|_{{\mathscr{L}}_0} = \1 \:. \]
\end{itemize}
Then in a block matrix representation corresponding to the direct sum decomposition~\eqref{dirsum2},
the operator~$V$ can be written as
\beq \label{Vans}
V = \begin{pmatrix} \1 & 0 & 0 \\ 0 & \Uran_{11} & \Uran_{12} \\ 0 & \Uran_{21} & \Uran_{22} \end{pmatrix}
\begin{pmatrix} 0 & \Wran & 0 \\ \1 & 0 & 0 \\ 0 & 0 & \1 \end{pmatrix} 
\begin{pmatrix} \1 & 0 & 0 \\ 0 & \Uran_{11} & \Uran_{12} \\ 0 & \Uran_{21} & \Uran_{22} \end{pmatrix}^{\!\!*} ,
\eeq
where
\beq \label{WUmap}
\Wran \in \U(\J_0, \I_0) \qquad \text{and} \qquad
\Uran \in \U({\mathscr{N}}_0)\:.
\eeq
\end{Lemma}
\Proof According to~(i), the subspace~${\mathscr{M}}_0 := V^{-1}(\I_0)$ is 
orthogonal to~$\I_0$ and is thus contained in~${\mathscr{N}}_0$.
Counting dimensions using~(ii), it follows that the subspace~${\mathscr{L}}_0$ coincides
with the orthogonal complement of~${\mathscr{M}}_0$ in~${\mathscr{N}}_0$.
Choosing an orthonormal basis~$(e_k)$ of~${\mathscr{M}}_0$
and choosing on~$\I_0$ the basis vectors~$V(e_k)$, 
in a block matrix notation corresponding to the direct sum
decomposition~${\mathscr{T}}_0 = \I_0 \oplus {\mathscr{M}}_0 \oplus {\mathscr{L}}_0$,
the operator~$V$ takes the form
\[ V = \begin{pmatrix} 0 & \Wran' & 0 \\ \1 & 0 & 0 \\ 0 & 0 & \1 \end{pmatrix} \]
with~$\Wran' \in \U({\mathscr{M}}_0, \I_0)$.
We now choose~$\Uran \in \U({\mathscr{N}}_0)$ such that it maps a given orthonormal basis of~$\J_0$
to the corresponding basis vectors~$e_k$ of~${\mathscr{M}}_0$.
\QED
Choosing orthonormal bases of~$\I_0$ and~$\J_0$, we shall always represent~$W$
as a unitary $n \times n$-matrix,
\[ \Wran \in \U(n)\:. \]
Moreover, we note that~\eqref{Vans} can also be written as
\beq \label{Vans2}
V = \pi_{\mathscr{N}_0} + \begin{pmatrix} \1 & 0 & 0 \\ 0 & \Uran_{11} & \Uran_{12} \\ 0 & \Uran_{21} & \Uran_{22} \end{pmatrix}
\begin{pmatrix} 0 & \Wran & 0 \\ \1 & -\1 & 0 \\ 0 & 0 & 0 \end{pmatrix} 
\begin{pmatrix} \1 & 0 & 0 \\ 0 & \Uran_{11} & \Uran_{12} \\ 0 & \Uran_{21} & \Uran_{22} \end{pmatrix}^{\!\!*} \:,
\eeq
where~$\pi_{\mathscr{N}_0}$ denotes the orthogonal projection to~${\mathscr{N}}_0$.
We will always choose the unitary operators~$V_\as$ in~\eqref{Pmixansatz} according
to~\eqref{Vans}.

The matrices~$\Wran$ and~$\Uran$ in our ansatz~\eqref{Vans} play a different role.
The matrix~$\Wran$ describes unitary transformations of the states in~$\I_0$.
Recalling that~$\I_0$ contains the free particle states, we can say that~$\Wran$
describes generalized phase transformations of the particle states.
The matrix~$\Uran$, on the other hand, determines with which states of the Dirac sea
the particle states are mixed. Using the same notation as in Section~\ref{secmixex}, we describe
the microscopic mixing by the collection of matrices
\[ \left\{ \begin{array}{ll}
\Uran_a \qquad & \text{for~$a=1,\ldots, \Lmix$} \\
\Wran_\alpha & \text{for~$\alpha=1,\ldots, \Lphase\:.$}
\end{array} \right. \]
The subsystems are then labeled by the corresponding composite index~$\as=(a \alpha)$.
Thus the total number~$L$ of subsystems~$M_\as$ is given by
\[ L = \Lmix\, \Lphase\:. \]
The number of subsystems should be large, as made precise by the
scaling~\eqref{Lrange}.
We choose~$\Wran_\alpha \in \U(n)$ as a random matrix,
taking the normalized Haar measure as the probability measure.
We postpone to specify the matrices~$\Uran_a$ until Section~\ref{secbackground}.

\subsection{Introducing the Interaction} \label{secinteract}
We now explain how the interaction is introduced in the presence of microscopic mixing.
Our starting point are the classical field equations~\eqref{field} and~\eqref{field2}.
In order to understand how microscopic mixing changes these equations, we need to
briefly reconsider their derivation in~\cite{sector}.
The field equations are obtained by evaluating the Euler-Lagrange equations corresponding to the causal action
principle weakly on the light cone. The important point for what follows is that this analysis involves
the fermionic projector~$P(x,y)$ away from the origin, i.e.\ for different arguments~$x \neq y$.
More precisely, the two space-time arguments of the fermionic projector have the following scaling,
\beq \label{xiscale}
\varepsilon \ll |x^0 - y^0|, |\vec{x}-\vec{y}| \ll \ell_\text{macro}
\eeq
(for details see~\cite[\S5.1]{sector}).
Let us consider what this means for the Dirac current~\eqref{Jdef}.
In an evaluation away from the origin, we need to replace the Dirac current by a
corresponding function of two arguments,
\[ J^i(y,x) := \sum_{k=1}^{\np} \overline{\tilde{\Psi}_k(y)} \gamma^i \tilde{\Psi}_k(x) 
- \sum_{l=1}^{\na} \overline{\tilde{\Phi}_l(y)} \gamma^i \tilde{\Phi}_l(x) \:. \]
If no microscopic mixing is present, the Dirac current~$J(y,x)$ is smooth in~$x$ and~$y$
and varies only on the macroscopic scale.
Hence the scaling~\eqref{xiscale} makes it possible to replace~$J(y,x)$ by the vector field~$J(x) =J(x,x)$,
up to errors of the order~$|\vec{\xi}|/\ell_\text{macro}$ (for detail see~\cite[Chapter~7]{sector}).
If microscopic mixing is present, however, the Dirac current does vary on the microscopic scale,
because it depends on the subsystems to which~$x$ and~$y$ belong.
Thus the Dirac current is fine-grained on the regularization scale. This implies that
the field equations~\eqref{field} and~\eqref{field2} must be modified. Qualitatively speaking, we must
build in an ``averaging process'' over the subsystems.

Before working out the resulting field equations quantitatively,
we note that, as the wave functions before microscopic mixing are macroscopic,
the resulting integrals over the subsystem can all be evaluated as in~\eqref{weights}
using relative weights of the subsystems. With this in mind, in the subsequent considerations
we may disregard the characteristic functions~$\chi_{M_{a}}$ of the subsystems.
Instead, we consider the subsystems separately, and then ``average over
the subsystems'' by taking weighted sums, similar as explained before~\eqref{delSab}.

We first explain our method in a first order perturbation expansion.
Recall that, in the vacuum and without microscopic mixing, the Dirac current~\eqref{Jdef}
can be expressed in terms of the fermionic projector~\eqref{P0def} as
\beq \label{curcomp}
J_k = -\Tr(\gamma^k P^{(0)}(x,x)) + \Tr(\gamma^k P^\text{sea}(x,x)) \:.
\eeq
With microscopic mixing, the corresponding Dirac current depends on a pair
of subsystems~$\as$ and~$\bs$. We denote it by~$J^{\bs|\as}$. 
In view of~\eqref{Pmixansatz}, we obtain similar to~\eqref{curcomp}
\beq \label{Jmix}
J^{\bs |\as }_k = \Tr_{\H_0} \!\big( V_\bs^* \, \Psi(x)^* \gamma_k \Psi(x) \, V_\as \big)
-\Tr_{{\mathscr{S}}_0} \!\big( \Psi(x)^* \gamma_k\Psi(x) \big) \:.
\eeq
If~$\as=\bs$, the unitary transformations drop out, and we obtain precisely~\eqref{curcomp}.
In the case~$\as \neq \bs$, however, we obtain a more complicated expression.
Note that, in view of our ansatz~\eqref{Vans2}, the operators~$V_\as$ and~$V_\bs$
are trivial except on a subspace of dimension~$2 n$.
This implies that in~\eqref{Jmix}, many terms drop out.
More precisely, at most~$5 n$ summands remain.
In particular, the differences of traces in~\eqref{Jmix} stays well-defined if the number of sea
states tends to infinity.

\begin{figure} %
\begin{picture}(0,0)%
\includegraphics{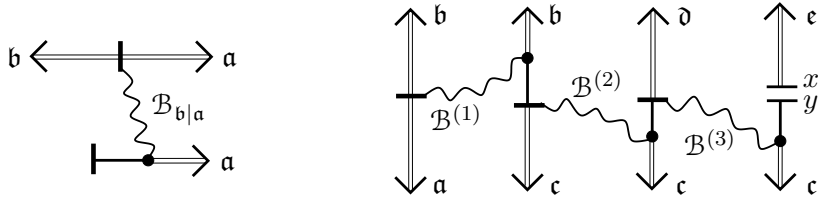}%
\end{picture}%
\setlength{\unitlength}{1533sp}%
\begingroup\makeatletter\ifx\SetFigFont\undefined%
\gdef\SetFigFont#1#2#3#4#5{%
  \reset@font\fontsize{#1}{#2pt}%
  \fontfamily{#3}\fontseries{#4}\fontshape{#5}%
  \selectfont}%
\fi\endgroup%
\begin{picture}(12951,3287)(-5459,-3383)
\put(-1995,-2840){\makebox(0,0)[lb]{\smash{{\SetFigFont{11}{13.2}{\rmdefault}{\mddefault}{\updefault}$\as$}}}}
\put(-1964,-1156){\makebox(0,0)[lb]{\smash{{\SetFigFont{11}{13.2}{\rmdefault}{\mddefault}{\updefault}$\as$}}}}
\put(-5444,-1171){\makebox(0,0)[lb]{\smash{{\SetFigFont{11}{13.2}{\rmdefault}{\mddefault}{\updefault}$\bs$}}}}
\put(-3108,-1936){\makebox(0,0)[lb]{\smash{{\SetFigFont{11}{13.2}{\rmdefault}{\mddefault}{\updefault}$\B_{\bs|\as}$}}}}
\put(3327,-3229){\makebox(0,0)[lb]{\smash{{\SetFigFont{11}{13.2}{\rmdefault}{\mddefault}{\updefault}${\mathfrak{c}}$}}}}
\put(5361,-3233){\makebox(0,0)[lb]{\smash{{\SetFigFont{11}{13.2}{\rmdefault}{\mddefault}{\updefault}${\mathfrak{c}}$}}}}
\put(7477,-3224){\makebox(0,0)[lb]{\smash{{\SetFigFont{11}{13.2}{\rmdefault}{\mddefault}{\updefault}${\mathfrak{c}}$}}}}
\put(1431,-3210){\makebox(0,0)[lb]{\smash{{\SetFigFont{11}{13.2}{\rmdefault}{\mddefault}{\updefault}$\as$}}}}
\put(1440,-471){\makebox(0,0)[lb]{\smash{{\SetFigFont{11}{13.2}{\rmdefault}{\mddefault}{\updefault}$\bs$}}}}
\put(3305,-485){\makebox(0,0)[lb]{\smash{{\SetFigFont{11}{13.2}{\rmdefault}{\mddefault}{\updefault}$\bs$}}}}
\put(5406,-490){\makebox(0,0)[lb]{\smash{{\SetFigFont{11}{13.2}{\rmdefault}{\mddefault}{\updefault}${\mathfrak{d}}$}}}}
\put(7475,-471){\makebox(0,0)[lb]{\smash{{\SetFigFont{11}{13.2}{\rmdefault}{\mddefault}{\updefault}${\mathfrak{e}}$}}}}
\put(1400,-2196){\makebox(0,0)[lb]{\smash{{\SetFigFont{11}{13.2}{\rmdefault}{\mddefault}{\updefault}$\B^{(1)}$}}}}
\put(3680,-1653){\makebox(0,0)[lb]{\smash{{\SetFigFont{11}{13.2}{\rmdefault}{\mddefault}{\updefault}$\B^{(2)}$}}}}
\put(5506,-2601){\makebox(0,0)[lb]{\smash{{\SetFigFont{11}{13.2}{\rmdefault}{\mddefault}{\updefault}$\B^{(3)}$}}}}
\put(7424,-1496){\makebox(0,0)[lb]{\smash{{\SetFigFont{11}{13.2}{\rmdefault}{\mddefault}{\updefault}$x$}}}}
\put(7427,-1840){\makebox(0,0)[lb]{\smash{{\SetFigFont{11}{13.2}{\rmdefault}{\mddefault}{\updefault}$y$}}}}
\end{picture}%
\caption{Examples of diagrams including microscopic mixing.}
\label{figmixing}
\end{figure}

In order to satisfy the Euler-Lagrange equations in the continuum limit~\eqref{field2} for the given pair~$(\as,\bs)$,
we need to perturb the fermionic projector by a bosonic potential~$\B_{\bs | \as}$, being a solution of the
corresponding field equations
\beq \label{fieldmix}
j_k[\B_{\bs | \as}] = \lambda \, J^{\bs | \as}_k(x) \:.
\eeq
Note that~$\B_{\bs | \as}$ is in general not symmetric, because~$(\B_{\bs | \as})^*=\B_{\as | \bs}$.
Nevertheless, we can perform the
perturbation expansion exactly explained in Section~\ref{seccoup}.
Taking into account that the adjoint also involves exchanging the subsystems~$\as \leftrightarrow \bs$,
the resulting fermionic projector will again be symmetric. In particular, we obtain to
first order
\[ \Delta P^{\as,\bs} = - s_m \,\B_{\bs|\as}\, P^{\as,\bs} - P^{\as,\bs} \,\B_{\bs|\as}\, s_m \:. \]
Decomposing~$P^{\as,\bs}$ according to~\eqref{Pmixansatz} into its bra and ket states, we find that
the states in the subsystem~$\as$, denoted for clarity by~$\psi^\as$, are perturbed by
\beq \label{delpsia}
\Delta \psi^\as = - s_m \,\B_{\bs|\as}\, \psi^\as \:.
\eeq
Combining~\eqref{delpsia} with the field equations~\eqref{fieldmix} involving the Dirac current~\eqref{Jmix},
we find that when a bosonic line couples to a ket-state~$\psi^\as$ in the subsystem~$\as$, then the
ket-state in the corresponding Dirac current is also in the subsystem~$\as$
(see the left of Figure~\ref{figmixing}).
We say that the two ket-states are {\em{synchronal}}.
The bra-state in the corresponding Dirac current, however, is in the subsystem~$\bs$,
and is therefore not synchronal to~$\psi^\as$.
The equation~\eqref{delpsia} suffers from the shortcoming that the right side depends on the
index~$\bs$, whereas the left side does not. Clearly, the wave function~$\psi^\as$ in the $\as^\text{th}$ subsystem
must be determined independent of the choice of the subsystem~$\bs$. 
For this reason, we can satisfy~\eqref{delpsia} only after taking the average over all subsystems~$\bs$,
\beq \label{delpsiaav}
\Delta \psi^\as = - \sum_{\bs \in {\mathfrak{M}}} c_\bs \,s_m \B_{\bs|\as} \psi^\as \:.
\eeq

In order to generalize the last construction to higher order perturbation theory, we
first note that with~\eqref{delpsiaav} we have fixed the effective potential in the subsystem~$B$ to be
\beq \label{Badef}
\B_{|\as} := \sum_{\bs \in {\mathfrak{M}}} c_\bs \, \B_{\bs|\as} \:.
\eeq
Thus the wave functions in the $\as^\text{th}$ subsystem can be obtained by
applying the corresponding unitary perturbation flow,
\[ \tilde{\psi}^\as = U[\B_{|\as}]\, \psi^\as \:, \]
where~$U$ is given as in~\eqref{Uflow}. Likewise, the fermionic projector is
obtained from that of the vacuum~\eqref{Pmixansatz} by inserting the perturbation flow,
\beq
P^{\as,\bs} = -U[\B_{|\as}] \:\Psi \, V_\as  \,V_\bs^* \, \Psi^*\: U[\B_{|\bs}]^* \:. \label{UPab}
\eeq
Employing similar to~\eqref{Bpower} the power ansatz
\beq \label{Bpower2}
\B_{|\as} = \sum_{p=0}^\infty \lambda^p\, \B^{(p)}_{|\as} \:,
\eeq
the field equations can be written similar to~\eqref{jiter} as
\begin{align} 
j_k[B^{(0)}_{|\as}] - M^2 A_k[\B^{(0)}_{|\as}] &= 0 \label{Bhom} \\
j_k[\B^{(p+1)}_{|\as}] - M^2 A_k[\B^{(p+1)}_{|\as}] &= -
\sum_{\bs \in {\mathfrak{M}}} c_\bs
\Tr_{\C^4} \!\big( \gamma_k (P^{(\as,\bs)})^{(p)}(x,x) \big) + \text{(s.c.)}\:,
\label{jiter2}
\end{align}
where~(s.c.) again denotes the singular contributions and the convolution terms.
Combining these effective field equations with the unitary perturbation flow~\eqref{UPab},
we obtain an iterative procedure for computing~$P^{\as,\bs}$ and~$\B_{|\as}$.
The resulting rules can be expressed graphically as follows: We start with the unfolded diagrams of the perturbation theory without microscopic mixing.
To every outer fermionic line we add an index~$\as, \bs, \ldots$ to denote the corresponding subsystem.
Next, one determines which outer lines are synchronal. The synchronal lines carry the same
subsystem index, whereas all asynchronal lines carry different subsystem indices.
An example of a resulting diagram is shown on the right of Figure~\ref{figmixing}.

\section{A Stochastic Bosonic Background Field} \label{secstoch}
In the power ansatz~\eqref{Bpower2} for~$\B_{|\as}$ we are still free to choose~$\B^{(0)}_{|\as}$
as a solution of the homogeneous field equation~\eqref{Bhom}.
Similar as described in~\cite[Section~4]{loop}, we choose~$\B^{(0)}_{|a}$ as a stochastic background field
with probability measure~${\mathfrak{D}} \B$.
The stochastic field may have contributions with a different dependence on the subsystems.
For example, there could be one contribution which is the same in all subsystems,
and another contribution which is stochastically independent in the subsystems.
Similar to~\cite[eq.~(4.5)]{loop}, this would mean that the covariance~${\mathcal{C}}(x,y)$
satisfies the equation
\beq \label{covsymm}
\int \B_{|\as}(x)\: \B_{|\bs}(y)\: {\mathfrak{D}} \B = {\mathcal{C}}(x,y) + \delta_{\as \bs}\: {\mathcal{C}}_\as(x,y)
\eeq
(where for ease in notation we omitted the spinor and/or tensor indices of~${\mathcal{C}}$
and~${\mathcal{C}}_\as$).
At this point, we do not need to specify the covariance any further
(for more details see Section~\ref{secbackground} below).
For clarity, we remark that our ansatz~\eqref{covsymm} only makes sense for an abelian gauge field.
In the non-abelian case, one could complement~\eqref{covsymm} by a non-linear term in~$\B$.
Alternatively, one could introduce the stochastic background field only for an abelian subgroup of
the gauge group.

\begin{figure} %
\begin{picture}(0,0)%
\includegraphics{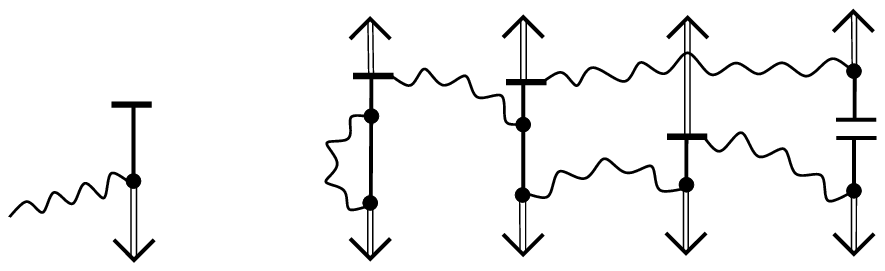}%
\end{picture}%
\setlength{\unitlength}{1533sp}%
\begingroup\makeatletter\ifx\SetFigFont\undefined%
\gdef\SetFigFont#1#2#3#4#5{%
  \reset@font\fontsize{#1}{#2pt}%
  \fontfamily{#3}\fontseries{#4}\fontshape{#5}%
  \selectfont}%
\fi\endgroup%
\begin{picture}(11035,3293)(-3543,-3389)
\put(3327,-3229){\makebox(0,0)[lb]{\smash{{\SetFigFont{11}{13.2}{\rmdefault}{\mddefault}{\updefault}$\as$}}}}
\put(5361,-3233){\makebox(0,0)[lb]{\smash{{\SetFigFont{11}{13.2}{\rmdefault}{\mddefault}{\updefault}$\as$}}}}
\put(7477,-3224){\makebox(0,0)[lb]{\smash{{\SetFigFont{11}{13.2}{\rmdefault}{\mddefault}{\updefault}$\as$}}}}
\put(1431,-3210){\makebox(0,0)[lb]{\smash{{\SetFigFont{11}{13.2}{\rmdefault}{\mddefault}{\updefault}$\as$}}}}
\put(1440,-471){\makebox(0,0)[lb]{\smash{{\SetFigFont{11}{13.2}{\rmdefault}{\mddefault}{\updefault}$\bs$}}}}
\put(3305,-485){\makebox(0,0)[lb]{\smash{{\SetFigFont{11}{13.2}{\rmdefault}{\mddefault}{\updefault}${\mathfrak{c}}$}}}}
\put(5406,-490){\makebox(0,0)[lb]{\smash{{\SetFigFont{11}{13.2}{\rmdefault}{\mddefault}{\updefault}${\mathfrak{d}}$}}}}
\put(7475,-471){\makebox(0,0)[lb]{\smash{{\SetFigFont{11}{13.2}{\rmdefault}{\mddefault}{\updefault}${\mathfrak{c}}$}}}}
\put(7424,-1496){\makebox(0,0)[lb]{\smash{{\SetFigFont{11}{13.2}{\rmdefault}{\mddefault}{\updefault}$x$}}}}
\put(7427,-1840){\makebox(0,0)[lb]{\smash{{\SetFigFont{11}{13.2}{\rmdefault}{\mddefault}{\updefault}$y$}}}}
\put(-1440,-3253){\makebox(0,0)[lb]{\smash{{\SetFigFont{11}{13.2}{\rmdefault}{\mddefault}{\updefault}$\as$}}}}
\put(1535,-1637){\makebox(0,0)[lb]{\smash{{\SetFigFont{11}{13.2}{\rmdefault}{\mddefault}{\updefault}$\B^{(3)}$}}}}
\put(5649,-2669){\makebox(0,0)[lb]{\smash{{\SetFigFont{11}{13.2}{\rmdefault}{\mddefault}{\updefault}$\B^{(2)}$}}}}
\put(-3528,-2177){\makebox(0,0)[lb]{\smash{{\SetFigFont{11}{13.2}{\rmdefault}{\mddefault}{\updefault}$\B^{(0)}_{|\as}$}}}}
\put(-56,-2125){\makebox(0,0)[lb]{\smash{{\SetFigFont{11}{13.2}{\rmdefault}{\mddefault}{\updefault}${\mathcal{C}}_\as$}}}}
\put(3771,-1781){\makebox(0,0)[lb]{\smash{{\SetFigFont{11}{13.2}{\rmdefault}{\mddefault}{\updefault}${\mathcal{C}}_\as$}}}}
\put(5976,-1323){\makebox(0,0)[lb]{\smash{{\SetFigFont{11}{13.2}{\rmdefault}{\mddefault}{\updefault}${\mathcal{C}}_{\mathfrak{c}}$}}}}
\end{picture}%
\caption{Diagrams involving the stochastic background field.}
\label{figbackground}
\end{figure}

In the Feynman diagrams, the stochastic background field is depicted either by bosonic lines~$B^{(0)}_{|\as}$
or by the covariance. Figure~\ref{figbackground} shows a few examples.
One should keep in mind that~$B^{(0)}_{|\as}$ couples only to the $\as^\text{th}$ subsystem.
Therefore, the covariance couples only to the fermionic legs being in the same subsystem.
In other words, just as explained on the left of Figure~\ref{figmixing} for a regular bosonic line,
also the covariance yields a {\em{synchronization}} of the legs of diagrams.

\section{Reduction to Anti-Symmetrized Synchronal Blocks} \label{secasb}
\subsection{Taking Averages over Subsystems} \label{secaverage}
The microscopic mixing gives rise to small-scale fluctuations of the fermionic projector.
This is described mathematically by the matrices~$V_\as$ in~\eqref{Pmixansatz},
which involve the random matrices~$\Wran_\alpha$ and~$\Uran_a$ in our ansatz~\eqref{Vans}.
The matrices~$V_\as$ can be understood similar to the stochastic bosonic background
field in Section~\ref{secstoch} as describing microscopic fluctuations for example on the Planck scale.
As a consequence of these microscopic fluctuations, many contributions to the fermionic
projector are highly oscillatory and thus become very small when taking averages over
the subsystems.
We now analyze this effect quantitatively for the generalized phase transformations~$\Wran_\alpha$
of the particle states.
We begin with a diagram as shown in Figures~\ref{figmixing} or~\ref{figbackground}.
Substituting the ansatz~\eqref{Vans} and multiplying out, we obtain a sum of expressions, each of which
involves a combination of products of the matrices~$\Wran_\alpha$, $\Wran_\beta$, \ldots
corresponding to the different subsystems. In order to average over the subsystems, we multiply
these expressions by~$c_\as, c_\bs, \ldots$, and sum over~$\as, \bs, \ldots$.
Since these operations can be carried out subsequently for the subsystem~$\as$, $\bs$, etc.,
it suffices to consider an expression of the form
\[ \sum_{\as \in {\mathfrak{M}}} c_\as \:(\Wran_\alpha)^{i_1}_{j_1} \cdots (\Wran_\alpha)^{i_p}_{j_p} \;\cdots \:. \]
In order to determine the relevant scalings, we may disregard the weight factors~$c_\as$
and consider instead the normalized counting measure. Thus our task is to analyze the
sum
\beq \label{prodel}
\frac{1}{\Lphase} \sum_{\alpha=1}^{\Lphase} (\Wran_\alpha)^{i_1}_{j_1} \cdots (\Wran_\alpha)^{i_p}_{j_p} \:.
\eeq

In the case~$p \ll n$, the expectation values and fluctuations of such products of random matrices
can be computed with moment methods (see~\cite{weingarten, collins+sniady}).
Unfortunately, these formulas are not good enough for our
purposes, because we need to cover the case~$p \sim n \gg 1$. For this purpose, we employ
a different method which uses the representation theory on the tensor product.
In preparation, it is helpful to note that, according to~\eqref{Vans} and~\eqref{WUmap},
the matrix elements in~\eqref{prodel} arise as linear mappings from~$\J_0$ to~$\I_0$.
Again choosing fixed bases, we can thus rewrite the product of matrix elements in~\eqref{prodel}
more abstractly in terms of the corresponding linear mapping on the tensor product
\beq \label{Walphap}
(\Wran_\alpha)^p \::\: \underbrace{\C^n \otimes \cdots \otimes \C^n}_\text{$p$ factors} \rightarrow
\underbrace{\C^n \otimes \cdots \otimes \C^n}_\text{$p$ factors}  \:.
\eeq
Since the matrix~$\Wran_\alpha \in \U(n)$ involves a random phase, it is obvious that~\eqref{Walphap}
vanishes in the statistical mean. In the next lemma, we estimate the fluctuations.

\begin{Prp} \label{prp51} In the case~$1 \leq p<n$, for any~$u, v \in \C^{np}$,
\beq \label{fluct1}
\bigg| \frac{1}{\Lphase} \sum_{\alpha=1}^{\Lphase} \big\langle u, (\Wran_\alpha)^p \,v \big\rangle \bigg|^2
\lesssim\frac{\|u\|^2\, \|v\|^2}{n \Lphase} \:.
\eeq
Here we identified the $p$-fold tensor product of~$\C^n$ with~$\C^{np}$, endowed with the canonical
scalar product~$\la .,. \ra$ and corresponding norm~$\|.\|$. The symbol~$\lesssim$ means that
the inequality holds for the mean value if the~$\Wran_\alpha$ are random matrices
distributed according to the normalized Haar measure on~$\U(n)$.
\end{Prp}
\Proof The mapping~$(\Wran_\alpha)^p$ in~\eqref{Walphap}
defines a representation~$U$ of the group~$\U(n)$ on the tensor product.
We decompose this representation into irreducible components,
\[ U = \bigoplus_{k=1}^k U_k \qquad \text{with} \qquad U_k \::\: I_k \rightarrow I_k \:, \]
where the~$I_k$ are mutually orthogonal subspaces of the tensor product,
\beq \label{reduce}
\underbrace{\C^n \otimes \cdots \otimes \C^n}_\text{$p$ factors} = \bigoplus_{k=1}^K I_k \:.
\eeq
To avoid confusion, we note that the group~$\U(n)$ factorizes as
\[ \U(n) = \U(1) \times \SU(n) \:. \]
Since the group~$\U(1)$ only gives a phase factor, it has no influence on irreducibility.
Hence the subspace~$I_k$ are just the irreducible components of the action of~$\SU(n)$
(for details see for example~\cite{broecker+tomdieck} or~\cite{schensted, sternberg}).
We also note that in the example of the group~$\SU(2)$, the irreducible components are the
well-known spin representation corresponding to spin weights~$0, \frac{1}{2}, 1, \frac{3}{2}, \ldots$.

We consider a fixed matrix element,
\[ f^i_j := \frac{1}{\Lphase} \sum_{\alpha=1}^{\Lphase} \big( U_k[\Wran_\alpha] \big)^i_j \:. \]
For random matrices~$\Wran_\alpha \in \U(n)$, the expectation value of this expression clearly vanishes.
The fluctuations are computed by
\begin{align}
|f^i_j|^2 &= \frac{1}{\Lphase^2} \sum_{\alpha, \beta=1}^{\Lphase} U_k[\Wran_\alpha]^i_j\;
\overline{U_k[\Wran_\beta]^i_j} \eqsim \frac{1}{\Lphase}\: \Big< \big|(U_k[W])^i_j \big|^2 \Big> \nonumber \\
&=\frac{1}{\Lphase\, \dim(I_k)} \sum_{\ell=1}^{\dim(I_k)} \Big< \big|(U_k[W])^i_\ell \big|^2 \Big>
=\frac{1}{\Lphase \, \dim(I_k)} \:, \label{fluctcomp}
\end{align}
where in the last line we used the symmetry in the index~$j$ together with the fact that
the matrix~$U_k$ is unitary. We finally use that all the representations~$U_k$
are non-trivial and have dimension at least~$n$ (see for example~\cite[Section~5]{sternberg}
or relation~\eqref{dimUl} at the beginning of Appendix~\ref{appyoung}, noting that the minimal dimension
is attained for the Young diagram with a single column).
\QED

We next consider the case~$p=n$. This case is special, because applying~$\Wran^p$ to
a totally antisymmetric vector gives the determinant,
\[ (\Wran \otimes \cdots \otimes \Wran)(\psi_1 \wedge \cdots \wedge \psi_n) 
= \det \Wran \;\psi_1 \wedge \cdots \wedge \psi_n\:. \]
This expression depends on~$\Wran$ only by a phase factor.
Computing the fluctuations of random phases, one obtains
\beq \label{fluct2}
\bigg| \frac{1}{\Lphase} \sum_{\alpha=1}^{\Lphase} e^{i \varphi_\alpha} \bigg|^2
= \frac{1}{\Lphase^2} \sum_{\alpha, \beta=1}^{\Lphase} e^{i (\varphi_\alpha-\varphi_\beta)}
\eqsim \frac{1}{\Lphase} \:.
\eeq
This formula resembles~\eqref{fluct1} in that it involves a factor~$1/\Lphase$.
However, it does not involve a factor~$1/n$. Keeping in mind that~$n$ should be at least as large
as the number of all fermions of the universe (for details see Section~\ref{secanti}), this
means that~\eqref{fluct2} is much larger
than~\eqref{fluct1}. The next proposition makes this argument more precise.
We denote the anti-symmetrization operator by~$\pi_\text{as}$,
\begin{align}
\pi_\text{as} \::\: \underbrace{\C^n \otimes \cdots \otimes \C^n}_\text{$n$ factors} &\rightarrow
\underbrace{\C^n \otimes \cdots \otimes \C^n}_\text{$n$ factors} \:, \nonumber \\
\pi_\text{as}(\psi_1 \otimes \cdots \otimes \psi_n) &=  \frac{1}{n!} \sum_{\sigma \in S_n}
(-1)^{\sign(\sigma)}\: \psi_{\sigma(1)} \otimes \cdots \otimes \psi_{\sigma(n)} \label{totas}
\end{align}

\begin{Prp} \label{prp52}
In the case~$p=n$, using the notation of Proposition~\ref{prp51}, for any~$u, v \in \C^{np}$ we have
the inequalities
\begin{align}
\bigg| \frac{1}{\Lphase} \sum_{\alpha=1}^{\Lphase} \det(\Wran_\alpha) \bigg|^2 &\eqsim \frac{1}{\Lphase}
\label{fluct3} \\
\bigg| \frac{1}{\Lphase} \sum_{\alpha=1}^{\Lphase} \Big\langle u,
\Big( (\Wran_\alpha)^n - \det(\Wran_\alpha)\: \pi_\text{as} \Big)\, v \Big\rangle \bigg|^2 
&\lesssim\frac{\|u\|^2\, \|v\|^2}{n \Lphase} \:. \label{fluct4}
\end{align}
\end{Prp}
\Proof The relation~\eqref{fluct3} is the same as~\eqref{fluct2}.
The estimate~\eqref{fluct4} follows exactly as the proof of Proposition~\ref{prp51}, keeping in mind that
the term~$-\det(\Wran)\: \pi_\text{as}$ cancels the totally antisymmetric representation of~$\Wran^n$,
and that all the other representations have dimension at least~$n$
(see again~\cite[Section~5]{sternberg} or relation~\eqref{dimUl} and use that
the minimal dimension is attained  for the Young diagram with two columns and
a single box in the second column).
\QED

The results of the previous Propositions~\ref{prp51} and~\ref{prp52} show that
the totally antisymmetric representation~\eqref{totas} dominates the
contribution by any other irreducible representation by a scaling factor~$\sqrt{n}$
(compare~\eqref{fluct3} with~\eqref{fluct4} and~\eqref{fluct1}).
However, this result is not quite satisfactory because the reduction~\eqref{reduce}
involves the direct sum of many different irreducible representations, coming with different
multiplicities. Thus, although each direct summand gives a small contribution, the
total contribution by all the direct summands might still be large.
In order to overcome this shortcoming, we need to take averages on the whole tensor product,
as we now explain. Our starting point is the formula~\eqref{reduce}.
On each of the subspaces~$I_k$ we have an irreducible representation of~$\U(n)$.
Clearly, the same representation can appear several times.
For ease in notation, we reorder the~$I_k$ such that the subspaces with the
totally antisymmetric representation~\eqref{totas} come last, i.e.\
\beq \label{reduce2}
\underbrace{\C^n \otimes \cdots \otimes \C^n}_\text{$p$ factors} = \Big( \bigoplus_{k=1}^L I_k \Big)
\oplus \Big( \bigoplus_{k=L+1}^K I^\text{\tiny{AS}}_k \Big) \:,
\eeq
where the subspaces~$I_k^\text{\tiny{AS}}$ are all one-dimensional and carry the
totally antisymmetric representation~\eqref{totas}.
In the case~$p<n$, where total anti-symmetrization is impossible, we know that~$L=K$.
In the case~$p=n$, on the other hand, the totally antisymmetric representation acts
only on the one-dimensional subspace spanned by the wedge product of the basis vectors of~$\C^n$,
and thus~$L=K-1$.

The fluctuations can be computed separately on every direct summand in~\eqref{reduce2}.
According to~\eqref{fluctcomp}, we pick up a factor~$1/\dim I_k$. 
Adding up the contributions by all the irreducible representations except for the
totally antisymmetric representation gives the expression
\beq \label{Iksum}
\sum_{k=1}^L \frac{1}{\dim I_k} \:.
\eeq
This should be small compared to the contribution by the totally antisymmetric
representations given by
\[ \frac{1}{\dim I^\text{AS}} = 1 \:. \]
The next proposition shows that this is indeed the case for large~$n$ in the case~$p \leq n$.
\begin{Prp} \label{prpcombi} For any~$1 \leq p \leq n$,
\[ \lim_{n \rightarrow \infty} \sum_{k=1}^L \frac{1}{\dim I_k} = 0 \qquad
\text{uniformly in~$p$}\:. \]
\end{Prp} \noindent
In order not to distract from the main ideas, the proof of this proposition is
postponed to Appendix~\ref{appyoung}.
In this appendix, we also say a few words on the case~$p>n$ which is not covered
by the above proposition.

\subsection{Anti-Symmetrized Synchronal Blocks} \label{secansyb}
The analysis of the previous section yields that to leading order in~$1/n$,
we may restrict attention to contributions which involve the random matrices~$\Wran_\alpha$
only to the $n^\text{th}$ power. Moreover, we should anti-symmetrize 
the corresponding indices of these matrices.
This strategy can be understood non-technically as follows:
As already mentioned at the beginning of the previous section, microscopic mixing describes
small-scale fluctuations of the fermionic projector. We can hope that
the effective macroscopic dynamics should be described
by contributions to the fermionic projector which are robust to microscopic mixing in the
sense that they are as far as possible independent of the choice of the random matrices.
The totally antisymmetric contribution of order~$n$ in~$\Wran_\alpha$ has the advantage
that in view of the formula
\beq \label{Wdet}
\epsilon^{j_1 \cdots j_n} \:\Wran^{i_1}_{j_1} \cdots \Wran^{i_n}_{j_n} = 
\det \Wran \: \epsilon_{i_1 \cdots i_n}\:,
\eeq
it depends on microscopic mixing only by a phase factor.
Indeed, this phase factor can be regarded as a physically
irrelevant normalization constant, as will be explained in Section~\ref{secbackground} below.

\begin{figure} %
\begin{picture}(0,0)%
\includegraphics{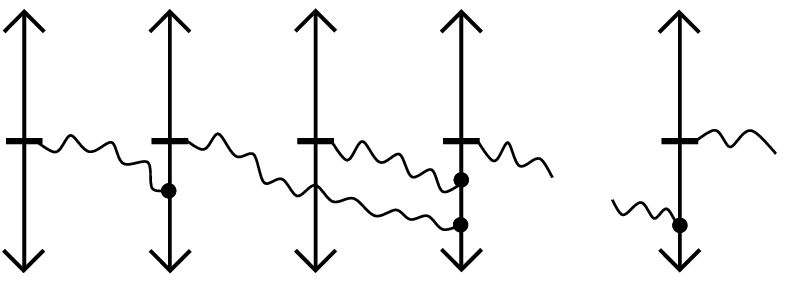}%
\end{picture}%
\setlength{\unitlength}{1533sp}%
\begingroup\makeatletter\ifx\SetFigFont\undefined%
\gdef\SetFigFont#1#2#3#4#5{%
  \reset@font\fontsize{#1}{#2pt}%
  \fontfamily{#3}\fontseries{#4}\fontshape{#5}%
  \selectfont}%
\fi\endgroup%
\begin{picture}(9619,4003)(602,-3983)
\put(1214,-3292){\makebox(0,0)[lb]{\smash{{\SetFigFont{11}{13.2}{\rmdefault}{\mddefault}{\updefault}$\as$}}}}
\put(3052,-3315){\makebox(0,0)[lb]{\smash{{\SetFigFont{11}{13.2}{\rmdefault}{\mddefault}{\updefault}$\as$}}}}
\put(4807,-3330){\makebox(0,0)[lb]{\smash{{\SetFigFont{11}{13.2}{\rmdefault}{\mddefault}{\updefault}$\as$}}}}
\put(6600,-3307){\makebox(0,0)[lb]{\smash{{\SetFigFont{11}{13.2}{\rmdefault}{\mddefault}{\updefault}$\as$}}}}
\put(9307,-3315){\makebox(0,0)[lb]{\smash{{\SetFigFont{11}{13.2}{\rmdefault}{\mddefault}{\updefault}$\as$}}}}
\put(1215,-359){\makebox(0,0)[lb]{\smash{{\SetFigFont{11}{13.2}{\rmdefault}{\mddefault}{\updefault}$\bs$}}}}
\put(3012,-372){\makebox(0,0)[lb]{\smash{{\SetFigFont{11}{13.2}{\rmdefault}{\mddefault}{\updefault}${\mathfrak{c}}$}}}}
\put(4836,-355){\makebox(0,0)[lb]{\smash{{\SetFigFont{11}{13.2}{\rmdefault}{\mddefault}{\updefault}${\mathfrak{d}}$}}}}
\put(6658,-359){\makebox(0,0)[lb]{\smash{{\SetFigFont{11}{13.2}{\rmdefault}{\mddefault}{\updefault}${\mathfrak{e}}$}}}}
\put(7471,-359){\makebox(0,0)[lb]{\smash{{\SetFigFont{11}{13.2}{\rmdefault}{\mddefault}{\updefault}$\cdots$}}}}
\put(4411,-3841){\makebox(0,0)[lb]{\smash{{\SetFigFont{11}{13.2}{\rmdefault}{\mddefault}{\updefault}$3$}}}}
\put(2611,-3841){\makebox(0,0)[lb]{\smash{{\SetFigFont{11}{13.2}{\rmdefault}{\mddefault}{\updefault}$2$}}}}
\put(6211,-3841){\makebox(0,0)[lb]{\smash{{\SetFigFont{11}{13.2}{\rmdefault}{\mddefault}{\updefault}$4$}}}}
\put(8911,-3841){\makebox(0,0)[lb]{\smash{{\SetFigFont{11}{13.2}{\rmdefault}{\mddefault}{\updefault}$n$}}}}
\put(7606,-1791){\makebox(0,0)[lb]{\smash{{\SetFigFont{11}{13.2}{\rmdefault}{\mddefault}{\updefault}$\cdots$}}}}
\put(811,-3841){\makebox(0,0)[lb]{\smash{{\SetFigFont{11}{13.2}{\rmdefault}{\mddefault}{\updefault}$1$}}}}
\put(7352,-3847){\makebox(0,0)[lb]{\smash{{\SetFigFont{11}{13.2}{\rmdefault}{\mddefault}{\updefault}$\cdots$}}}}
\end{picture}%
\caption{A diagram with synchronal ket states.}
\label{figsync}
\end{figure}

Let us analyze systematically how the matrix~$\Wran_\alpha$ comes up in the
perturbation expansion. First, according to~\eqref{UPab} and~\eqref{Pmixansatz},
every free state in~$\I_0$ comes with a factor~$\Wran_\alpha$.
In order to obtain contributions to the perturbation expansion which involve~$n$ factors~$\Wran_\alpha$,
one can consider a diagram involving~$n$ fermionic lines, where all the ket-states are
synchronal due to the fact that all the bosonic lines couple to the ket states.
A typical example is shown in Figure~\ref{figsync}.
We note that in this and the following figures, for notational simplicity we do not
write double lines for the on-shell contributions, but use simple lines for all factors~$s_m$,
$p_m$ and~$k_m$. Then, according to~\eqref{UPab}, the ket state of every fermionic line involves the
microscopic mixing matrix~$V_\as$.
If we always take the contribution to~\eqref{Vans} which involves~$\Wran$, the resulting expression
involves~$n$ factors~$\Wran_\alpha$. Anti-symmetrizing and applying~\eqref{Wdet},
the dependence on~$\Wran_\alpha$ reduces to a phase factor.

The construction so far has the shortcoming that, since the bosonic lines
all synchronize the ket states, the bra states of fermionic states are not synchronal
(as indicated in Figure~\ref{figsync} by the bra indices~$\bs, {\mathfrak{c}}, {\mathfrak{d}}, \ldots$).
Let us deduce that that the diagram is {\em{not}} uniquely defined up to a complex prefactor:
In order to get rid of the dependence on the matrices~$\Wran_\beta, \Wran_\gamma, \ldots$
and~$\Uran_b, \Uran_c, \ldots$, we need to take the first summand in~\eqref{Vans2}, i.e.\
we must set~$\Uran_b, \Uran_c, \ldots$ equal to~$\pi_{\mathscr{N}_0}$.
Then each fermionic line gives the analytic expression
\[ \Tr_{\T_0} \bigg\{ (U[\B_{|\bs}] \Psi)^*(z)\:(U[\B_{|\as}] \:\Psi)(z)
\begin{pmatrix} 0 & \Wran_\alpha (\Uran_a)_{11}^* & \Wran_\alpha (\Uran_a)_{21}^* \\ 0 & 0 & 0 \\ 0 & 0 & 0 \end{pmatrix}
\bigg\} \:, \]
(where~$z$ is the space-time point where the outgoing bosonic line begins).
Taking the anti-symmetrized product and using~\eqref{Wdet}, the matrix~$\Uran_a$ still
describes a non-trivial mixing of the vector spaces~$\I_0$, $\J_0$ and~${\mathscr{K}}_0$.

\begin{figure} %
\begin{picture}(0,0)%
\includegraphics{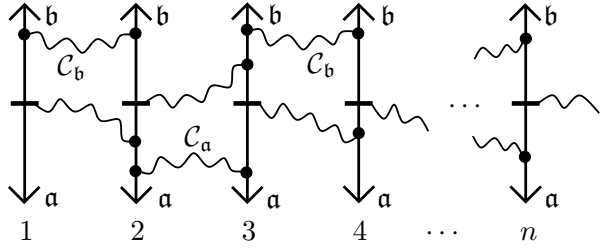}%
\end{picture}%
\setlength{\unitlength}{1533sp}%
\begingroup\makeatletter\ifx\SetFigFont\undefined%
\gdef\SetFigFont#1#2#3#4#5{%
  \reset@font\fontsize{#1}{#2pt}%
  \fontfamily{#3}\fontseries{#4}\fontshape{#5}%
  \selectfont}%
\fi\endgroup%
\begin{picture}(9619,3996)(602,-3982)
\put(1214,-3292){\makebox(0,0)[lb]{\smash{{\SetFigFont{11}{13.2}{\rmdefault}{\mddefault}{\updefault}$\as$}}}}
\put(3052,-3315){\makebox(0,0)[lb]{\smash{{\SetFigFont{11}{13.2}{\rmdefault}{\mddefault}{\updefault}$\as$}}}}
\put(4807,-3330){\makebox(0,0)[lb]{\smash{{\SetFigFont{11}{13.2}{\rmdefault}{\mddefault}{\updefault}$\as$}}}}
\put(6600,-3307){\makebox(0,0)[lb]{\smash{{\SetFigFont{11}{13.2}{\rmdefault}{\mddefault}{\updefault}$\as$}}}}
\put(9307,-3315){\makebox(0,0)[lb]{\smash{{\SetFigFont{11}{13.2}{\rmdefault}{\mddefault}{\updefault}$\as$}}}}
\put(1215,-359){\makebox(0,0)[lb]{\smash{{\SetFigFont{11}{13.2}{\rmdefault}{\mddefault}{\updefault}$\bs$}}}}
\put(3012,-372){\makebox(0,0)[lb]{\smash{{\SetFigFont{11}{13.2}{\rmdefault}{\mddefault}{\updefault}$\bs$}}}}
\put(4836,-355){\makebox(0,0)[lb]{\smash{{\SetFigFont{11}{13.2}{\rmdefault}{\mddefault}{\updefault}$\bs$}}}}
\put(6658,-359){\makebox(0,0)[lb]{\smash{{\SetFigFont{11}{13.2}{\rmdefault}{\mddefault}{\updefault}$\bs$}}}}
\put(9291,-359){\makebox(0,0)[lb]{\smash{{\SetFigFont{11}{13.2}{\rmdefault}{\mddefault}{\updefault}$\bs$}}}}
\put(2611,-3841){\makebox(0,0)[lb]{\smash{{\SetFigFont{11}{13.2}{\rmdefault}{\mddefault}{\updefault}$2$}}}}
\put(4411,-3841){\makebox(0,0)[lb]{\smash{{\SetFigFont{11}{13.2}{\rmdefault}{\mddefault}{\updefault}$3$}}}}
\put(6211,-3841){\makebox(0,0)[lb]{\smash{{\SetFigFont{11}{13.2}{\rmdefault}{\mddefault}{\updefault}$4$}}}}
\put(8911,-3841){\makebox(0,0)[lb]{\smash{{\SetFigFont{11}{13.2}{\rmdefault}{\mddefault}{\updefault}$n$}}}}
\put(811,-3841){\makebox(0,0)[lb]{\smash{{\SetFigFont{11}{13.2}{\rmdefault}{\mddefault}{\updefault}$1$}}}}
\put(7382,-3846){\makebox(0,0)[lb]{\smash{{\SetFigFont{11}{13.2}{\rmdefault}{\mddefault}{\updefault}$\cdots$}}}}
\put(7749,-1783){\makebox(0,0)[lb]{\smash{{\SetFigFont{11}{13.2}{\rmdefault}{\mddefault}{\updefault}$\cdots$}}}}
\put(5474,-1138){\makebox(0,0)[lb]{\smash{{\SetFigFont{11}{13.2}{\rmdefault}{\mddefault}{\updefault}${\mathcal{C}}_\bs$}}}}
\put(1432,-1184){\makebox(0,0)[lb]{\smash{{\SetFigFont{11}{13.2}{\rmdefault}{\mddefault}{\updefault}${\mathcal{C}}_\bs$}}}}
\put(3487,-2324){\makebox(0,0)[lb]{\smash{{\SetFigFont{11}{13.2}{\rmdefault}{\mddefault}{\updefault}${\mathcal{C}}_\as$}}}}
\end{picture}%
\caption{An anti-symmetrized synchronal block (\ansyb).}
\label{figasb}
\end{figure}

A method to overcome this shortcoming is to consider diagrams where
all the bra states {\em{and}} all the ket states are synchronal. This can be achieved
by coupling the stochastic background field and using that the covariance also
synchronizes the corresponding fermionic legs (for a typical example see Figure~\ref{figasb},
where for simplicity we set the covariance~${\mathcal{C}}$ in~\eqref{covsymm} equal to zero).
Now we can consider the contribution which involves~$n$ factors~$\Wran_\alpha$ as
well as~$n$ factors~$\Wran_\beta^*$.
After anti-symmetrizing, we can apply~\eqref{Wdet} both to the bra and to the ket states.
Then the dependence on both~$\Wran_\alpha$ and~$\Wran_\beta$ reduces to a phase.
Since each fermionic line involves the matrices~$\Uran_a$ and~$\Uran_b$ only in the
combination~$\pi_{\J_0} \Uran_a^* \Uran_b \pi_{\J_0}$,
the dependence on these matrices is again described by a complex number, namely
\beq \label{comfact}
\det \left( \pi_{\J_0} \Uran_a^*\, \Uran_b |_{\J_0} \right) .
\eeq
In this way, we have found an expression which depends on microscopic
mixing only via a complex prefactor. We refer to this expression as
an {\em{anti-symmetrized synchronal block}} (\ansyb).

\begin{figure} %
\begin{picture}(0,0)%
\includegraphics{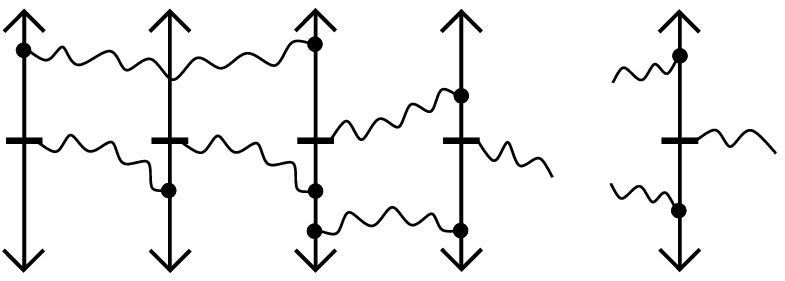}%
\end{picture}%
\setlength{\unitlength}{1533sp}%
\begingroup\makeatletter\ifx\SetFigFont\undefined%
\gdef\SetFigFont#1#2#3#4#5{%
  \reset@font\fontsize{#1}{#2pt}%
  \fontfamily{#3}\fontseries{#4}\fontshape{#5}%
  \selectfont}%
\fi\endgroup%
\begin{picture}(9619,3991)(602,-3977)
\put(1214,-3292){\makebox(0,0)[lb]{\smash{{\SetFigFont{11}{13.2}{\rmdefault}{\mddefault}{\updefault}$\as$}}}}
\put(4807,-3330){\makebox(0,0)[lb]{\smash{{\SetFigFont{11}{13.2}{\rmdefault}{\mddefault}{\updefault}$\as$}}}}
\put(6600,-3307){\makebox(0,0)[lb]{\smash{{\SetFigFont{11}{13.2}{\rmdefault}{\mddefault}{\updefault}$\as$}}}}
\put(9307,-3315){\makebox(0,0)[lb]{\smash{{\SetFigFont{11}{13.2}{\rmdefault}{\mddefault}{\updefault}$\as$}}}}
\put(1215,-359){\makebox(0,0)[lb]{\smash{{\SetFigFont{11}{13.2}{\rmdefault}{\mddefault}{\updefault}$\bs$}}}}
\put(4836,-355){\makebox(0,0)[lb]{\smash{{\SetFigFont{11}{13.2}{\rmdefault}{\mddefault}{\updefault}$\bs$}}}}
\put(6658,-359){\makebox(0,0)[lb]{\smash{{\SetFigFont{11}{13.2}{\rmdefault}{\mddefault}{\updefault}$\bs$}}}}
\put(9291,-359){\makebox(0,0)[lb]{\smash{{\SetFigFont{11}{13.2}{\rmdefault}{\mddefault}{\updefault}$\bs$}}}}
\put(4411,-3841){\makebox(0,0)[lb]{\smash{{\SetFigFont{11}{13.2}{\rmdefault}{\mddefault}{\updefault}$2$}}}}
\put(6211,-3841){\makebox(0,0)[lb]{\smash{{\SetFigFont{11}{13.2}{\rmdefault}{\mddefault}{\updefault}$3$}}}}
\put(8911,-3841){\makebox(0,0)[lb]{\smash{{\SetFigFont{11}{13.2}{\rmdefault}{\mddefault}{\updefault}$n$}}}}
\put(811,-3841){\makebox(0,0)[lb]{\smash{{\SetFigFont{11}{13.2}{\rmdefault}{\mddefault}{\updefault}$1$}}}}
\put(7749,-1783){\makebox(0,0)[lb]{\smash{{\SetFigFont{11}{13.2}{\rmdefault}{\mddefault}{\updefault}$\cdots$}}}}
\put(3029,-3307){\makebox(0,0)[lb]{\smash{{\SetFigFont{11}{13.2}{\rmdefault}{\mddefault}{\updefault}$\as$}}}}
\put(7367,-3839){\makebox(0,0)[lb]{\smash{{\SetFigFont{11}{13.2}{\rmdefault}{\mddefault}{\updefault}$\cdots$}}}}
\put(1749,-1144){\makebox(0,0)[lb]{\smash{{\SetFigFont{11}{13.2}{\rmdefault}{\mddefault}{\updefault}${\mathcal{C}}_\bs$}}}}
\put(9129,-1396){\makebox(0,0)[lb]{\smash{{\SetFigFont{11}{13.2}{\rmdefault}{\mddefault}{\updefault}$x$}}}}
\put(5347,-2333){\makebox(0,0)[lb]{\smash{{\SetFigFont{11}{13.2}{\rmdefault}{\mddefault}{\updefault}${\mathcal{C}}_\as$}}}}
\end{picture}%
\caption{An \ansyb\ involving sea states.}
\label{figasbsea}
\end{figure}

Before going on, we generalize the construction to involve sea states (which will later
give rise to fermion loops). To this end, we begin with a diagram which consists of~$p>n$
fermionic lines. At $p-n$ of these lines, both at the bra and at the ket state we choose the
contribution~$\pi_{\mathscr{N}_0}$ in~\eqref{Vans2}. Then these so-called {\em{sea lines}}
are independent of microscopic mixing; they simply involve the trace over~${\mathscr{N}_0}$.
We assume that at the remaining~$n$ fermionic lines, all the bra and ket states are synchronal
(for a typical example see Figure~\ref{figasbsea}).
Anti-symmetrizing at these~$n$ so-called {\em{particle lines}}, the dependence on the microscopic
mixing again reduces to a complex factor. We refer to the resulting diagram as an {\em{\ansyb\ involving sea states}}.

\section{The Dynamics of an Anti-Symmetrized Synchronal Block} \label{secasbdyn}
\subsection{Cutting the Fermionic Lines} \label{seccut}
We now want to factorize an \ansyb\ (possibly involving sea states)
into the inner product of two $n$-particle wave functions.
Our method is to cut the $n$ particle lines (i.e.\ all the fermionic lines except for the sea lines). We
proceed inductively from the right to the left.
At the line on the very right, we will cut both the ket and the bra state.
Let us begin with the ket state. Fixing a free ket state~$\phi \in \I_0$
and denoting the point where the outgoing bosonic line begins as in Figure~\ref{figasbsea}
by~$x$, the ket state can be described by a wave function~$\psi(x)$.
This wave function coincides with a certain contribution
to the perturbation series as defined by the unitary perturbation flow,
\beq \label{Uphi}
\tilde{\phi}(x) := \big( U[\B] \phi \big)(x) \;=\; \cdots + \psi(x) + \cdots \:.
\eeq
Here the bosonic potential~$\B$ is the sum of~$\B_{|\as}^{(0)}$ with all the bosonic
potentials generated to the left of the particle line under consideration.
Which contribution to~$\tilde{\phi}$ to take is determined by the order in~$\B$
(i.e.\ the number of bosonic lines entering the ket state) and the ordering in which
the different contributions to~$\B$ couple to the ket state.
In order to see the underlying structure, we prefer to work instead of~$\psi$
with the wave function~$\tilde{\phi}$. By expanding in powers of~$\B$ and
selecting the contributions with a certain ordering in the different contributions to~$\B$,
our results for~$\tilde{\phi}$ can immediately be translated to a corresponding result on~$\psi$.

By construction of the unitary perturbation flow, the wave function~$\tilde{\phi}$ is a solution
of the Dirac equation
\beq \label{dirB}
(i \Pdd + \B - m) \,\tilde{\phi} = 0 \:.
\eeq
Evaluating this wave function at some time~$t_0$ gives a spatial wave
function~$\tilde{\phi}_0(\vec{x}) := \tilde{\phi}(t_0, \vec{x})$.
Taking~$\tilde{\phi}_0$ as the initial data and solving the Cauchy problem for the Dirac
equation~\eqref{dirB},
\beq \label{cauchyt}
(i \Pdd + \B - m) \,\phi = 0 \:,\qquad \phi|_{t_0} = \tilde{\phi}_0 \:,
\eeq
we clearly get back the wave function~$\tilde{\phi}$.
An explicit representation for the solution of this Cauchy problem
can be given in terms of the causal fundamental solution, as we now
recall. 
In the vacuum, the {\em{advanced}} and {\em{retarded Green's functions}} are defined in momentum
space by
\[ s^{\lor}_{m}(k) = \lim_{\varepsilon \searrow 0}
\frac{\slashed{k} + m}{k^{2}-m^{2}-i \varepsilon k^{0}}
\qquad {\mbox{and}} \qquad
s^{\land}_{m}(k) = \lim_{\varepsilon \searrow 0}
\frac{\slashed{k} + m}{k^{2}-m^{2}+i \varepsilon k^{0}} \:, \]
respectively (with the limit~$\varepsilon \searrow 0$ taken in the distributional sense).
Computing the Fourier transform with residues, one finds that they are {\em{causal}} in the sense that
their supports lie in the upper and lower light cone, respectively.
In the presence of the external potential~$\B$, the advanced and retarded
Green's functions, which we denote for clarity with a tilde, are defined perturbatively by
\beq \tilde{s}_m^{\vee}=\sum_{n=0}^{\infty}(-s_m^{\vee} \,\mathscr{B})^n \,s_m^{\vee}\;, \qquad
\tilde{s}_m^{\wedge}=\sum_{n=0}^{\infty}(-s_m^{\wedge}\, \mathscr{B})^n \,s_m^{\wedge}\:,
\label{series-scaustilde}
\eeq
where the operator products involving  the potential $\mathscr{B}$ are defined as follows,
\[ (s_m^{\vee} \,\mathscr{B} \,s_m^{\vee})(x,y):=\int d^4z \:s_m^{\vee}(x,z)
\,\mathscr{B}(z) \,s_m^{\vee}(z,y) \:. \]
Using the causal support properties of the vacuum Green's functions,
one verifies inductively that every summand of the perturbation expansion~\eqref{series-scaustilde}
is again supported in the future respectively past light cone. 
The {\em{causal fundamental solution}} is defined by
\beq \label{kmgreen}
\tilde{k}_m = \frac{1}{2 \pi i} \left(\tilde{s}_m^\lor - \tilde{s}_m^\land \right) .
\eeq

\begin{Prp} \label{prpcauchy} The solution~$\tilde{\phi}$
of the Cauchy problem~\eqref{cauchyt} has the representation
\beq \label{Bcrep}
\tilde{\phi}(t, \vec{x}) = 2 \pi \int_{\R^3} \tilde{k}_m(t,\vec{x}; t_0, \vec{y})\, \gamma^0\, \tilde{\phi}_0(\vec{y})\: d^3y\:.
\eeq
\end{Prp}
\Proof Since~$\tilde{\phi}$ and~$\tilde{k}_m$ satisfy the Dirac equation, it suffices to prove
the proposition in the case~$t>t_0$. In this case, the formula~\eqref{Bcrep} 
simplifies in view of~\eqref{kmdef} to
\[ \tilde{\phi}(x) = i \int_{\R^3} \tilde{s}^\wedge_m(x,y)\, \gamma^0\, \tilde{\phi}_0(y) \big|_{y=(t_0, \vec{y})}
\: d^3y\:, \]
where we set~$x=(t, \vec{x})$.
This identity is derived as follows: We choose a non-negative function~$\eta \in C^\infty(\R)$
with~$\eta|_{[t_0, t]} \equiv 1$ and~$\eta_{(-\infty, t_0-1)} \equiv 0$.
We also consider~$\eta$ as a function on the time variable in space-time. Then
\beq \label{seta}
\tilde{\phi}(x) = (\eta \tilde{\phi})(x) = \tilde{s}^\wedge_m \big( (i \Pdd + \B - m) (\eta \tilde{\phi}) \big)
= \tilde{s}^\wedge_m \big( i \gamma^0 \,\dot{\eta}\, \tilde{\phi}) \big) \:,
\eeq
where we used the defining equation of the Green's function~$\tilde{s}_m^\wedge (i \Pdd_x +\B- m)=\1$
together with the fact that~$\tilde{\phi}$ is a solution of the Dirac equation.
To conclude the proof, we choose a sequence~$\eta_l$ such that
the sequence of derivatives~$\dot{\eta}_l$ converges as~$l \rightarrow \infty$ in the
distributional sense to the $\delta$-distribution~$\delta_{t_0}$ supported at~$t_0$. Then
\begin{align*}
\tilde{s}^\wedge_m \big( i \gamma^0 \,\dot{\eta}\, \tilde{\phi}) \big)(x)
&= \int \left( \tilde{s}^\wedge_m(x,y) \big( i \gamma^0 \, \dot{\eta}(y^0)\, \tilde{\phi}(y) \big) \right) d^4y \\
&\rightarrow \int_{\R^3} \left( \tilde{s}^\wedge_m(x,y) \big( i \gamma^0 \tilde{\phi}) \right) 
\big|_{y=(t_0, \vec{y})} \:d^3y \:,
\end{align*}
giving the result.
\QED

\begin{figure} %
\begin{picture}(0,0)%
\includegraphics{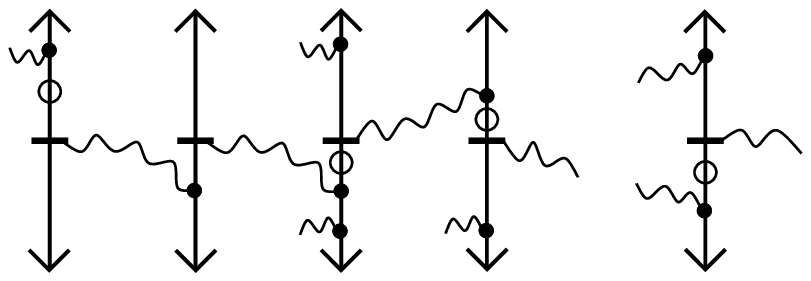}%
\end{picture}%
\setlength{\unitlength}{1533sp}%
\begingroup\makeatletter\ifx\SetFigFont\undefined%
\gdef\SetFigFont#1#2#3#4#5{%
  \reset@font\fontsize{#1}{#2pt}%
  \fontfamily{#3}\fontseries{#4}\fontshape{#5}%
  \selectfont}%
\fi\endgroup%
\begin{picture}(10728,3991)(-507,-3977)
\put(4807,-3330){\makebox(0,0)[lb]{\smash{{\SetFigFont{11}{13.2}{\rmdefault}{\mddefault}{\updefault}$\as$}}}}
\put(6600,-3307){\makebox(0,0)[lb]{\smash{{\SetFigFont{11}{13.2}{\rmdefault}{\mddefault}{\updefault}$\as$}}}}
\put(9307,-3315){\makebox(0,0)[lb]{\smash{{\SetFigFont{11}{13.2}{\rmdefault}{\mddefault}{\updefault}$\as$}}}}
\put(1215,-359){\makebox(0,0)[lb]{\smash{{\SetFigFont{11}{13.2}{\rmdefault}{\mddefault}{\updefault}$\bs$}}}}
\put(4836,-355){\makebox(0,0)[lb]{\smash{{\SetFigFont{11}{13.2}{\rmdefault}{\mddefault}{\updefault}$\bs$}}}}
\put(6658,-359){\makebox(0,0)[lb]{\smash{{\SetFigFont{11}{13.2}{\rmdefault}{\mddefault}{\updefault}$\bs$}}}}
\put(9291,-359){\makebox(0,0)[lb]{\smash{{\SetFigFont{11}{13.2}{\rmdefault}{\mddefault}{\updefault}$\bs$}}}}
\put(4411,-3841){\makebox(0,0)[lb]{\smash{{\SetFigFont{11}{13.2}{\rmdefault}{\mddefault}{\updefault}$2$}}}}
\put(6211,-3841){\makebox(0,0)[lb]{\smash{{\SetFigFont{11}{13.2}{\rmdefault}{\mddefault}{\updefault}$3$}}}}
\put(8911,-3841){\makebox(0,0)[lb]{\smash{{\SetFigFont{11}{13.2}{\rmdefault}{\mddefault}{\updefault}$n$}}}}
\put(811,-3841){\makebox(0,0)[lb]{\smash{{\SetFigFont{11}{13.2}{\rmdefault}{\mddefault}{\updefault}$1$}}}}
\put(7749,-1783){\makebox(0,0)[lb]{\smash{{\SetFigFont{11}{13.2}{\rmdefault}{\mddefault}{\updefault}$\cdots$}}}}
\put(7367,-3839){\makebox(0,0)[lb]{\smash{{\SetFigFont{11}{13.2}{\rmdefault}{\mddefault}{\updefault}$\cdots$}}}}
\put(1214,-3292){\makebox(0,0)[lb]{\smash{{\SetFigFont{11}{13.2}{\rmdefault}{\mddefault}{\updefault}$\as$}}}}
\put(4732,-2069){\makebox(0,0)[lb]{\smash{{\SetFigFont{11}{13.2}{\rmdefault}{\mddefault}{\updefault}$z_2$}}}}
\put(1116,-1199){\makebox(0,0)[lb]{\smash{{\SetFigFont{11}{13.2}{\rmdefault}{\mddefault}{\updefault}$z_1$}}}}
\put(6517,-1492){\makebox(0,0)[lb]{\smash{{\SetFigFont{11}{13.2}{\rmdefault}{\mddefault}{\updefault}$z_3$}}}}
\put(9217,-2183){\makebox(0,0)[lb]{\smash{{\SetFigFont{11}{13.2}{\rmdefault}{\mddefault}{\updefault}$z_n$}}}}
\put(-492,-731){\makebox(0,0)[lb]{\smash{{\SetFigFont{11}{13.2}{\rmdefault}{\mddefault}{\updefault}$\B^{(0)}_{|\bs}$}}}}
\put(3115,-684){\makebox(0,0)[lb]{\smash{{\SetFigFont{11}{13.2}{\rmdefault}{\mddefault}{\updefault}$\B^{(0)}_{|\bs}$}}}}
\put(4929,-2868){\makebox(0,0)[lb]{\smash{{\SetFigFont{11}{13.2}{\rmdefault}{\mddefault}{\updefault}$\B^{(0)}_{|\as}$}}}}
\put(3137,-2884){\makebox(0,0)[lb]{\smash{{\SetFigFont{11}{13.2}{\rmdefault}{\mddefault}{\updefault}$\B^{(0)}_{|\as}$}}}}
\end{picture}%
\caption{An \ansyb\ with insertions.}
\label{figinsert}
\end{figure}

Applying this proposition to the function~$\tilde{\phi}$ in~\eqref{Uphi} gives the identity
\beq \label{glue}
(U[\B] \phi \big)(x) = 2 \pi \int_{\R^3} \tilde{k}_m(x; t_0, \vec{z})\, \gamma^0\, 
(U[\B] \phi \big)(t_0, \vec{z})\: d^3 z \:.
\eeq
Graphically, this identity allows us to ``glue together'' a solution of the Dirac equation
with the corresponding fundamental solution. Therefore, we refer to~\eqref{glue} as
the {\bf{glueing identity}}. Expanding this identity to the desired order in~$\B$ and
selecting the contributions with a certain ordering in the different contributions to~$\B$
(as explained after~\eqref{Uphi}), we can decompose the bra state in
a particle line into a sum of new lines with the same ingoing bosonic lines but
with an additional dependence on a space-time point~$z:=(t_0, \vec{z})$. The construction ensures that
after carrying out the spatial integral over~$\vec{z}$, we get back the original line.
In the Feynman diagrams, we denote the ``insertions'' by a small circle
(see Figure~\ref{figinsert}).

The insertion procedure just described for the bra state at the very right can be
carried out similarly for the ket state and for all the other particle lines.
We insert one space-time point~$z_l$ in the $l^\text{th}$ particle line.
At each particle line, we have the freedom to insert either in the bra state or in the
ket state. In order to determine which insertion to choose, we proceed inductively
from the right to the left, applying the following rules:
\begin{itemize}
\item[(1)] At the particle line to the very right, we take the mean value of the
bra insertion and the ket insertion.
\item[(2)] Assume that the particle lines labelled by~$l+1, \ldots, n$ already carry an insertion.
We then consider the bosonic line generated at the $l^\text{th}$ particle line and follow
it to the right. If it enters a sea line, we go to the bosonic line which leaves this sea line
and again follow it to the right. We proceed until the bosonic line enters one of
the particle lines~$l+1, \ldots, n$.
\item[(3)] If this bosonic line ends below the insertion,
at the $l^\text{th}$ particle line we take the bra insertion. Otherwise, we take the ket insertion.
\end{itemize}
These rules ensure that a bosonic line generated below an insertion also ends below
an insertion. Similarly, every bosonic line generated above also ends above
the insertion. Note that no insertions are introduced in the sea lines.
This construction is illustrated in Figure~\ref{figinsert}.

\begin{figure} %
\begin{picture}(0,0)%
\includegraphics{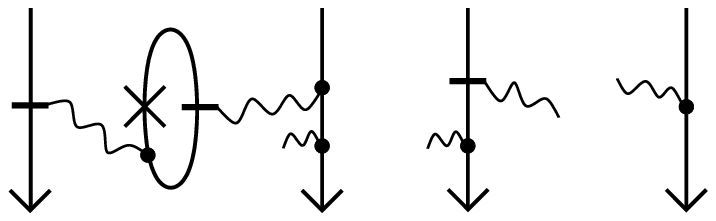}%
\end{picture}%
\setlength{\unitlength}{1533sp}%
\begingroup\makeatletter\ifx\SetFigFont\undefined%
\gdef\SetFigFont#1#2#3#4#5{%
  \reset@font\fontsize{#1}{#2pt}%
  \fontfamily{#3}\fontseries{#4}\fontshape{#5}%
  \selectfont}%
\fi\endgroup%
\begin{picture}(13313,3261)(-3991,-3466)
\put(1214,-3292){\makebox(0,0)[lb]{\smash{{\SetFigFont{11}{13.2}{\rmdefault}{\mddefault}{\updefault}$\as$}}}}
\put(4807,-3330){\makebox(0,0)[lb]{\smash{{\SetFigFont{11}{13.2}{\rmdefault}{\mddefault}{\updefault}$\as$}}}}
\put(6600,-3307){\makebox(0,0)[lb]{\smash{{\SetFigFont{11}{13.2}{\rmdefault}{\mddefault}{\updefault}$\as$}}}}
\put(9307,-3315){\makebox(0,0)[lb]{\smash{{\SetFigFont{11}{13.2}{\rmdefault}{\mddefault}{\updefault}$\as$}}}}
\put(7546,-2526){\makebox(0,0)[lb]{\smash{{\SetFigFont{11}{13.2}{\rmdefault}{\mddefault}{\updefault}$\cdots$}}}}
\put(1051,-1037){\makebox(0,0)[lb]{\smash{{\SetFigFont{11}{13.2}{\rmdefault}{\mddefault}{\updefault}$\alpha_1$}}}}
\put(4344,-556){\makebox(0,0)[lb]{\smash{{\SetFigFont{11}{13.2}{\rmdefault}{\mddefault}{\updefault}$z_2$}}}}
\put(6144,-556){\makebox(0,0)[lb]{\smash{{\SetFigFont{11}{13.2}{\rmdefault}{\mddefault}{\updefault}$z_3$}}}}
\put(8837,-556){\makebox(0,0)[lb]{\smash{{\SetFigFont{11}{13.2}{\rmdefault}{\mddefault}{\updefault}$z_n$}}}}
\put(7592,-556){\makebox(0,0)[lb]{\smash{{\SetFigFont{11}{13.2}{\rmdefault}{\mddefault}{\updefault}$\cdots$}}}}
\put(-3976,-2056){\makebox(0,0)[lb]{\smash{{\SetFigFont{11}{13.2}{\rmdefault}{\mddefault}{\updefault}$\Psi^{\alpha_1 \cdots \alpha_n}(z_1, \ldots, z_n)=$}}}}
\put(9121,-1074){\makebox(0,0)[lb]{\smash{{\SetFigFont{11}{13.2}{\rmdefault}{\mddefault}{\updefault}$\alpha_n$}}}}
\put(781,-556){\makebox(0,0)[lb]{\smash{{\SetFigFont{11}{13.2}{\rmdefault}{\mddefault}{\updefault}$z_1$}}}}
\put(6436,-1074){\makebox(0,0)[lb]{\smash{{\SetFigFont{11}{13.2}{\rmdefault}{\mddefault}{\updefault}$\alpha_3$}}}}
\put(4673,-1074){\makebox(0,0)[lb]{\smash{{\SetFigFont{11}{13.2}{\rmdefault}{\mddefault}{\updefault}$\alpha_2$}}}}
\put(3174,-2689){\makebox(0,0)[lb]{\smash{{\SetFigFont{11}{13.2}{\rmdefault}{\mddefault}{\updefault}$\B^{(0)}_{|\as}$}}}}
\put(4943,-2697){\makebox(0,0)[lb]{\smash{{\SetFigFont{11}{13.2}{\rmdefault}{\mddefault}{\updefault}$\B^{(0)}_{|\as}$}}}}
\end{picture}%
\caption{An $n$-particle wave function.}
\label{figwave}
\end{figure}

Having introduced the insertions, we cut the diagram at each insertion
dropping the matrix~$\gamma^0$, i.e.\ symbolically
\[ \overline{\psi(z)} \gamma^0 \phi(z) \rightarrow \overline{\psi(z)} \text{ and } \phi(z) \:. \]
Then all the ket states give rise to an $n$-particle wave function
\beq \label{Psiintro}
\Psi^{\alpha_1 \cdots \alpha_n}(z_1, \ldots, z_n)\:.
\eeq
Moreover, we ``fold'' the sea lines to obtain fermion loops.
This is illustrated in Figure~\ref{figwave}.

\subsection{The Fock-Krein Space, Unitarity of the Time Evolution} \label{seckrein}
We would like to endow the $n$-particle wave functions~\eqref{Psiintro} (like in Figure~\ref{figwave})
with an inner product. Taking the inner product of two $n$-particle
wave functions should give back the corresponding diagrams before cutting (like in Figure~\ref{figasbsea}).
In order to accomplish this goal, one should keep in mind that  before cutting, at every particle line
exactly one bosonic lines begins (see the thick vertical lines in Figure~\ref{figasbsea}).
After cutting, however, the particle line may or may not involve an outgoing bosonic line
(see Figure~\ref{figwave}). Thus when forming the inner product, we must make sure
that when pairing two particle lines, exactly one of these lines should involve an outgoing bosonic line.
In order keep track of the outgoing bosonic lines, we introduce an additional index~$b_l$
which takes the values
\beq \label{bldef}
b_l = \left\{ \!\! \begin{array}{c} -1 \\ 1 \end{array} \!\! \right\} \text{ if the $l^\text{th}$ particle line has }
\left\{ \!\! \begin{array}{c} \text{no} \\ \text{one} \end{array} \!\! \right\} \text{ outgoing bosonic line}\:.
\eeq
Thus we write the $n$-particle wave function~\eqref{Psiintro} as
\beq \label{psibl}
\Psi^{\alpha_1 \cdots \alpha_n}_{b_1 \cdots b_n}(z_1, \ldots, z_n)\:.
\eeq
Choosing all the space-time points~$z_1, \ldots z_n$ at the same time~$t$, we obtain a vector in
the following Hilbert space,
\beq \label{L2top}
\Psi^{\alpha_1 \cdots \alpha_n}_{b_1 \cdots b_n}(t; \vec{z}_1, \ldots, \vec{z}_n)
\in L^2(\R^3, \C^8)^n \cong L^2(\R^{3n}, \C^{8^n})
\eeq
(the eight components come about as four Dirac components, taken twice because of
the two possible values of~$b_l$). The inner product of two such wave functions is defined by
\beq \label{kreinip}
\begin{split}
(\Psi | \Phi)|_t &:=
\int_{\R^3} \!\!\! d^3z_1 \cdots\! \int_{\R^3} \!\!\! d^3z_n\; \sum_{\alpha_1, \ldots, \alpha_n=1}^4 \;
\sum_{b_1, \ldots, b_n= \pm 1} \; \sum_{b'_1, \ldots, b'_n= \pm 1} \\
&\qquad \times \Psi^{\alpha_1 \cdots \alpha_n}_{b_1 \cdots b_n}(t; \vec{z}_1, \ldots, \vec{z}_n)^\dagger
\;{\mathfrak{t}}^{b_1}_{b'_1} \cdots {\mathfrak{t}}^{b_n}_{b'_n}\;
\Phi^{\alpha_1 \cdots \alpha_n}_{b'_1 \cdots b'_n}(t; \vec{z}_1, \ldots, \vec{z}_n) \:,
\end{split}
\eeq
where the matrix~${\mathfrak{t}}$ only has off-diagonal contributions,
\beq \label{Tdef}
{\mathfrak{t}}^b_{b'} = 1-\delta^b_{b'}\:.
\eeq
Hence apart from from the matrix~${\mathfrak{t}}$,
at every particle line we take the usual spatial scalar product on Dirac wave functions
(cf.~\eqref{print}).
The matrix~${\mathfrak{t}}$ ensures that every paired particle line involves exactly one outgoing bosonic
line.

We point out that in general, the $n$-particle wave function~\eqref{L2top} is {\em{not totally antisymmetric}}.
It is totally antisymmetric in the free particle states in~$\I_0$ entering the bra
of each particle line (see Figure~\ref{figwave}). However, since the interacting lines
are not symmetric under permutations of the particle lines (as is obvious in Figure~\ref{figwave}),
the resulting wave function is no longer totally antisymmetric.
This loss of anti-symmetry can be understood from the fact that the particle lines
in our diagrams are always ordered such that the bosonic lines go from the left to the right.
This ordering is inherent in the perturbation expansion of the nonlinear system of the Dirac equation coupled to
a classical bosonic field equation~\eqref{UPab}--\eqref{jiter2}, where~$\B^{(n)}$ couples to~$P^{(n')}$
only if~$n'>n$. To avoid confusion, we note that the $n$ fermions are nevertheless
indistinguishable, because~$\Psi$ is still totally antisymmetric in the states of~$\I_0$
(hence interchanging any two particles gives a minus sign).
The point is that the arguments~$z_1, \ldots, z_n$ do not refer to the different particles, but
are ordered with respect to the order in perturbation theory. For this reason, the wave function~$\Psi$
is not antisymmetric in these arguments.

Moreover, we note that,
since the matrix~$S$ has one positive and one negative eigenvalue, the inner
product~\eqref{kreinip} is {\em{not positive definite}}. It merely is an indefinite inner product.
Together with the $L^2$-topology in~\eqref{L2top}, we obtain a Krein space,
the so-called {\bf{Fock-Krein space}}~$(\Fock^\text{\tiny{\rm{Krein}}}_n, (.|.))$.

\begin{figure} %
\begin{picture}(0,0)%
\includegraphics{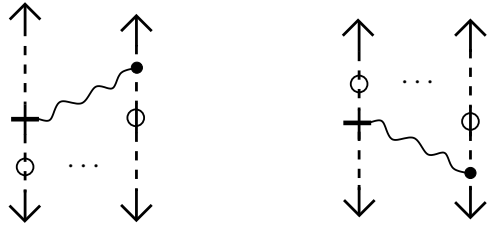}%
\end{picture}%
\setlength{\unitlength}{1533sp}%
\begingroup\makeatletter\ifx\SetFigFont\undefined%
\gdef\SetFigFont#1#2#3#4#5{%
  \reset@font\fontsize{#1}{#2pt}%
  \fontfamily{#3}\fontseries{#4}\fontshape{#5}%
  \selectfont}%
\fi\endgroup%
\begin{picture}(7787,3595)(605,-3236)
\put(1576,-2415){\makebox(0,0)[lb]{\smash{{\SetFigFont{11}{13.2}{\rmdefault}{\mddefault}{\updefault}$\cdots$}}}}
\put(6976,-1058){\makebox(0,0)[lb]{\smash{{\SetFigFont{11}{13.2}{\rmdefault}{\mddefault}{\updefault}$\cdots$}}}}
\end{picture}%
\caption{Symmetry giving current conservation.}
\label{figconserve}
\end{figure}

\begin{Thm} {\bf{(unitarity)}} \label{thmunitary}
For any two wave functions~$\Psi, \Phi \in \Fock^\text{\tiny{\rm{Krein}}}_n$ obtained by cutting the
\ansybs\ according to the construction in Section~\ref{seccut},
the inner product~\eqref{kreinip} is independent of the time~$t$.
\end{Thm}
\Proof We consider the diagrams and proceed inductively from the
left to the right. At the space-time point~$z_1$, both the bra and the ket states are solutions of the Dirac
equation in the external field~$\B^{(0)}_\as$. Hence by current conservation, it is clear
that the integral over~$\vec{z}_1$ is independent of~$t$ (keeping the space-time
points~$z_2, \ldots, z_n$ fixed). After carrying out the integral over~$\vec{z}_1$,
the bra and ket states at~$z_1$ are ``glued together'' according to~\eqref{glue}.
Then the bra and ket states at the position~$z_2$ are solutions of the Dirac equation
in the external field~$\B$ being the sum of~$\B^{(0)}_\as$ and the bosonic field leaving the
first particle line. Hence we again have current conservation, so that the integral over~$\vec{z}_2$
is again time independent (keeping~$z_3, \ldots, z_n$ fixed). Proceeding inductively, we conclude
that all the spatial integrals, and thus also the inner product~\eqref{kreinip}, are time independent.
\QED
We remark that the current conservation used in the above proof can also be verified
to every order in perturbation theory. It then means that at every particle line,
the same bosonic lines enter above and below the cutting points. This is ensured by
the symmetry in our cutting rules as depicted in Figure~\ref{figconserve}.
We also point out that it is crucial for the proof that in~\eqref{kreinip}, the $l^\text{th}$ argument
of~$\Psi$ is paired to the $l^\text{th}$ argument of~$\Phi$. In particular, if we had totally
anti-symmetrized~$\Psi$ and~$\Phi$, unitarity would have been lost.

\subsection{The Freedom in Choosing the Green's Functions} \label{secgreen}
We can now clarify a point which was disregarded so far: the {\em{freedom in choosing the Green's functions}}.
The {\em{fermionic}} Green's functions are determined by the causal perturbation expansion,
which gives a concise combinatorics for the Green's functions and fundamental solutions
in the operator product expansion~\eqref{Pexform}.
We remark that the combinatorial details depend on which normalization condition is imposed on the
fermionic projector, but this is of no relevance here (for details see~\cite{norm}).
For the particle lines, it may not be convenient to work with the Green's functions of the causal
perturbation expansion. For example to describe a scattering process, it is preferable to choose the
wave functions of the particle lines initially, and then to study the retarded time evolution into the future.
In such situations, the general procedure is to simply choose the Green's functions of the particle lines
in the most suitable way. The Green's functions of the sea lines, however, must always be chosen
according to the causal perturbation expansion. For a detailed derivation and discussion of this
procedure we refer to the paper~\cite{norm}.

The {\em{bosonic}} Green's functions can be chosen arbitrarily, except for the following important
condition: For the above current conservation to hold, it is essential that to every particle line,
the same bosonic field couples above and below the cutting points.
As is illustrated in Figure~\ref{figconserve}, the bosonic lines which couples above the cutting
point is generated by the bra to the left, whereas the bosonic line which enters the below the cutting point
is generated by the ket. Since the inner product~\eqref{kreinip} involves a Hermitian conjugation of the
bra, the bosonic fields will be equal only if the bosonic Green's function is Hermitian, meaning that
\beq \label{S0real}
S_0(x,y)^\dagger = S_0(x,y) \:.
\eeq
For a scalar bosonic field, this condition simply means that the bosonic Green's function is real-valued.
Similarly, for vector bosons every vector component should be real-valued.
The condition~\eqref{S0real} is satisfied for the retarded and advanced Green's functions.
But it is violated for the Feynman propagator, which is complex-valued.
A particular choice for~$S_0$ which complies with~\eqref{S0real} is the retarded Green's
function~$S_0^\wedge$, which for a massless scalar field takes the form
\beq \label{S0def}
S_0^\wedge(x,y) = \lim_{\varepsilon \searrow 0} \int \frac{d^4k}{(2 \pi)^4} \:
\frac{1}{k^{2}+i \varepsilon k^{0}} \: e^{-ik (x-y)} \:.
\eeq
This Green's function is the canonical choice when considering the initial-value problem and
evolving the system to the future. In other situations, however, one may prefer not to distinguish
a direction of time. Then it is natural to choose the mean of the advanced and retarded Green's function,
\[ S_0 = \frac{1}{2} \left( S_0^\wedge + S_0^\vee \right) \]
(where the advanced Green's function~$S_0^\vee$ is obtained by flipping the sign
of~$\varepsilon$ in~\eqref{S0def}).
This choice is also compatible with~\eqref{S0def}. Clearly, there are many other choices which
satisfy~\eqref{S0def}. In what follows, we make the physically reasonable assumption that
the Green's function~$S_0(x,y)$ should be causal in the sense that it vanishes if~$x$ and~$y$
have space-like separation. This reduces the freedom to the ansatz
\beq \label{S0caus}
S_0 = \tau \,S_0^\wedge + (1-\tau) \,S_0^\vee \qquad \text{with~$\tau \in \R$}\:,
\eeq
involving one free parameter~$\tau$.

\subsection{Explanatory Remarks}
The cutting of the particle lines and the Fock-Krein space require some explanation.
We first reconsider the construction of cutting the diagrams to obtain many-particle wave functions
(see Figures~\ref{figinsert} and~\ref{figwave}).
Before cutting, in every particle line one can distinguish a bra
and a ket state, which are separated by the outgoing bosonic line (thus in Figure~\ref{figasbsea},
the part above the thick horizontal line is the bra, and the part below is the ket).
It is the starting point of our construction that this distinction between bra and ket
does not seem accessible to an observer. Namely, a measurement device, which takes part
in the interaction, couples in the same way to the bra and the ket of each particle line.
Our cutting procedure specifies the possible points in each particle line where a ``virtual observer'' might
couple to the line. Our cutting procedure ensures that if we integrate over the positions, we recover the
diagrams before cutting.

In order to make this picture of a ``virtual observer'' more precise, one could for example
add a test charge~$q$ to the system and consider how this test charge couples to the
particle line. This could be described quantitatively by taking {\em{variational derivatives}}
with respect to the electromagnetic potential, i.e.\ symbolically
\[ \frac{\delta}{\delta \B(y_1)} \cdots \frac{\delta}{\delta \B(y_n)} (\ansyb) \:. \]
Taking such variational derivatives is in fact similar to our cutting procedure. However,
there is the major difference that a variational derivative introduces a Green's function.
More specifically, in the example of an electromagnetic potential, one has
\beq \label{varder}
\frac{\delta}{\delta A_0(y)} \phi(x) = \tilde{s}_m(x,y) \,\gamma^0\, \phi(y) \:.
\eeq
This differs from our cutting rule~\eqref{glue} in that the fundamental solution~$\tilde{k}_m$
has been replaced by the Green's function~$\tilde{s}_m$.

This difference is in fact irrelevant if we consider a scattering process. Namely, in this
situation, the ``virtual observer'' is in the past or future of the interaction region.
As a consequence, we know that the argument~$y$ of the Green's function lies to the
past respectively future of the argument~$x$. Thus we may use~\eqref{kmgreen}
to rewrite the Green's functions in terms of the fundamental solution.
Then the variational derivative~\eqref{varder} goes over to the cutting rule~\eqref{glue}
(up to irrelevant prefactors, which we do not want to discuss here).
But it is important that~\eqref{varder} and~\eqref{glue} do {\em{not agree
for intermediate times}} in a scattering process (or, more generally, if an interaction is present for all times).
Namely, in this case, the computation
\[ \frac{\partial}{\partial y^j} \tilde{s}_m(x,y) \gamma^j \phi(y) = i \delta^4(x-y) \phi(y) \]
shows that the vector field corresponding to~\eqref{varder} is not divergence-free.
As a consequence, the spatial integral of~\eqref{varder} will in general depend on time,
so that the unitarity statement analogous to Theorem~\ref{thmunitary} would no longer hold if
we worked with variational derivatives~\eqref{varder}.

We conclude that in a scattering process, our cutting procedure is equivalent to working with
variational derivatives~\eqref{varder} corresponding to ``virtual test charges''.
This justifies the physical picture that the cutting points are coupled to ``virtual observers.''
However, this physical picture fails at intermediate times or if an interaction is present
at all times. In this general situation, the cutting procedure has the great advantage that
the quantum field is well-defined even at intermediate times.

One might ask why every particle line involves {\em{exactly only one
cutting point}}. First, working with fewer cutting points would have the major disadvantage that the
diagrams would not split into two separate many-particle wave functions, making it
impossible to introduce many-particle states. Moreover, since by integrating over a
position~$\vec{z}_l$ we can always remove the corresponding cut, it seems no loss in generality
to consider exactly~$n$ cutting points.
The fact that every particle line should not involve more than one cut can be understood
from the causality in a scattering process. Namely, in a scattering process the time evolution
is causal, meaning that in every particle line we work with the retarded Green's function.
If our virtual observer acts at some fixed time~$t$ at the distinct positions~$\vec{z}_1, \ldots, \vec{z}_n$,
then the corresponding space-time points~$(t, \vec{z}_l)$ are spacelike separated.
As a consequence, more than one cutting point at one particle line is impossible due to finite
propagation speed.

A final point which might need clarification is why we were
so careful that every bosonic line belongs either completely to the bra or completely the ket state
(see the cutting rule~(2)). Why is not possible to cut the particle lines once above and once
below the bosonic line? Clearly, this would make it necessary to also cut the bosonic line.
But why is it not possible to do so?
The answer to this question is that the bosonic line is described by a Green's function~$S_0$,
and that it is impossible for principal reasons to cut a Green's function.
To see the problem, suppose that we had cut a Green's function~$S_0(x,y)$
in such a way that a spatial integral over the cutting point gives back the Green's function, i.e.\ symbolically
\beq \label{S0dec}
S_0(x,y) = \int_{\R^3} A \big( x,(t,\vec{z}) \big) \,B \big( (t, \vec{z}),y \big) \: d^3z \:.
\eeq
In order for the cutting procedure to be applicable to a scattering process, this formula
should hold for any sufficiently large~$t$.
Here~$A$ and~$B$ should be suitable Green's functions or fundamental solutions.
Then for given~$x$ and~$y$, we can choose~$t$ in the future of~$x$ and~$y$.
Applying the wave operator acting on~$x$ to the right side of~\eqref{S0def},
we get a contribution only if~$x=(t, \vec{z})$. However, as~$t$ lies in the future of~$x$,
this contribution vanishes. Hence~$S_0(x,y)$ is a solution of the homogeneous
wave equation, in contradiction to the properties of the Green's function.
In other words, in the decomposition~\eqref{S0dec} it is impossible to encode the
fact that~$S_0$ is a solution of the {\em{inhomogeneous}} wave equation with
inhomogeneity~$\delta^4(x-y)$.

\section{Recombination of Anti-Symmetrized Synchronal Blocks} \label{secrecombine}
In Section~\ref{secstoch} we introduced a stochastic background field
which led to a synchronization of the legs of diagrams (see Figure~\ref{figbackground}).
We now analyze the effect of the stochastic background field more quantitatively.

\subsection{Stochastic Coupling to the Sea} \label{secstocup}
\begin{figure} %
\begin{picture}(0,0)%
\includegraphics{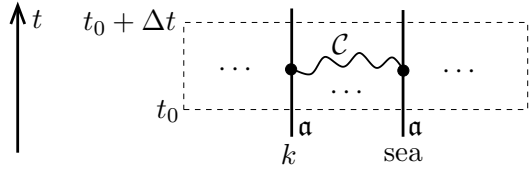}%
\end{picture}%
\setlength{\unitlength}{1533sp}%
\begingroup\makeatletter\ifx\SetFigFont\undefined%
\gdef\SetFigFont#1#2#3#4#5{%
  \reset@font\fontsize{#1}{#2pt}%
  \fontfamily{#3}\fontseries{#4}\fontshape{#5}%
  \selectfont}%
\fi\endgroup%
\begin{picture}(8422,2817)(-3684,-3477)
\put(724,-3341){\makebox(0,0)[lb]{\smash{{\SetFigFont{11}{13.2}{\rmdefault}{\mddefault}{\updefault}$k$}}}}
\put(2389,-3301){\makebox(0,0)[lb]{\smash{{\SetFigFont{11}{13.2}{\rmdefault}{\mddefault}{\updefault}sea}}}}
\put(999,-2867){\makebox(0,0)[lb]{\smash{{\SetFigFont{11}{13.2}{\rmdefault}{\mddefault}{\updefault}$\as$}}}}
\put(-3296,-1129){\makebox(0,0)[lb]{\smash{{\SetFigFont{11}{13.2}{\rmdefault}{\mddefault}{\updefault}$t$}}}}
\put(1514,-2296){\makebox(0,0)[lb]{\smash{{\SetFigFont{11}{13.2}{\rmdefault}{\mddefault}{\updefault}$\cdots$}}}}
\put(2779,-2877){\makebox(0,0)[lb]{\smash{{\SetFigFont{11}{13.2}{\rmdefault}{\mddefault}{\updefault}$\as$}}}}
\put(-286,-1921){\makebox(0,0)[lb]{\smash{{\SetFigFont{11}{13.2}{\rmdefault}{\mddefault}{\updefault}$\cdots$}}}}
\put(3334,-1951){\makebox(0,0)[lb]{\smash{{\SetFigFont{11}{13.2}{\rmdefault}{\mddefault}{\updefault}$\cdots$}}}}
\put(-1284,-2586){\makebox(0,0)[lb]{\smash{{\SetFigFont{11}{13.2}{\rmdefault}{\mddefault}{\updefault}$t_0$}}}}
\put(-2471,-1174){\makebox(0,0)[lb]{\smash{{\SetFigFont{11}{13.2}{\rmdefault}{\mddefault}{\updefault}$t_0+\Delta t$}}}}
\put(1557,-1556){\makebox(0,0)[lb]{\smash{{\SetFigFont{11}{13.2}{\rmdefault}{\mddefault}{\updefault}${\mathcal{C}}$}}}}
\end{picture}%
\caption{Coupling by the stochastic background field.}
\label{figstocoup}
\end{figure}

In order to analyze the interaction of the fermionic states via the stochastic background field
in a concrete example, we consider the effect of an electromagnetic field to first order in
perturbation theory. Then the interaction of the $k^\text{th}$ particle state~$\Psi_k$ with
a sea state~$\psi_\text{sea}$ is described by the Feynman diagram in Figure~\ref{figstocoup}
corresponding to the analytic expression
\beq \label{firstorder}
\iint_{M \times M} d^4x \,d^4y \;\big( \overline{\Psi_k^\text{out}(x)} \gamma^j \Psi_k^\text{in}(x) \big)
\: {\mathcal{C}}^{ij}(x-y) \:\big( \overline{\psi^\text{out}_\text{sea}(y)} \gamma^j \psi_\text{sea}^\text{in}(y) \big) \:,
\eeq
were~${\mathcal{C}}$ is the covariance in~\eqref{covsymm} (which for simplicity we assume to
be homogeneous, so that it only depends on the difference vector~$x-y$).
We want to study on which time scale the interaction comes into play. To this end, we
localize the interaction to the time interval~$[t_0, t_0+\Delta t]$ (see Figure~\ref{figstocoup}).
In order to have a smooth cutoff, we choose a test function~$\eta \in C^\infty_0((t, t+\Delta t))$
and insert it into the space-time integrals in~\eqref{firstorder} by the replacement
\[ \iint_{M \times M} d^4x \,d^4y  \longrightarrow
\int_{\R^4} \eta(t)\, dt\: d^3x \int_{\R^4} \eta(t')\, dt'\: d^3y \:. \]
Moreover, it is preferable to arrange discrete fermionic states by considering the system
in finite three-volume. Replacing space by a three-dimensional torus 
of length~$\ell$ (just as in~\cite[Section~2.1]{veltman} or similarly in~\cite[\S2.6]{PFP}),
the momenta (denoted in what follows by~$\vec{p}$, $\vec{q}$ and~$\vec{r}$) are on the lattice
\[ {\mathfrak{L}} := \Big( \frac{2 \pi}{\ell}\: \Z \Big)^3 \:. \]
Rewriting the spatial integrals in~\eqref{firstorder} in momentum space, we obtain the expression
\beq \label{express}
\begin{split}
\frac{1}{\ell^9} &\sum_{\vec{p}, \vec{q}, \vec{r} \in {\mathfrak{L}}}
\int_{-\infty}^\infty \eta(t) \,dt \int_{-\infty}^\infty \eta(t')\, dt' \\
&\times  \big( \overline{\Psi_k^\text{out}(t; \vec{p}+\vec{q})} \gamma^j \Psi_k^\text{in}(t; \vec{p}) \big)
\: {\mathcal{C}}^{ij}(t-t'; \vec{q}) \:\big( \overline{\psi^\text{out}_\text{sea}(t';\vec{r}-\vec{q})} \gamma^j
\psi_\text{sea}^\text{in}(t'; \vec{r}) \big) \:.
\end{split}
\eeq
Here~$\vec{q}$ is the momentum of the bosonic line, and~$\vec{p}$ and~$\vec{r}$ are the momenta
of the incoming fermionic states. Note that, due to momentum conservation,
this determines the outgoing momenta.

Next, we use that the bosonic field is massless, meaning that~${\mathcal{C}}$ is supported on the
mass cone. Thus we can write~${\mathcal{C}}$ as
\beq \label{Cansatz}
{\mathcal{C}}^{ij}(t-t'; \vec{q}) = \sum_\pm f^{ij}_\pm(\vec{q})\: e^{\pm i |\vec{q}| (t-t')}
\eeq
for suitable functions~$f^{ij}_\pm$. Using this ansatz, we can write~\eqref{express} as
\beq \label{scatter}
\begin{split}
\sum_\pm \frac{1}{\ell^9} &\sum_{\vec{p}, \vec{q}, \vec{r} \in {\mathfrak{L}}} f^{ij}_\pm(\vec{q})
\int_{-\infty}^\infty \eta(t) \,dt \int_{-\infty}^\infty \eta(t')\, dt' \:  e^{\pm i |\vec{q}| (t-t')} \\
&\times  \big( \overline{\Psi_k^\text{out}(t; \vec{p}+\vec{q})} \gamma^j \Psi_k^\text{in}(t; \vec{p}) \big)
\; \big( \overline{\psi^\text{out}_\text{sea}(t';\vec{r}-\vec{q})} \gamma^j
\psi_\text{sea}^\text{in}(t'; \vec{r}) \big) \:.
\end{split}
\eeq
A typical example for the covariance is to choose a multiple times the causal fundamental solution,
\beq \label{Covscale}
{\mathcal{C}}^{ij}(q) = \nu \,g^{ij}\: \delta(q^2)\: \epsilon(q^0)
\eeq
with a coupling constant~$\nu \in \R$. In this case,
\[ {\mathcal{C}}^{ij}(t-t'; \vec{q})
= \nu\, g^{ij} \int_{-\infty}^\infty \frac{d\omega}{2 \pi}\: \delta(\omega^2-|\vec{q}\,|^2)\: \epsilon(\omega)\:
e^{-i \omega(t-t')} = \pm \frac{\nu g^{ij} }{4 \pi\, |\vec{q}\,|}\: e^{\mp i |\vec{q}\,| (t-t')} \]
and thus
\beq \label{fijscale}
f^{ij}_\pm(\vec{q}) = \mp \frac{\nu g^{ij} }{4 \pi\, |\vec{q}\,|} \:.
\eeq

\begin{figure} %
\begin{picture}(0,0)%
\includegraphics{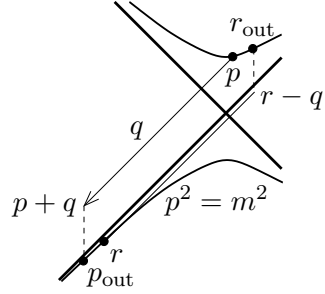}%
\end{picture}%
\setlength{\unitlength}{1533sp}%
\begingroup\makeatletter\ifx\SetFigFont\undefined%
\gdef\SetFigFont#1#2#3#4#5{%
  \reset@font\fontsize{#1}{#2pt}%
  \fontfamily{#3}\fontseries{#4}\fontshape{#5}%
  \selectfont}%
\fi\endgroup%
\begin{picture}(4414,4711)(-5261,-3828)
\put(-4041,-3674){\makebox(0,0)[lb]{\smash{{\SetFigFont{11}{13.2}{\rmdefault}{\mddefault}{\updefault}$p_\text{out}$}}}}
\put(-3681,-3369){\makebox(0,0)[lb]{\smash{{\SetFigFont{11}{13.2}{\rmdefault}{\mddefault}{\updefault}$r$}}}}
\put(-1801,-444){\makebox(0,0)[lb]{\smash{{\SetFigFont{11}{13.2}{\rmdefault}{\mddefault}{\updefault}$p$}}}}
\put(-1251,-809){\makebox(0,0)[lb]{\smash{{\SetFigFont{11}{13.2}{\rmdefault}{\mddefault}{\updefault}$r-q$}}}}
\put(-3356,-1299){\makebox(0,0)[lb]{\smash{{\SetFigFont{11}{13.2}{\rmdefault}{\mddefault}{\updefault}$q$}}}}
\put(-5246,-2549){\makebox(0,0)[lb]{\smash{{\SetFigFont{11}{13.2}{\rmdefault}{\mddefault}{\updefault}$p+q$}}}}
\put(-2786,-2524){\makebox(0,0)[lb]{\smash{{\SetFigFont{11}{13.2}{\rmdefault}{\mddefault}{\updefault}$p^2=m^2$}}}}
\put(-1746,366){\makebox(0,0)[lb]{\smash{{\SetFigFont{11}{13.2}{\rmdefault}{\mddefault}{\updefault}$r_\text{out}$}}}}
\end{picture}%
\caption{A typical exchange process.}
\label{figscatter}
\end{figure}

Similar as in a scattering process, we now consider the situation when the in- and outgoing
fermionic states are on-shell. Denoting the incoming momenta by~$p$ and~$r$,
the outgoing momenta~$p_\text{out}$ and~$q_\text{out}$ are given by
\[ p_\text{out} = \big( \pm \sqrt{|\vec{p}+\vec{q}|^2+m^2}, \vec{p}+\vec{q} \big) \qquad \text{and} \qquad
r_\text{out} = \big( \pm \sqrt{|\vec{r}-\vec{q}|^2+m^2}, \vec{r}-\vec{q} \big) \:. \]
Substituting the corresponding plane-wave solutions into~\eqref{scatter}, we can rewrite the time integrals
in terms of the Fourier transforms of~$\eta$. We thus obtain, up to irrelevant constants, the
expression
\beq \label{Lint}
\frac{1}{\ell^3} \sum_{\vec{q} \in {\mathfrak{L}}} f^{ij}_\pm(\vec{q})\;
\hat{\eta}\big(p_\text{out}^0 - p^0 \pm |\vec{q}| \big)  \:
\hat{\eta}\big(r_\text{out}^0 - r^0 \mp |\vec{q}| \big) \:.
\eeq
In the limiting case of an infinite interaction time, the function~$\eta$ becomes constant,
so that~$\hat{\eta}$ goes over to a multiple times the $\delta$ distribution.
Then the factors~$\hat{\eta}$ in~\eqref{Lint} express the conservation of energy at the two vertices.
But the conservation of energy and momentum prevents any nontrivial choices of~$p$, $q$ and~$r$
(see Figure~\ref{figscatter}).
The situation becomes more interesting if the interaction time~$\Delta t$ is finite. Then the
functions~$\hat{\eta}$ decay on the scale~$1/(\Delta t)$. As shown in Figure~\ref{figscatter},
this gives rise to contributions to~\eqref{Lint} for large momenta~$q$ if~$p$ lies on the upper and~$r$
lies on the lower mass shell. The outgoing momenta~$q_\text{out}$ and~$r_\text{out}$ lie on the
lower and upper mass shell, respectively. Thus the particle and the sea state exchange their roles,
motivating the name {\em{exchange process}}. For this process to occur, the arguments of the
functions~$\hat{\eta}$ in~\eqref{Lint} (denoted in Figure~\ref{figscatter} by dashed lines) are of the
order of the Compton scale, meaning that
\beq \label{Cscale}
\Delta t \lesssim m^{-1} \:.
\eeq
Let us consider the amplitude of the exchange process. Since~$q$ can be chosen arbitrarily large,
the vector~$\vec{q}$ can be an arbitrary vector of the lattice~${\mathfrak{L}}$.
As one sees in the example~\eqref{fijscale}, the function~$f^{ij}_\pm(\vec{q})$ decays typically
on the order~$1/|\vec{q}|$. As a consequence, the sum in~\eqref{Lint}
diverges like~$|\vec{q}|^2$. This means that the amplitude of the process
is infinite, even if the coupling constant~$\nu$ is very small.
In other words, even for an arbitrarily weak stochastic background field, the exchange process
is relevant, because the particle state~$\Psi_k$ may interact with many different states of the sea.

\subsection{Recombination}
The general conclusion of the analysis in Section~\ref{secstocup} is that the stochastic background field,
even if very weak, leads to a process on the Compton scale where the particle states
are exchanged with states of the Dirac sea.
This effect should be taken into account in the microscopic mixing procedure in that the
mixing subspace~$\I_0$ should not be kept fixed, but it should be adapted continually
to the present configuration of the wave functions.
We now explain how this intuitive picture can be made precise.
We first recall that, according to our ansatz~\eqref{Pchar} and~\eqref{Pmixansatz},
the space-time arguments~$x$ and~$y$ of the fermionic projector determine
in which subsystem~$M_\as$ we are.
So far, the considered subsystems~$M_\as$ were spread out over all of space-time
and distributed uniformly in space-time (see Section~\ref{secmixex}).
We now generalize this picture by considering {\em{subsystems}} which are {\em{localized in bounded
macroscopic space-time regions}}.
This gives us the freedom to choose the subsystems 
such that the microscopic mixing in different space-time regions involves different mixing subspaces.

In preparation, we {\em{generalize the form of~$V_\as$}} in such a way
that the effective mixing subspace depends on the index~$\as$.
Note that, according to~\eqref{Vans}, the matrix~$\Uran$ leaves~$\I_0$ invariant, whereas~$\Wran$
maps the subspaces~$\I_0$ and~$\J_0$ unitarily to each other. Hence the transformation~$V$,
\eqref{Vans}, maps the subspace of~$\I_0$ unitarily to the subspace~$\Uran \,\J_0\, \Uran^*$.
We now generalize the form of~\eqref{Vans} and~\eqref{WUmap} to
\beq \label{Vans3}
V_{\as} = \Uran_a \begin{pmatrix} 0 & \Wran_\alpha & 0 \\ \1 & 0 & 0 \\ 0 & 0 & \1 \end{pmatrix} \Uran_a^*
\eeq
with
\beq \label{WUmap2}
\Wran_\alpha \in \U(\J_0, \I_0) \qquad \text{and} \qquad
\Uran_a \in \U({\mathscr{H}}_0) \:.
\eeq
Now~$\Uran_a$ no longer needs to leave~$\I_0$ invariant. Consequently, the transformation~$V_\as$
maps the subspaces~$\Uran_a \,\I_0$ and~$\Uran_a \,\J_0$
unitarily to each other. Setting
\beq \label{IJadef}
\I_a = \Uran_a \,\I_0 \qquad \text{and} \qquad \J_a = \Uran_a \,\J_0 \:,
\eeq
we can interpret~$\I_a$ as the effective mixing space used in the subsystem~$M_\as$.
Considering~$\Wran_\alpha$ as a random matrix describes a microscopic
mixing of the subspace~$\I_a$ with the subspace~$\I_b$.

\begin{figure} %
\begin{picture}(0,0)%
\includegraphics{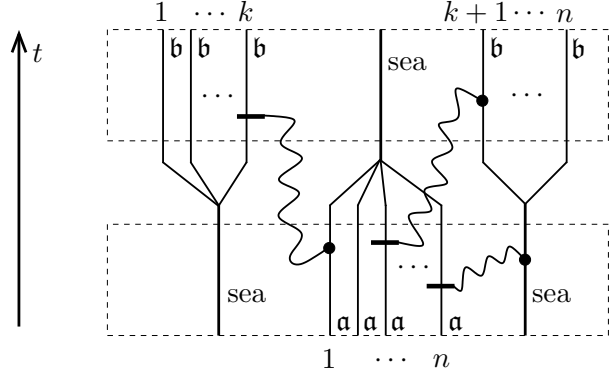}%
\end{picture}%
\setlength{\unitlength}{1533sp}%
\begingroup\makeatletter\ifx\SetFigFont\undefined%
\gdef\SetFigFont#1#2#3#4#5{%
  \reset@font\fontsize{#1}{#2pt}%
  \fontfamily{#3}\fontseries{#4}\fontshape{#5}%
  \selectfont}%
\fi\endgroup%
\begin{picture}(9707,6100)(-4284,-3512)
\put(2809,-2761){\makebox(0,0)[lb]{\smash{{\SetFigFont{11}{13.2}{\rmdefault}{\mddefault}{\updefault}$\as$}}}}
\put(764,-3376){\makebox(0,0)[lb]{\smash{{\SetFigFont{11}{13.2}{\rmdefault}{\mddefault}{\updefault}$1$}}}}
\put(1981,-1861){\makebox(0,0)[lb]{\smash{{\SetFigFont{11}{13.2}{\rmdefault}{\mddefault}{\updefault}$\cdots$}}}}
\put(1616,-3366){\makebox(0,0)[lb]{\smash{{\SetFigFont{11}{13.2}{\rmdefault}{\mddefault}{\updefault}$\cdots$}}}}
\put(2564,-3366){\makebox(0,0)[lb]{\smash{{\SetFigFont{11}{13.2}{\rmdefault}{\mddefault}{\updefault}$n$}}}}
\put(-3909,1578){\makebox(0,0)[lb]{\smash{{\SetFigFont{11}{13.2}{\rmdefault}{\mddefault}{\updefault}$t$}}}}
\put(-1709,1649){\makebox(0,0)[lb]{\smash{{\SetFigFont{11}{13.2}{\rmdefault}{\mddefault}{\updefault}$\bs$}}}}
\put(-1259,1649){\makebox(0,0)[lb]{\smash{{\SetFigFont{11}{13.2}{\rmdefault}{\mddefault}{\updefault}$\bs$}}}}
\put(-359,1649){\makebox(0,0)[lb]{\smash{{\SetFigFont{11}{13.2}{\rmdefault}{\mddefault}{\updefault}$\bs$}}}}
\put(-1317,2234){\makebox(0,0)[lb]{\smash{{\SetFigFont{11}{13.2}{\rmdefault}{\mddefault}{\updefault}$\cdots$}}}}
\put(-587,2234){\makebox(0,0)[lb]{\smash{{\SetFigFont{11}{13.2}{\rmdefault}{\mddefault}{\updefault}$k$}}}}
\put(-1944,2234){\makebox(0,0)[lb]{\smash{{\SetFigFont{11}{13.2}{\rmdefault}{\mddefault}{\updefault}$1$}}}}
\put(2727,2242){\makebox(0,0)[lb]{\smash{{\SetFigFont{11}{13.2}{\rmdefault}{\mddefault}{\updefault}$k+1$}}}}
\put(3474,1649){\makebox(0,0)[lb]{\smash{{\SetFigFont{11}{13.2}{\rmdefault}{\mddefault}{\updefault}$\bs$}}}}
\put(4794,1634){\makebox(0,0)[lb]{\smash{{\SetFigFont{11}{13.2}{\rmdefault}{\mddefault}{\updefault}$\bs$}}}}
\put(3811,929){\makebox(0,0)[lb]{\smash{{\SetFigFont{11}{13.2}{\rmdefault}{\mddefault}{\updefault}$\cdots$}}}}
\put(-1189,917){\makebox(0,0)[lb]{\smash{{\SetFigFont{11}{13.2}{\rmdefault}{\mddefault}{\updefault}$\cdots$}}}}
\put(3852,2249){\makebox(0,0)[lb]{\smash{{\SetFigFont{11}{13.2}{\rmdefault}{\mddefault}{\updefault}$\cdots$}}}}
\put(4557,2228){\makebox(0,0)[lb]{\smash{{\SetFigFont{11}{13.2}{\rmdefault}{\mddefault}{\updefault}$n$}}}}
\put(-761,-2311){\makebox(0,0)[lb]{\smash{{\SetFigFont{11}{13.2}{\rmdefault}{\mddefault}{\updefault}sea}}}}
\put(4169,-2311){\makebox(0,0)[lb]{\smash{{\SetFigFont{11}{13.2}{\rmdefault}{\mddefault}{\updefault}sea}}}}
\put(1836,1444){\makebox(0,0)[lb]{\smash{{\SetFigFont{11}{13.2}{\rmdefault}{\mddefault}{\updefault}sea}}}}
\put(1879,-2761){\makebox(0,0)[lb]{\smash{{\SetFigFont{11}{13.2}{\rmdefault}{\mddefault}{\updefault}$\as$}}}}
\put(999,-2761){\makebox(0,0)[lb]{\smash{{\SetFigFont{11}{13.2}{\rmdefault}{\mddefault}{\updefault}$\as$}}}}
\put(1431,-2761){\makebox(0,0)[lb]{\smash{{\SetFigFont{11}{13.2}{\rmdefault}{\mddefault}{\updefault}$\as$}}}}
\end{picture}%
\caption{Recombination of two \ansybs.}
\label{figrecombine}
\end{figure}

We next consider two subsystems~$M_\as$ and~$M_\bs$ localized in different 
macroscopic space-time regions, such that~$M_\bs$ lies to the future of~$M_\as$
(see Figure~\ref{figrecombine}).
The time evolution from~$M_\as$ to~$M_\bs$ involves the coupling to the
stochastic background field as considered in Section~\ref{secstocup}.
This means graphically that all the particle lines in Figure~\ref{figrecombine}
should be connected to the sea lines by bosonic lines as in Figure~\ref{figstocoup};
these lines were omitted in Figure~\eqref{figrecombine} for the sake of clarity.
As a consequence of this interaction, the particle states~$\Psi_1, \ldots, \Psi_n$
in the subsystem~$M_\as$ take part in exchange processes, meaning that
in the subsystem~$M_\bs$ they have converted to sea states.
Conversely, some sea states in the subsystem~$M_\as$ will have become particle
states. As a consequence, the particle subspace~$\I_\bs$ in the later subsystem~$M_\bs$ will be
different from the particle subspace~$\I_\as$ in the earlier subsystem~$M_\as$.

In the corresponding \ansybs, the particle states (i.e.\ the states in~$I_\as$ respectively~$I_\bs$)
are again totally anti-symmetrized. The sea states, on the other hand, are not anti-symmetrized.
In Figure~\ref{figrecombine}, all the sea states were again combined to one fermionic line
which does not carry an index~$\as$ or~$\bs$, and which for clarity we labelled by ``sea.''
Different~\ansybs\ may interact with each other by exchanging bosonic lines.
More precisely, a particle line in an \ansyb\ may have at most one outgoing bosonic line,
again denoted by a bar. However, it may have an arbitrary number of incoming bosonic lines.
A bosonic line may begin and and in the same \ansyb\ or just as well in two different \ansybs.
We refer to this mechanism as the {\em{recombination}} of \ansybs.
The relevant scales will be specified and discussed in the next section.

\subsection{Scalings and Background Synchronization} \label{secbackground}
In this section, we make a few explanatory remarks on the recombination and specify the relevant scalings.
We first note that by considering~$\Uran_a$ of the special form
\beq \label{Uspecial}
\Uran = \begin{pmatrix} \1 & 0 & 0 \\ 0 & \Uran_{11} & \Uran_{12} \\ 0 & \Uran_{21} & \Uran_{22} \end{pmatrix}\qquad \text{with} \qquad \Uran \in \U({\mathscr{N}}_0) \:,
\eeq
we get back to the ansatz~\eqref{Vans}. The operator~\eqref{Uspecial} transforms~$\J_0$ to~$\J_a$,
whereas the more general operator in~\eqref{WUmap2} also transforms~$\I_0$ to~$\I_a$.
Conceptually, the transformations of~$\J_0$ and ~$\I_0$ are independent of each other;
in particular, they could enter the microscopic mixing on different microscopic scales or
with a different macroscopic space-time dependence. However, since the matrix~$\Uran$
played no role in the dynamics of an \ansyb\ as worked out in Sections \ref{secasb}
and~\ref{secasbdyn} (it only gave rise to a complex prefactor~\eqref{comfact}), 
it seems sufficient four our purposes not to treat the transformation of~$\J_0$ separately,
but to always combine it with the transformation of~$\I_0$. This is why
we denote both~$\I_a$ and~$\J_a$ by the same index~$a$
and describe the transformations of~$\I_0$ and~$\J_0$ by the same operator~$\Uran_a$.

We next point out that the lower indices~$a$ and~$\alpha$ play a very
different role, as we now explain. The matrices~$\Wran_\alpha$ have the purpose of introducing
the microscopic mixing
to avoid the divergences of the action (just as explained in Sections~\ref{secwhy}
and~\ref{secmixex}). There is no point in introducing a macroscopic space-time dependence
of these matrices.
Thus we assume throughout that~$\Wran_\alpha$ is microscopically mixed, and that it can be
described by a random matrix with the probability measure as given by the normalized Haar measure.
The transformation~$\Uran_a$, on the other hand, has the purpose of keeping track 
of the particle states. It can involve a microscopic mixing, corresponding to a physical system
being a superposition of different particle configurations in the same space-time region.
But more importantly, it has a macroscopic dependence on at least the Compton scale~\eqref{Cscale},
which takes into account the dynamics and the mixing of all the wave functions due to
exchange processes.

In order to make this concept precise in the simplest possible way, we
assume that the space-time regions can be written as
\[ M_\as = M_{(a \alpha)} = M_a^\text{macro} \cap M_\alpha^\text{micro} \:, \]
where the sets~$M^\text{macro}_a$ and~$M_\alpha^\text{micro}$ have the following properties:
\begin{itemize}
\item[(a)] The sets~$M^\text{micro}_\alpha$ are fine-grained and uniformly distributed in space-time,
meaning that in analogy to~\eqref{weights},
\[ \int_{M^\text{micro}_\alpha} f(x)\, d^4x = c_\alpha \int_M f(x)\, d^4x + \text{(higher orders in~$\varepsilon/\ell_\text{macro}$)} \]
with non-negative coefficients~$c_\alpha$ and~$\sum_{\alpha=1}^{\Lphase} c_\alpha = 1$.
\item[(b)] The sets~$M^\text{macro}_a$ may also be fine-grained, but have an additional
macroscopic space-time dependence, i.e.
\[ \int_{M^\text{macro}_a} f(x)\, d^4x = \int_M c_a(x)\:f(x)\, d^4x + \text{(higher orders in~$\varepsilon/\ell_\text{macro}$)} \]
with non-negative functions~$c_a(x)$ and~$\sum_{a=1}^\Lmix c_a(x) = 1$.
The functions~$c_a(x)$ depend only on the macroscopic scale~$\ell_\text{macro}$,
and should be localized on a scale~$\ell_\text{mix}$ which is at most the Compton scale, i.e.
\[ \supp c_a \quad \text{has size} \quad \leq \ell_\text{mix} \lesssim \frac{1}{m} \:. \]
Here the ``size'' could be defined for example as the
maximal radius of an Euclidean ball contained in~$\supp c_a$, where the
maximum is taken over the coordinates of all possible reference frames.
\end{itemize}
Clearly, the ansatz~\eqref{Vans3} and~\eqref{WUmap2} as well as the above
assumptions~(a) and~(b) are ad-hoc and could be modified in many ways.
We do not claim that this ansatz describes the microscopic structure of physical space-time
(which we expect to have a much more complicated form). But the above description of microscopic 
mixing seems to be the simplest possible ansatz which models all the relevant effects.

We next discuss the strength of the bosonic background field.
The above assumption~(b) has the consequence that an \ansyb\ synchronized with~$\as=(a \alpha)$,
the outgoing bosonic lines must be created in a space-time region~$M_a$ localized on the Compton scale.
Similarly, the covariance~${\mathcal{C}}_\as$ in~\eqref{covsymm} only couples to the
fermions in the subsystem~$M_\as$, again giving a localization on the Compton scale.
The covariance~${\mathcal{C}}$ in~\eqref{covsymm}, on the other hand, couples to all subsystems
in the same way, so that we get no restriction on the space-time region in which the interaction must
take place. This difference in interaction times implies that the effect of the covariance~${\mathcal{C}}$
on the wave functions is by a scaling factor $m T$ larger than that of~${\mathcal{C}}_\as$
(where the interaction time~$T$ could be as large as the lifetime of the universe).

This argument using the interaction times is very useful in connection with the synchronization
of the particle lines. Namely, suppose that we consider a covariance which depends only on
the index~$\alpha$, i.e.\ in modification of~\eqref{covsymm}
\beq \label{cova}
\int \B_{|\as}(x)\: \B_{|\bs}(y)\: {\mathfrak{D}} \B = \delta_{\alpha \beta}\: {\mathcal{C}}_\alpha(x,y) \:.
\eeq
Then, in view of our above assumption~(a), we do not get any restriction on the interaction time,
so that the effect of the stochastic background field grows linearly in~$T$.
This makes it possible to synchronize the particle lines even with an arbitrarily weak
stochastic background field.
We also recall that for the recombination, the stochastic background field could
be arbitrary small (see~\eqref{Covscale} and the argument thereafter).
This leads us to introduce {\bf{background-synchronized systems}}
as follows:
\begin{quote}
We consider a stochastic bosonic background field with the covariance~\eqref{cova}.
We assume that the covariance is so small that it can be neglected in the computation of
physical processes. Nevertheless, the stochastic background field triggers the recombination,
and it synchronizes the bosonic lines as in shown in Figure~\ref{figbackground}.
Even more, we assume that the particle lines are connected by so many covariances
that the synchronization by the outgoing and incoming bosonic lines (as shown in Figure~\ref{figmixing})
is irrelevant.
\end{quote}
The last assumption is a major simplification of the combinatorics because every outgoing
bosonic line can couple to any particle line to its right, without constraints coming from the
condition that all~$n$ particle lines must be synchronized.
It will be used in Section~\ref{secfock} below.

Working in a background-synchronized system also simplifies the treatment of the
phase factors~$\det \Wran_\alpha$ of the random matrices~$\Wran_\alpha$
as well as of the matrices~$\Uran_a$, as we now explain.
Recall that in Section~\ref{secaverage} we summed over phase factors to obtain
fluctuations of the order~$1/\sqrt{\Lphase}$ (see Propositions~\ref{prp51} and~\ref{prp52}).
The matrices~$\Uran_a$, on the other hand, involved our construction via the prefactors~\eqref{comfact}.
In a background-synchronized system, the stochastic background field does not only synchronize
the particle lines in one \ansyb\, but also the particle lines belonging to different \ansybs.
In the example of Figure~\ref{figrecombine}, this means that the indices~$\as$ and~$\bs$
involve the same random matrix~$\Wran_\alpha$, i.e.
\[ \as = (a \alpha) \qquad \text{and} \qquad \bs=(b \alpha) \:. \]
In fact, as the bosonic field only synchronizes the index~$\alpha$,
it is a-priori not clear that the index~$a$ is the same for all particle lines in the
lower \ansyb\ in Figure~\ref{figrecombine}. Likewise, in the upper \ansyb\ the indices~$b$
might be different. The synchronization of these ``macroscopic'' indices can be deduced as
follows: According to~\eqref{IJadef}, the matrices~$\Uran_a$ in~\eqref{Vans3} map~$\J_0$ to
the space~$\J_a \subset \H_0$. Since the choice of~$\J_a$ is arbitrary and the 
dimension of~$\H_0$ is much larger than that of~$\J_0$,
it is likely that the subspace~$\J_a$ is almost orthogonal to the subspaces~$\I_b$ 
and~$\J_b$ for~$b \neq a$. More precisely, we have the typical scaling
\[ \big\| \pi_{\J_0} \Uran^*_b \,\Uran_a |_{\J_0} \big\|\:, \big\| \pi_{\I_0} \Uran^*_b \,\Uran_a |_{\J_0} \big\|
\lesssim \sqrt{n/f}\:. \]
Since every fermionic line involves these operators,
in Figure~\ref{figrecombine} we only need to take into account the contributions
where in the lower \ansyb\ we always have the same index~$a$, in the upper \ansyb\ we
always have the same index~$b$, and we have a total anti-symmetrization in the $n$ particle
lines in each \ansyb.

More generally, this argument shows that in a background-synchronized system,
all the coupled \ansybs\ have the same index~$\alpha$.
This means that we no longer get sums of the phase factors~$\det \Wran_\alpha$.
In particular, the factors~$1/\Lphase$ in Propositions~\ref{prp51} and~\ref{prp52}
no longer describe the physically correct scaling.
Instead, the index~$\alpha$ is fixed, and when considering several \ansybs,
we simply get powers of the phase factor~$\det \Wran_\alpha$.
As a consequence, the factor~$\det \Wran_\alpha$ can be regarded simply as a 
physically irrelevant phase.

\subsection{Anti-Particles and the Choice of the Mixing Space} \label{secanti}
We now analyze the freedom in choosing the mixing space~$\I_\as$ and
explain how our description respects the concept of antiparticles.
Recall that the state stability analysis for plane-wave solutions in Section~\ref{secwhy} yields 
that a state on the lower mass shell gives a finite contribution to the action, whereas 
the contribution of a state on the upper mass shell diverges (see~\eqref{Sfinite} and~\eqref{Sdiv}).
This observation led us to introduce the microscopic mixing as a method to remove this
divergence (see~\eqref{gain2}).
At first sight, this consideration seems to apply only to the states~$\Psi_l$ on the upper mass shell.
For the states~$\Phi_k$ on the lower mass shell, however, the contribution to the action
is finite~\eqref{Sfinite}, making it unnecessary to introduce a microscopic mixing.
In other words, the microscopic mixing should only affect the particle but not the
anti-particle states. The resulting asymmetry between the treatment of particles and anti-particles
is puzzling and seems to contradict physical observations.

Fortunately, the picture changes completely as soon as the interaction is taken into account.
Namely, in this case we can no longer speak of solutions on the upper and lower mass shell. Instead, every
solution is necessarily a superposition components of positive and negative frequency.
In order to work in a simple example, consider the superposition of plane waves
\[ \Psi(t,\vec{x}) = c_+ e^{i k_+ x} \chi_+ + c_- e^{i k_- x} \chi_- \:, \]
where the momenta~$k_+$ and~$k_-$ are on the upper and lower mass shell, respectively.
The corresponding contribution to the action~\eqref{dS} is computed similar to~\eqref{Sdiv} by
\[ (\delta \Sact)[\Psi_l] \simeq |c_+|^2\: m^3\, \varepsilon^{-2} \:. \]
This shows that we get a divergent contribution to the action, unless~$c_+=0$.
Since in a physical situation~$c_+$ will never be exactly zero, we see that
also the anti-particle states give rise to divergences.
This leads to the important conclusion that the mixing space~$\I_0$ should also
include states on the lower mass shell.

More detailed information on the possible choices of the mixing space is obtained by the following
scaling argument: In Section~\ref{secbackground} we argued that the subsystems $M_\as$
should be localized on the Compton scale. This implies that the momenta of
the states restricted to the subsystems are ``smeared out'' on the scale~$\ell_\text{mix}^{-1}$.
Likewise, the macroscopic interaction mixes momenta on the scale~$\ell_\text{macro}^{-1}$.
Thus we can say that all states on the lower mass shell have a significant
contribution of positive frequency, provided that their energy is of the order~$\lesssim 1/\ell_\text{macro}$.
This leads us to impose that
\begin{quote}
$\I_\as$ should include all occupied states of positive and negative frequency
with energy of the order~$\lesssim 1/ \min(\ell_\text{macro}, \ell_\text{mix})$.
\end{quote}
Moreover, $\I_\as$ should be chosen that the ensemble of states of the mixing space
does not contribute to the field equations. For example, this can be accomplished
by arranging that the corresponding currents coincide in all sectors of the fermionic projector.
Apart from these general constraints, the form of~$\I_\as$ is unknown.

Under the above assumptions, anti-particles are treated properly. Namely, an anti-particle is
described by a ``hole'' in the sea of particles in~$\I_\as$. Since without the hole, the
states in~$\I_\as$ do not contribute to the field equations, the hole is a
quasi-particle of positive energy and positive electric charge.

\subsection{The Limiting Case of an Instantaneous Recombination} \label{secfockeff}
\begin{figure} %
\begin{picture}(0,0)%
\includegraphics{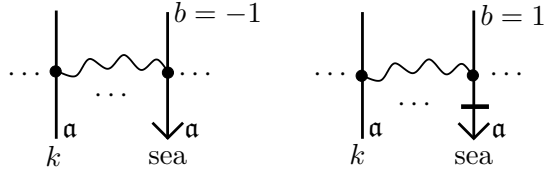}%
\end{picture}%
\setlength{\unitlength}{1533sp}%
\begingroup\makeatletter\ifx\SetFigFont\undefined%
\gdef\SetFigFont#1#2#3#4#5{%
  \reset@font\fontsize{#1}{#2pt}%
  \fontfamily{#3}\fontseries{#4}\fontshape{#5}%
  \selectfont}%
\fi\endgroup%
\begin{picture}(7916,2804)(99,-3477)
\put(1514,-2296){\makebox(0,0)[lb]{\smash{{\SetFigFont{11}{13.2}{\rmdefault}{\mddefault}{\updefault}$\cdots$}}}}
\put(5644,-3311){\makebox(0,0)[lb]{\smash{{\SetFigFont{11}{13.2}{\rmdefault}{\mddefault}{\updefault}$k$}}}}
\put(2799,-1012){\makebox(0,0)[lb]{\smash{{\SetFigFont{11}{13.2}{\rmdefault}{\mddefault}{\updefault}$b=-1$}}}}
\put(7759,-1057){\makebox(0,0)[lb]{\smash{{\SetFigFont{11}{13.2}{\rmdefault}{\mddefault}{\updefault}$b=1$}}}}
\put(5054,-1966){\makebox(0,0)[lb]{\smash{{\SetFigFont{11}{13.2}{\rmdefault}{\mddefault}{\updefault}$\cdots$}}}}
\put(2874,-1966){\makebox(0,0)[lb]{\smash{{\SetFigFont{11}{13.2}{\rmdefault}{\mddefault}{\updefault}$\cdots$}}}}
\put(7909,-1966){\makebox(0,0)[lb]{\smash{{\SetFigFont{11}{13.2}{\rmdefault}{\mddefault}{\updefault}$\cdots$}}}}
\put(999,-2867){\makebox(0,0)[lb]{\smash{{\SetFigFont{11}{13.2}{\rmdefault}{\mddefault}{\updefault}$\as$}}}}
\put(5941,-2851){\makebox(0,0)[lb]{\smash{{\SetFigFont{11}{13.2}{\rmdefault}{\mddefault}{\updefault}$\as$}}}}
\put(114,-1966){\makebox(0,0)[lb]{\smash{{\SetFigFont{11}{13.2}{\rmdefault}{\mddefault}{\updefault}$\cdots$}}}}
\put(6414,-2451){\makebox(0,0)[lb]{\smash{{\SetFigFont{11}{13.2}{\rmdefault}{\mddefault}{\updefault}$\cdots$}}}}
\put(7936,-2871){\makebox(0,0)[lb]{\smash{{\SetFigFont{11}{13.2}{\rmdefault}{\mddefault}{\updefault}$\as$}}}}
\put(2969,-2872){\makebox(0,0)[lb]{\smash{{\SetFigFont{11}{13.2}{\rmdefault}{\mddefault}{\updefault}$\as$}}}}
\put(724,-3341){\makebox(0,0)[lb]{\smash{{\SetFigFont{11}{13.2}{\rmdefault}{\mddefault}{\updefault}$k$}}}}
\put(2389,-3301){\makebox(0,0)[lb]{\smash{{\SetFigFont{11}{13.2}{\rmdefault}{\mddefault}{\updefault}sea}}}}
\put(7339,-3301){\makebox(0,0)[lb]{\smash{{\SetFigFont{11}{13.2}{\rmdefault}{\mddefault}{\updefault}sea}}}}
\end{picture}%
\caption{Symmetrization in the lower index~$b$.}
\label{figsymm}
\end{figure}

We now describe an effective description of the dynamics which takes into account the recombination of
\ansybs. In preparation, we begin with a Fock-Krein state~$\Psi$ of the form~\eqref{L2top} and
consider the corresponding Fock-Krein state~$\Psi_\text{rec}$ obtained if all particle states undergo an
exchange process, but no other interaction takes places. 
In this situation, the particle states after recombination are not coupled to each other by outgoing
bosonic lines. Then, as explained in Section~\ref{seckrein}, the anti-symmetrization in the
mixing space implies that the wave function~$\Psi_\text{rec}$ is totally antisymmetric in the spatial 
and spinor indices. Moreover, as shown in Figure~\ref{figsymm},
the stochastic coupling can take place similarly with a sea state whose lower index
is plus or minus one. Likewise, the interaction is independent of whether the lower index
of the $k^\text{th}$ particle line is plus or minus one.
Hence the wave function~$\Psi_\text{rec}$ will not depend on the lower indices.
We conclude that the wave function after recombination is obtained from the
original Fock-Krein wave function by anti-symmetrizing in the spatial 
and spinor indices and by summing over the lower indices, i.e.\
\beq \label{psirec}
\begin{split}
(&\Psi_\text{rec})^{\alpha_1 \cdots \alpha_n}_{b_1 \cdots b_n}(t; \vec{z}_1, \ldots, \vec{z}_n) \\
&\quad= \frac{1}{2^n \,n!} \sum_{b'_1, \ldots, b'_n = \pm 1} \;\sum_{\sigma \in S_n}
(-1)^{\sign(\sigma)}\: 
\Psi^{\alpha_{\sigma(1)} \cdots \alpha_{\sigma(n)}}_{b'_1 \cdots b'_n}(t; \vec{z}_{\sigma(1)}, \ldots,
\vec{z}_{\sigma(n)}) \:.
\end{split}
\eeq
Since this wave function does not depend on the lower indices, we often omit them and simply
write~$\Psi_\text{rec}^{\alpha_1 \cdots \alpha_n}(t; \vec{z}_1, \ldots, \vec{z}_n)$.
Taking the inner product~\eqref{kreinip} of two such recombined wave functions,
we can carry out the sums over the lower indices to obtain
\beq \label{Feffsp}
\begin{split}
(\Psi | \Phi)|_t &:= 2^n
\int_{\R^3} \!\!\! d^3z_1 \cdots\! \int_{\R^3} \!\!\! d^3z_n\; \sum_{\alpha_1, \ldots, \alpha_n=1}^4 \\
&\qquad \times \Psi^{\alpha_1 \cdots \alpha_n}_\text{rec}(t; \vec{z}_1, \ldots, \vec{z}_n)^\dagger
\Phi^{\alpha_1 \cdots \alpha_n}_\text{rec}(t; \vec{z}_1, \ldots, \vec{z}_n) \:.
\end{split}
\eeq
Note that every spatial integral simply is the integral over the usual probability density.
In particular, the inner product~\eqref{Feffsp} is positive definite and thus defines a
scalar product on the wave functions of the form~\eqref{psirec}.
We denote the space of wave functions of the form~\eqref{psirec} together with this
scalar product as the {\em{effective Fock space}}~$(\Fock_n, (.|.)_n)$.
Forming its completion gives a Hilbert space. At the same time, $\Fock_n$ is a
positive definite subspace of the Fock-Krein space~$(\Fock_n^\text{\tiny{Krein}}, (.|.)_n)$.
The mapping which to any~$\Psi \in \Fock_n^\text{\tiny{Krein}}$
associates the corresponding wave function~$\Psi_\text{rec} \in \Fock_n$ as
given by~\eqref{psirec} is an orthogonal projection operator, which we denote by
\beq \label{proj}
\Pi \::\: \Fock_n^\text{\tiny{\rm{Krein}}} \rightarrow \Fock_n \:, \qquad
\Pi \Psi = \Psi_\text{rec} \:.
\eeq

It is important to observe that the effective Fock space is {\em{not invariant}} under
the Fock-Krein dynamics introduced in Section~\ref{seckrein}.
To see how this comes about, let us consider the Fock-Krein wave function in~\eqref{figwave},
taking initially a wave function~$\Psi \in \Fock_n$ in the effective Fock space.
As soon as one of the particle lines emits a bosonic line and another line absorbs it,
the fact that the bosonic lines go from the left to the right destroys the anti-symmetry
of the Fock-Krein wave functions. For a detailed description of the Fock-Krein dynamics
we refer to Section~\ref{secfock1} below.

As explained above, the recombination of the \ansybs\ can be described effectively
by the projection~\eqref{Feffsp} to the effective Fock space.
We have in mind that the recombination takes place continually on the Compton scale,
which in many situations is much smaller than the relevant macroscopic length scales of the physical system.
Therefore, it seems a good approximation to project repeatedly to the effective Fock space,
and to consider the limit when the time intervals between the projections tends to zero.
Denoting the unitary time evolution of Section~\ref{seckrein} from time~$t$ to time~$t'$ by
by~$U_{t, t'}^\text{\tiny{\rm{Krein}}}$, we thus introduce the {\em{effective time evolution}} operator by
\beq \label{Uproj}
U^\text{\tiny{eff}}_{T, 0} := \lim_{N \rightarrow \infty} \Pi \:U^\text{\tiny{\rm{Krein}}}_{T, \frac{N-1}{N}\:T}\: \Pi
\:U^\text{\tiny{\rm{Krein}}}_{\frac{N-1}{N}\:T, \frac{N-2}{N}\:T}\: \Pi \cdots
U^\text{\tiny{\rm{Krein}}}_{\frac{T}{N}, 0}\: \Pi \;:\; \Fock_n \rightarrow \Fock_n \:.
\eeq
The repeated projections confine the dynamics to the effective Fock space.
Similar as in the well-known description of adiabatic processes, the effective time evolution is again unitary.
We thus obtain a description of the dynamics by a unitary operator on a Hilbert space.
We refer to this method as the limiting case of an {\em{instantaneous recombination}} of \ansybs.

\section{Description in the Fock Space Formalism} \label{secfock}
In order to get a closer connection to the standard formulation of quantum field theory,
we now reformulate the dynamics of \ansybs\ in the formalism
of Fock spaces. We first consider the dynamics of one \ansyb\
in the Fock-Krein space (Section~\ref{secfock1}), and then analyze the 
effective dynamics obtained in the limiting case of an instantaneous recombination (Section~\ref{secfock2}).
We assume throughout this section that the system is background-synchronized
(as introduced in Section~\ref{secbackground}).

\subsection{The Dynamics of an Anti-Symmetrized Synchronal Block} \label{secfock1}
\subsubsection{\bf{The Field Operators}} \label{secfield}
We begin with the free dynamics of an \ansyb.
A difference to the standard formalism of quantum field theory is that the fermionic lines
are numbered from~$1, \ldots n$ (see Figure~\ref{figwave}). Moreover, every fermionic
line carries a lower index~$b=\pm 1$ (see~\eqref{bldef} and~\eqref{psibl}). In order to take
these differences into account, we introduce fermionic field
operators~$\hat{\Psi}_{[l,b]}(t,\vec{x})$ which satisfy the equal time anti-commutation relations
\beq \label{ac}
\begin{split}
\{ \hat{\Psi}^\alpha_{[l,b]}(t,\vec{x}), \hat{\Psi}^\beta_{[l',b']}(t,\vec{y})^\dagger \} &= 
\delta_{l l'}\,{\mathfrak{t}}^b_{b'}\, \delta^\alpha_\beta\: \delta^3(\vec{x}-\vec{y}) \\
\{ \hat{\Psi}^\alpha_{[l,b]}(t,\vec{x}), \hat{\Psi}^\beta_{[l',b']}(t,\vec{y}) \} &= 0 =
\{ \hat{\Psi}^\alpha_{[l,b]}(t,\vec{x})^\dagger, \hat{\Psi}^\beta_{[l',b']}(t,\vec{y})^\dagger \} \:,
\end{split}
\eeq
where~${\mathfrak{t}}$ is again the matrix~\eqref{Tdef}
($\alpha, \beta$ denote the spinor indices and~$b \in \{ \pm 1\}$, $l \in \{1,\ldots, n\}$).
Apart from the additional indices~$l,l'$ and~$b,b'$, these are usual canonical anti-commutation
relations (see for example~\cite[eq.~(13.53)]{bjorken2} or~\cite[eq.~(3.102)]{peskin+schroeder}).
Imposing that these field operators satisfy the free Dirac equation, we obtain the
anti-commutation relations for arbitrary times
\beq \label{ac2}
\begin{split}
\{ \hat{\Psi}^\alpha_{[l,b]}(x), \hat{\Psi}^\beta_{[l',b']}(y)^\dagger \} &= 2 \pi\,
\delta_{l l'}\,{\mathfrak{t}}^b_{b'}\: \big( k_m(x,y) \,\gamma^0 \big)^\alpha_\beta \\
\{ \hat{\Psi}^\alpha_{[l,b]}(x), \hat{\Psi}^\beta_{[l',b']}(y) \} &= 0 =
\{ \hat{\Psi}^\alpha_{[l,b]}(x)^\dagger, \hat{\Psi}^\beta_{[l',b']}(y)^\dagger \} \:,
\end{split}
\eeq
where~$k_m$ denotes the causal fundamental solution~\eqref{kmdef}
(note that, specializing the result of Proposition~\ref{prpcauchy} to the non-interacting situation,
one sees that~$2 \pi k_m(t,\vec{x}; t,\vec{x}') \gamma^0 = \delta^3(\vec{x}-\vec{x}')\, \1$).

The free fermionic states can be built up with the usual Fock space construction.
To this end, we introduce a vacuum state~$|0\ket$ with the properties
\beq \label{vacprop}
\bra 0 | 0 \ket = 1 \qquad \text{and} \qquad \hat{\Psi}_{[l,b]}(x) \,|0\ket = 0 \qquad
\text{for all~$b$, $l$ and~$x$}\:.
\eeq
Suppose we want to build up a quantum state which at time~$t_0$
is a (not necessarily totally antisymmetric) product of one-particle states
\[ \Psi^{\alpha_1 \cdots \alpha_n}_{b1 \cdots b_n}(\vec{z}_1, \ldots, \vec{z}_n)
= (\psi_1)_{b_1}^{\alpha_1}(\vec{z}_1) \cdots (\psi_n)_{b_n}^{\alpha_n}(\vec{z}_n) \:, \]
each carrying a lower index~$b_l \in \{0,1\}$. We set
\beq \label{Psi1}
\begin{split}
|\Psi & \ket = (-1)^{[\frac{n}{2}]} \sum_{\alpha_1, \ldots, \alpha_n=1}^4 \\
&\times \int_{\R^3} (\psi_1)_{b_1}^{\alpha_1}(\vec{z}_1)\,
\hat{\Psi}^{\alpha_1}_{[1,-b_1]}(t_0, \vec{z}_1)^\dagger \cdots
\int_{\R^3} (\psi_1)_{b_n}^{\alpha_n}(\vec{z}_n)\,
\hat{\Psi}^{\alpha_n}_{[n,-b_n]}(t_0, \vec{z}_n)^\dagger \;|0\ket \:.
\end{split}
\eeq
Then a short computation using the anti-commutation relations shows that
the many-particle wave function can be recovered as an expectation value,
\beq \label{Psi2}
\Psi^{\alpha_1 \cdots \alpha_n}_{b1 \cdots b_n}(z_1, \ldots, z_n) 
= \bra 0 | \hat{\Psi}^{\alpha_1}_{[1,b_1]}(z_1) \cdots \hat{\Psi}^{\alpha_n}_{[n,b_n]}(z_n) 
| \Psi \ket \:.
\eeq
By taking linear combination, this construction of~$|\Psi \ket$ immediately extends
to general wave functions in the Fock-Krein space~\eqref{L2top}.
Thus we can represent general vectors of the Fock-Krein space by states~$\Psi$.

Combining the anti-commutation relations~\eqref{ac2} with the vacuum property~\eqref{vacprop},
one can also compute time-ordered products. For example,
\beq \label{Tpsi}
\begin{split}
{\rm{T}} \big( \hat{\Psi}^\alpha_{[l,b]}(x) \, \hat{\Psi}^\beta_{[l',b']}(y)^\dagger \big)
\,|\,0 \ket &= 2 \pi \,\Theta(x^0-y^0)\:
\delta_{l l'}\,{\mathfrak{t}}^b_{b'}\: \big( k_m(x,y) \,\gamma^0 \big)^\alpha_\beta\,|\,0 \ket \\
&= i \,\delta_{l l'}\,{\mathfrak{t}}^b_{b'}\: \big( s_m^\wedge(x,y) \,\gamma^0 \big)^\alpha_\beta\,|\,0 \ket \:,
\end{split}
\eeq
where in the last step we applied~\eqref{kmgreen}
(and~$\Theta$ is again the Heaviside function). 

We point out that, due to the factors~${\mathfrak{t}}^b_{b'}$ in~\eqref{ac},
the Fock space generated by acting with our field operators on the vacuum
automatically carries an {\em{indefinite}} inner product which agrees with the
inner product~\eqref{kreinip}. This can be seen inductively from the computation
\beq \label{Focksp}
\begin{split}
\bra \big( &\hat{\Psi}^\alpha_{[l,b]}(t, \vec{x})^\dagger\,\cdots\, 0 \big) \:|\:
\big( \hat{\Psi}^\beta_{[l',b']}(t, \vec{y})^\dagger\,\cdots\, 0 \big) \ket \\
&= \bra 0 \,|\, \cdots \hat{\Psi}^\alpha_{[l,b]}(t, \vec{x}) \, \hat{\Psi}^\beta_{[l',b']}(t, \vec{y})^\dagger\cdots \,|\, 0 \ket \\
&= \delta_{l l'}\,{\mathfrak{t}}^b_{b'}\, \delta^\alpha_\beta\: \delta^3(\vec{x}-\vec{y})\; \bra 0 \,|\, \cdots  \,|\, 0 \ket \:.
\end{split}
\eeq
Integrating over the spatial variables gives agreement with~\eqref{kreinip}.

For describing the bosonic lines, we must take into account that the interaction takes place from
the left to the right (see Figure~\ref{figwave}). To this end, we introduce field operators~$\Bu_{[l]}(t,\vec{x})$
(for the outgoing bosonic lines) and~$\Bd_{[l]}(x)$ (for the incoming bosonic lines).
Constructing representations of the bosonic field operators is a somewhat subtle issue
if~$\B$ is a gauge field, because one must fix the gauge and/or one must treat the gauge freedom
with ghost fields. Since we do not want to specify the form of~$\B$, we simply disregard
these well-known issues and construct a representation as if the usual Fock space
construction worked. Moreover, for ease in notation we shall omit the possible tensor indices
of~$\B$. In order to obtain the correct description of the Fock-Krein
dynamics, the vacuum expectation value of a time-ordered product should be given by
\beq \label{Tab}
\bra 0 \,|\, {\rm{T}} \big( \Bd_{[k]}(x) \, \Bu_{[l]}(y) \big) \,|\,0 \ket =  i\, \Theta(k-l+1) \,S_0(x,y)
\eeq
with~$S_0$ according to~\eqref{S0caus}.
Here the Heaviside function gives a contribution
only if~$k>l$, corresponding to the fact that the bosonic lines go from the left  to the right.
The relation for the vacuum expectation value of the time-ordered product implies the
commutator relations, as we now explain. First, taking the adjoint of~\eqref{Tab} and
using~\eqref{S0real}, we obtain
\beq \label{Tabad}
\bra 0 \,|\, {\rm{T}} \big( \Bd_{[k]}(x) \, \Bu_{[l]}(y) \big)^\dagger \,|\,0 \ket =  -i\, \Theta(k-l+1) \,S_0(x,y)
\eeq
Next, we write the time-ordered product as
\begin{align*}
{\rm{T}} \big( \Bd_{[k]}(x) \, \Bu_{[l]}(y) \big) &=
\Theta(x^0 - y^0)\, \Bd_{[k]}(x) \, \Bu_{[l]}(y) + \Theta(y^0-x^0)\,  \Bu_{[l]}(y) \, \Bd_{[k]}(x) \\
\intertext{and take the adjoint,}
{\rm{T}} \big( \Bd_{[k]}(x) \, \Bu_{[l]}(y) \big)^\dagger &=
\Theta(x^0 - y^0)\, \Bu_{[l]}(y)\, \Bd_{[k]}(x) + \Theta(y^0-x^0)\,   \Bd_{[k]}(x)\, \Bu_{[l]}(y) \:.
\end{align*}
Subtracting these formulas gives
\[ {\rm{T}} \big( \Bd_{[k]}(x) \, \Bu_{[l]}(y) \big) - {\rm{T}} \big( \Bd_{[k]}(x) \, \Bu_{[l]}(y) \big)^\dagger
= \epsilon(x^0-y^0)\: \big[\Bd_{[k]}(x), \Bu_{[l]}(y)\big] \:. \]
Comparing this formula with~\eqref{Tab} and~\eqref{Tabad}, we obtain the commutator relations
\beq \label{Bfield}
\begin{split}
[\Bd_{[k]}(x), \Bu_{[l]}(y)] &= 2 i\,\Theta(k-l+1)
\big( \tau \,S_0^\wedge(x,y) - (1-\tau) \,S_0^\vee(x,y) \big) \\
[\Bd_{[k]}(x), \Bd_{[l]}(y)] &= 0 = [\Bu_{[k]}(x), \Bu_{[l]}(y)] \:.
\end{split}
\eeq

The vacuum expectation value~\eqref{Tab} can be realized in several ways. One
method is to decompose the bosonic field operators into creation and annihilation parts,
\beq \label{Bdecomp}
\Bd_{[k]}(x) = a_{[k]}(x) + a^\dagger_{[k]}(x) \:,\qquad
\Bu_{[l]}(x) = b_{[l]}(x) + b^\dagger_{[l]}(x)
\eeq
and to assume that the annihilation operators vanish on the vacuum,
\beq \label{abvac}
a_{[k]}(x) \,|0 \ket = 0 = b_{[l]}(x) \,|0 \ket \qquad \text{for all~$k, l$ and~$x$}\:.
\eeq
Moreover, we impose the commutation relations
\beq \label{abc}
\big[ a_{[k]}(x), b_{[l]}^\dagger(y) \big] = i \,\Theta(k-l+1)
\big( \tau \,S_0^\wedge(x,y) - (1-\tau) \,S_0^\vee(x,y) \big)
\eeq
(and all other commutators vanish). Taking the adjoint, we obtain
\[ \big[ b_{[l]}(y), a_{[k]}^\dagger(x) \big] = -i\,\Theta(k-l+1)
\big( \tau \,S_0^\wedge(x,y) - (1-\tau) \,S_0^\vee(x,y) \big) \:. \]
A short computation using these commutator relations gives
\begin{align*}
\bra & 0 \,|\, {\rm{T}} \big( \Bd_{[k]}(x) \, \Bu_{[l]}(y) \big) \,|\,0 \ket \\
&= \bra 0 \,|\, \Big( \Theta(x^0-y^0) \:a_{[k]}(x) \, b^\dagger_{[l]}(y) 
+ \Theta(y^0-x^0) \:b_{[l]}(y) \,a^\dagger_{[k]}(x) \Big) \,|\,0 \ket \\
&= \Theta(x^0-y^0) \big[ a_{[k]}(x), b^\dagger_{[l]}(y) \big]
+ \Theta(y^0-x^0) \big[ b_{[l]}(y), a^\dagger_{[k]}(x)\big] \\
&= i\,\Theta(k-l+1) \left( \tau\, S_0^\wedge(x,y) + (1-\tau)\, S_0^\vee(x,y) \right)
= i\,\Theta(k-l+1) \,S_0(x,y) \:,
\end{align*}
giving agreement with~\eqref{Tab} and~\eqref{S0caus}.

By applying the fermionic and bosonic creation operators to the vacuum,
we obtain a Fock space. The inner product on this Fock space is determined by
the commutation and anti-commutation relations as well as the properties
of the vacuum state (just as explained for the fermions in~\eqref{Focksp}).
A short computation using~\eqref{abvac} and~\eqref{abc} shows that this inner product
is also indefinite on the bosonic sector.
This indefiniteness does not cause any conceptual difficulties if one keeps in mind that
the bosonic field operators~$\Bd$ and~$\Bu$ are merely a mathematical device
which makes it possible to describe the interaction by a classical field in the language
of Fock spaces.

\subsubsection{\bf{Building up the Initial State}} \label{secinitial}
\begin{figure} %
\begin{picture}(0,0)%
\includegraphics{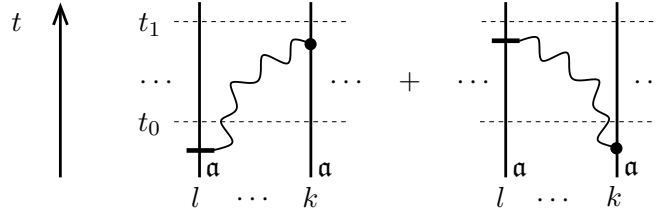}%
\end{picture}%
\setlength{\unitlength}{1533sp}%
\begingroup\makeatletter\ifx\SetFigFont\undefined%
\gdef\SetFigFont#1#2#3#4#5{%
  \reset@font\fontsize{#1}{#2pt}%
  \fontfamily{#3}\fontseries{#4}\fontshape{#5}%
  \selectfont}%
\fi\endgroup%
\begin{picture}(10412,3500)(-2154,-3517)
\put(999,-2867){\makebox(0,0)[lb]{\smash{{\SetFigFont{11}{13.2}{\rmdefault}{\mddefault}{\updefault}$\as$}}}}
\put(2809,-2872){\makebox(0,0)[lb]{\smash{{\SetFigFont{11}{13.2}{\rmdefault}{\mddefault}{\updefault}$\as$}}}}
\put(-96,-2147){\makebox(0,0)[lb]{\smash{{\SetFigFont{11}{13.2}{\rmdefault}{\mddefault}{\updefault}$t_0$}}}}
\put(-91,-527){\makebox(0,0)[lb]{\smash{{\SetFigFont{11}{13.2}{\rmdefault}{\mddefault}{\updefault}$t_1$}}}}
\put(764,-3376){\makebox(0,0)[lb]{\smash{{\SetFigFont{11}{13.2}{\rmdefault}{\mddefault}{\updefault}$l$}}}}
\put(2569,-3381){\makebox(0,0)[lb]{\smash{{\SetFigFont{11}{13.2}{\rmdefault}{\mddefault}{\updefault}$k$}}}}
\put(1484,-3366){\makebox(0,0)[lb]{\smash{{\SetFigFont{11}{13.2}{\rmdefault}{\mddefault}{\updefault}$\cdots$}}}}
\put(-76,-1466){\makebox(0,0)[lb]{\smash{{\SetFigFont{11}{13.2}{\rmdefault}{\mddefault}{\updefault}$\cdots$}}}}
\put(5699,-3381){\makebox(0,0)[lb]{\smash{{\SetFigFont{11}{13.2}{\rmdefault}{\mddefault}{\updefault}$l$}}}}
\put(7474,-3366){\makebox(0,0)[lb]{\smash{{\SetFigFont{11}{13.2}{\rmdefault}{\mddefault}{\updefault}$k$}}}}
\put(6344,-3371){\makebox(0,0)[lb]{\smash{{\SetFigFont{11}{13.2}{\rmdefault}{\mddefault}{\updefault}$\cdots$}}}}
\put(7919,-1466){\makebox(0,0)[lb]{\smash{{\SetFigFont{11}{13.2}{\rmdefault}{\mddefault}{\updefault}$\cdots$}}}}
\put(-2139,-526){\makebox(0,0)[lb]{\smash{{\SetFigFont{11}{13.2}{\rmdefault}{\mddefault}{\updefault}$t$}}}}
\put(2999,-1466){\makebox(0,0)[lb]{\smash{{\SetFigFont{11}{13.2}{\rmdefault}{\mddefault}{\updefault}$\cdots \quad + \quad \cdots$}}}}
\put(5954,-2882){\makebox(0,0)[lb]{\smash{{\SetFigFont{11}{13.2}{\rmdefault}{\mddefault}{\updefault}$\as$}}}}
\put(7764,-2862){\makebox(0,0)[lb]{\smash{{\SetFigFont{11}{13.2}{\rmdefault}{\mddefault}{\updefault}$\as$}}}}
\end{picture}%
\caption{An initial state involving a bosonic field.}
\label{figdyn}
\end{figure}

We already saw in~\eqref{Psi1} and~\eqref{Psi2} how the fermionic $n$-particle wave
function of the form~\eqref{psibl} can be realized as a Fock state. However,
this $n$-particle wave function is not sufficient to describe the dynamics, as we now explain.
Suppose we consider the process where a bosonic field generated at the $l^\text{th}$ particle line
ends at the $k^\text{th}$ particle line (see Figure~\ref{figdyn}).
Suppose that we want to prescribe initial data at time~$t_0$, and we want to follow the dynamics
up to a later time~$t_1$ (as indicated in the figure by the arrow of time).
Then we need to take into account a contribution where the bosonic field was created {\em{before}}
the time~$t_0$ (see the left of Figure~\ref{figdyn}).
Moreover, since we allow the bosonic Green's function~\eqref{S0caus} to have an advanced component,
it is also possible that a bosonic line is created after the time~$t_0$, but it propagates to the past
and ends before the time~$t_0$ (as shown on the right of Figure~\ref{figdyn}).
In both cases, the dynamics is not determined by the fermionic wave function alone, but we
must keep track of the bosonic field at the initial time~$t_0$.

This can be accomplished by acting with the wave operators~$\Bu(t_0, \vec{x})$ and~$\Bd(t_0, \vec{x})$
on the vacuum.
When doing so, one must keep in mind that if a bosonic field has been emitted, the index~$b$ of the
corresponding particle line must have the value one (see~\eqref{bldef}). Thus for the
preparation of the initial state, we first set
\beq \label{Psi0}
\Psi_0 = G_1\, G_2\, \cdots G_n |0\ket \:,
\eeq
where for each of the operators~$G_l$ there are the following possible choices:
\begin{gather}
G_l = \int_{\R^3} (\psi_l)_{-1}^{\alpha_l}(\vec{z})\, \hat{\Psi}^{\alpha_l}_{[l,1]}(t_0, \vec{z})^\dagger\, d^3z 
\label{pos1} \\
G_l = \int_{\R^3} (\psi_l)_{1}^{\alpha_l}(\vec{z})\, \hat{\Psi}^{\alpha_l}_{[l,-1]}(t_0, \vec{z})^\dagger\, d^3z
\label{pos2} \\
G_l = \Big( \int_{\R^3} (\psi_l)_{1}^{\alpha_l}(\vec{z})\, \hat{\Psi}^{\alpha_l}_{[l,-1]}(t_0, \vec{z})^\dagger\, d^3z
\Big) \Big( \int_{\R^3} \B_l(\vec{z})\, \Bu_{[l]}(t_0, \vec{z}) \, d^3z \Big) . \label{pos3}
\end{gather}
With~\eqref{pos1} one generates a wave with index~$b=-1$, which has not yet
emitted a bosonic line. The fermionic wave generated by~\eqref{pos2} has already emitted a bosonic line,
but no bosonic field is generated (which means that the bosonic line was already absorbed
before the time~$t_0$). In~\eqref{pos3} the fermionic wave has already emitted a bosonic line
corresponding to the bosonic field~$\B_l(\vec{z})$ (this is the case shown on the left of Figure~\ref{figdyn}).
Moreover, we can generate insertions where a bosonic line ends before the time~$t_0$
(as shown on the right of Figure~\ref{figdyn}). To this end, we multiply~$\Psi_0$ by an arbitrary
number of operators of the form
\beq \label{pos4}
\int_{\R^3} \B_k(\vec{z})\, \Bd_{[k]}(t_0, \vec{z}) \, d^3z \:.
\eeq
We denote the resulting Fock state by~$\Psi$.

\subsubsection{\bf{The Hamiltonian}}
We are now in the position to describe the dynamics. We first consider the
situation without the fermionic loop diagrams, which will be treated afterwards
(see Proposition~\ref{prploop} below).

\begin{Prp} \label{prpH} Suppose that the Fock state~$\Psi \in \Fock^\text{\tiny{\rm{Krein}}}_n$
describes the initial data at time~$t_0$ (as explained above). Moreover, assume that the
system is background-synchronized (see Section~\ref{secstoch}).
Then, omitting the diagrams involving fer\-mionic loop diagrams, the Fock-Krein state at a later time~$t$ is
the expectation value of the time-ordered product
\beq \label{Texp}
\begin{split}
\Psi&^{\alpha_1 \cdots \alpha_n}_{b1 \cdots b_n}(t; z_1, \ldots, z_n) \\
&\quad= \bra 0 \,|\, \hat{\Psi}^{\alpha_1}_{[1,b_1]}(z_1) \cdots \hat{\Psi}^{\alpha_n}_{[n,b_n]}(z_n)
\Texp \Big( -i \int_{t_0}^t \Hint(t) \Big) \,|\, \Psi \ket \:,
\end{split}
\eeq
were~$\Hint$ is the Hamiltonian
\begin{align}
\Hint(t) &= \sum_{k=1}^n \sum_{b=\pm1} \int_{\R^3} \hat{\Psi}_{[k,b]}^\dagger(t,\vec{x})\,
\gamma^0 \Bd_{[k]}(t,\vec{x})  \, \hat{\Psi}_{[k,-b]}(t,\vec{x})\: d^3x \label{Hint1} \\
&\quad+ \lambda
\sum_{l=1}^n \int_{\R^3} \hat{\Psi}_{[l,-1]}^\dagger(t,\vec{x})\, \gamma^0 \Bu_{[l]}(t,\vec{x})
\, \hat{\Psi}_{[l,-1]}(t,\vec{x})\: d^3x \:. \label{Hint2}
\end{align}
\end{Prp} \noindent
Before coming to the proof, we make two explanatory remarks.
First, the index ``int'' at the Hamiltonian indicates that we are in the interaction picture.
In fact, as the free dynamics is taken care of by the commutation and anti-commutation relations~\eqref{ac2}
and~\eqref{Bfield}, the operator~$H_\text{int}$ only involves the interaction terms.
The second remark concerns the symmetry of the Hamiltonian and its connection to the
unitarity of the time evolution. Obviously, the above Hamiltonian is
symmetric on the joint fermionic and bosonic Fock space constructed in Section~\ref{secfield}.
This also implies that the ordered exponential in~\eqref{Texp} is a unitary operator on this Fock space.
However, this unitarity is {\em{not}} directly related to the unitary of the time evolution in
the Fock-Krein space~$(\Fock^\text{\tiny{\rm{Krein}}}_n, (.|.))$ as shown in Theorem~\ref{thmunitary}.
To see the difference, one should keep in mind that the inner product~$(.|.)$ on~$\Fock^\text{\tiny{\rm{Krein}}}_n$
only involves the fermionic component, whereas the inner product on the Fock space
constructed in Section~\ref{secfield} also involves a bosonic component.
This difference also becomes apparent in the expectation value~\eqref{Texp},
which removes the bosonic component. These different notions of unitarity also come about for different
reasons: The unitarity of the time evolution in~$(\Fock^\text{\tiny{\rm{Krein}}}_n, (.|.))$ is based on
the physical concepts of the conservation of the Dirac current (for a classical interaction)
and likewise the conservation of probability for a quantum mechanical particle. The symmetry of the
Hamiltonian~$\Hint$, however, merely is a consequence of how we set up the bosonic
field operators in~\eqref{Bdecomp}, \eqref{abvac} and~\eqref{abc}.

\Proof[Proof of Proposition~\ref{prpH}]
Expanding the time-ordered exponential in powers of~$\Hint$ and
expanding the initial state using~\eqref{Psi0} and~\eqref{pos4}, we obtain
a sum of operator products acting on the vacuum.
We first show that any such operator product vanishes whenever it involves
at least two factors~$\Bu_{[l]}$ for some~$l \in \{1,\ldots, n\}$.
To this end, suppose that an operator product involves two factors~$\Bu_{[l]}$.
According to~\eqref{pos3} and~\eqref{Hint2}, the first factor~$\Bu_{[l]}$ (counting from right to left
after time ordering) comes with a factor~$\hat{\Psi}^\dagger_{[l,-1]}$, so that the $l^\text{th}$ particle
line has the lower index~$b=1$. Subsequent factors of~\eqref{Hint1} do not change the
lower indices of the particle lines (because the factor~$\hat{\Psi}_{[k,b]}$
generates a state with the same lower index as is annihilated by~$\hat{\Psi}_{[k,-b]}$).
Therefore, the operator~\eqref{Hint2} which involves the second factor~$\Bu_{[l]}$
acts on a fermionic state where the~$l^\text{th}$ particle line has the lower index~$b=1$.
But then the factor~$\hat{\Psi}_{[l,-1]}$ in~\eqref{Hint2} gives zero.

We next consider the expectation value in~\eqref{Texp} for the above operator products.
Denoting the factors~$\Bu_{[.]}$ in such an operator product by~$l_1, \ldots, l_p$,
we know from the previous consideration that the indices~$l_j$ are pairwise distinct.
Moreover, it follows by construction of the bosonic
fields~\eqref{Bdecomp}--\eqref{abc} that the vacuum expectation value vanishes
unless every operator~$\Bu_{[l]}$ comes with a corresponding operator~$\Bd_{[k]}$
with~$k>l$. Hence we can write the vacuum expectation value as
\[ \bra 0 \,|\, T \left( \cdots \big( \Bd_{[k_1]}(x_1) \cdots \Bd_{[k_p]}(x_k) \big)
\big( \Bu_{[l_1]}(y_1) \cdots \Bu_{[l_p]}(y_1) \big)
\cdots \right) \,|\, 0 \ket \:, \]
where the dots include the fermionic field operators and space-time integrals.
Similar to the calculation after~\eqref{abc}, this expectation value can be computed by substituting
the decomposition~\eqref{Bdecomp} and by commuting all annihilation operators to the right with
the help of~\eqref{abc}. Similar as in~\eqref{Tab}, we thus obtain factors of the bosonic Green's functions
of the form
\beq \label{ihp}
i^p\, S_0(x_{\sigma(1)}, y_1) \cdots S_0(x_{\sigma(p)}, y_p) \:,
\eeq
where~$\sigma$ runs over all permutations of~$\{1, \ldots, p\}$ for which~$k_\sigma(j) > l_j$
for all~$j$. In this way, we obtain the bosonic lines in the Fock-Krein dynamics
(see Figure~\ref{figwave}), where we sum over all possible combinations for which all
bosonic lines go from the left to the right. This implements precisely the
combinatorics for a background-synchronized \ansyb.

It remains to compute the vacuum expectation value of the fermionic field operators.
Since the field operators anti-commute unless they have the same lower index~$[.]$
(see~\eqref{ac2}), we can compute the expectation value for each
fixed lower index separately. This amounts to restricting attention to the~$l^\text{th}$
particle line and to considering only the contributions to~\eqref{pos1}--\eqref{pos3}
and~\eqref{Hint1}, \eqref{Hint2} for this fixed~$l$ and~$k=l$.
In all these formulas, the creation and annihilation operators alternate, and two adjacent factors
have the same space-time dependence. We thus obtain expressions of the form
\beq \label{Tord}
\text{T} \big( \hat{\Psi}^\dagger_\bullet(x_1) \hat{\Psi}_\bullet(x_1)
\;\cdots\; \hat{\Psi}^\dagger_\bullet(x_q) \hat{\Psi}_\bullet(x_q)\;
\hat{\Psi}^\dagger_\bullet(y) \big) | 0 \ket \:.
\eeq
If we reorder the operators such that~$x_1$ lies in the future of~$x_2$,
and so on, and~$x_q$ lies in the future of~$y$, then we can use
the anti-commutator relations similar to~\eqref{Tpsi} to simplify~\eqref{Tord} to
\beq \label{iHfact}
i^q \,\hat{\Psi}^\dagger_{[l, b_1l]}(x_1)  | 0 \ket \;
s^\wedge_m(x_1, x_2) \cdots s^\wedge_m(x_{q-1}, x_q)\: s^\wedge_m(x_q, y) \:,
\eeq
where for clarity we omitted the factors~${\mathfrak{t}}^b_{b'}$.
As desired, we thus obtain the retarded Green's functions of the fermions.

Let us consider the lower indices~$b$. Similar to~\eqref{Tpsi}, the contractions in~\eqref{Tord} give
rise to factors~${\mathfrak{t}}^{b'_j}_{b_{j+1}}$. This means that the summand~\eqref{Hint1}
preserves the lower factor. The summand~\eqref{Hint2}, on the other hand,
changes the lower index from~$-1$ to~$+1$, in agreement with the requirement~\eqref{bldef}
that the lower index changes when a bosonic line is emitted. This shows that the
lower index~$b$ is indeed handled in agreement with the dynamics in the Fock-Krein space.

Finally, we must collect all the prefactors. In order to keep track of the factors~$i$, one must
keep in mind that the Hamiltonian in~\eqref{Texp} is multiplied by~$-i$.
This compensates precisely the factors~$i$ in~\eqref{iHfact}. Comparing~\eqref{ihp}
with the classical equation~\eqref{jiter}, one sees that every bosonic line
in the expectation value~\eqref{Texp} carries an additional factor~$i$.
Next, every outgoing bosonic line comes with
an insertion, along which the particle line is cut (cf.\ Figures~\ref{figinsert} and~\ref{figwave}).
In a retarded time evolution in the Fock-Krein space, the glueing identity~\eqref{glue} can be written as
\beq \label{glue2}
(U[\B] \phi \big)(x) = -i \int_{\R^3} \tilde{s}^\vee_m(x; t_0, \vec{z})\, \gamma^0\, 
(U[\B] \phi \big)(t_0, \vec{z})\: d^3 z
\eeq
(in order to understand the appearance of the advanced Green's function, one should
keep in mind that the cutting surface at time~$t_0$ lies to the future of the considered interaction).
The factors~$-i$ in~\eqref{glue2} precisely compensate the factor~$i^p$ in~\eqref{ihp}.
This concludes the proof.
\QED

\begin{figure} %
\begin{picture}(0,0)%
\includegraphics{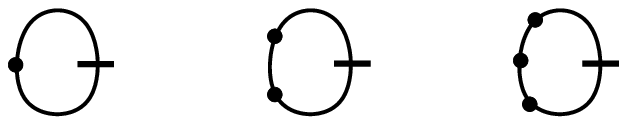}%
\end{picture}%
\setlength{\unitlength}{1533sp}%
\begingroup\makeatletter\ifx\SetFigFont\undefined%
\gdef\SetFigFont#1#2#3#4#5{%
  \reset@font\fontsize{#1}{#2pt}%
  \fontfamily{#3}\fontseries{#4}\fontshape{#5}%
  \selectfont}%
\fi\endgroup%
\begin{picture}(8027,2413)(-1211,-3500)
\put(-1196,-2171){\makebox(0,0)[lb]{\smash{{\SetFigFont{11}{13.2}{\rmdefault}{\mddefault}{\updefault}$y$}}}}
\put(3763,-2151){\makebox(0,0)[lb]{\smash{{\SetFigFont{11}{13.2}{\rmdefault}{\mddefault}{\updefault}$x$}}}}
\put(6801,-2151){\makebox(0,0)[lb]{\smash{{\SetFigFont{11}{13.2}{\rmdefault}{\mddefault}{\updefault}$x$}}}}
\put(551,-2161){\makebox(0,0)[lb]{\smash{{\SetFigFont{11}{13.2}{\rmdefault}{\mddefault}{\updefault}$x$}}}}
\put(4888,-2065){\makebox(0,0)[lb]{\smash{{\SetFigFont{11}{13.2}{\rmdefault}{\mddefault}{\updefault}$y_2$}}}}
\put(5072,-1474){\makebox(0,0)[lb]{\smash{{\SetFigFont{11}{13.2}{\rmdefault}{\mddefault}{\updefault}$y_1$}}}}
\put(1869,-1726){\makebox(0,0)[lb]{\smash{{\SetFigFont{11}{13.2}{\rmdefault}{\mddefault}{\updefault}$y_1$}}}}
\put(1862,-2564){\makebox(0,0)[lb]{\smash{{\SetFigFont{11}{13.2}{\rmdefault}{\mddefault}{\updefault}$y_2$}}}}
\put(1576,-3346){\makebox(0,0)[lb]{\smash{{\SetFigFont{11}{13.2}{\rmdefault}{\mddefault}{\updefault}${\mathcal{L}}_2(x,y_1, y_2)$}}}}
\put(4411,-3346){\makebox(0,0)[lb]{\smash{{\SetFigFont{11}{13.2}{\rmdefault}{\mddefault}{\updefault}${\mathcal{L}}_3(x,y_1,y_2,y_3)$}}}}
\put(4979,-2602){\makebox(0,0)[lb]{\smash{{\SetFigFont{11}{13.2}{\rmdefault}{\mddefault}{\updefault}$y_3$}}}}
\put(-989,-3346){\makebox(0,0)[lb]{\smash{{\SetFigFont{11}{13.2}{\rmdefault}{\mddefault}{\updefault}${\mathcal{L}}_1(x,y)$}}}}
\end{picture}%
\caption{The kernels~${\mathcal{L}}_\ell$ describing the fermion loops.}
\label{figloop}
\end{figure}

We now come to the description of the fermionic loop diagrams which contribute to the
Fock-Krein dynamics (see Figure~\ref{figwave}).
In our description, the fermionic loop diagrams are a graphical notation for
the corrections to the field equations coming from low- and high-energy contributions to the fermionic
projector (see~\cite[Sections~8.2 and~8.4]{sector}).
Here, we describe these contributions by kernels~${\mathcal{L}}_\ell(x, y_1, \ldots, y_\ell)$
which have one outgoing and~$\ell$ incoming bosonic lines.
The kernels can be depicted by truncated diagrams involving one fermion loop,
as is illustrated in Figure~\ref{figloop}
(where for simplicity we omitted the marks~$\times$ for the bra/kets inside the loops).

\begin{Prp} \label{prploop}
Suppose that the Fock state~$\Psi$ describes the initial data at time~$t_0$
of a Fock-Krein state (as explained above). Moreover, assume that the
system is background-synchronized (see Section~\ref{secstoch}).
Then the Fock-Krein state at a later time~$t$ is the time-ordered expectation value
\[ \Psi^{\alpha_1 \cdots \alpha_n}_{b1 \cdots b_n}(t; z_1, \ldots, z_n)
= \bra 0 \,|\, \hat{\Psi}^{\alpha_1}_{[1,b_1]}(z_1) \cdots \hat{\Psi}^{\alpha_n}_{[n,b_n]}(z_n)
\,|\, U^\text{\tiny{\rm{Krein}}}_{t, t_0} \,\Psi \ket \]
with
\begin{align*}
U^\text{\tiny{\rm{Krein}}}_{t, t_0} =\,& \Texp \bigg( -i \int_{t_0}^t \Hint(t)  \\
&+ \lambda \sum_{\ell=1}^n \;\sum_{p=1}^n \:\sum_{k_1,\ldots, k_\ell < p} \int d^4x \int d^4y_1 \cdots \int d^4y_\ell \\
& \hspace*{3cm} \times \Bu_{[p]}(x) \:{\mathcal{L}}_\ell(x,y_1, \ldots, y_\ell)\:\Bd_{[k_1]}(y_1) \cdots \Bd_{[k_\ell]}(y_\ell)
\bigg)
\end{align*}
(and~$\Hint$ as in Proposition~\ref{prpH}).
\end{Prp}
\Proof Exactly as explained in the proof of Proposition~\eqref{prpH}, the
bosonic field operators can be contracted with a Wick rule to give factors of~$S_0$.
Then the kernels~${\mathcal{L}}_\ell$ give rise to the fermion loops (see for example Figure~\ref{figwave};
we omit the combinatorial details, because they will not be needed later on).
\QED

The description of the fermion loops by the kernels~${\mathcal{L}}_\ell$ is a major difference to
standard quantum field theory.
In particular, we do not describe the fermion loops in the fermionic Fock space formalism.
The reason is that the method of taking averages over subsystems described
in Section~\ref{secaverage} only gives an anti-symmetrization of the $n$ particle lines
in the \ansyb, but the sea states are not anti-symmetrized.
Therefore, fermionic Fock spaces do not seem appropriate for describing the sea states.
In particular, the fermion loops cannot be described in the fermionic Fock space formalism
for principal reasons.

We also point out that the kernels~${\mathcal{L}}_\ell$ are ultraviolet finite, so that there is no need to
subtract counter terms or to renormalize the fermion loops. This difference can be understood from the fact that
the singular contributions to the fermionic loop diagrams drop out of the Euler-Lagrange equations corresponding
to the causal action principle. We refer the interested reader to the survey article~\cite{srev}.

\subsection{The Effective Dynamics with Instantaneous Recombination}  \label{secfock2}
Let us recall the different fermionic Fock spaces and how they are related.
In Section~\ref{seckrein} we introduced the {\em{Fock-Krein space}}~$(\Fock^\text{\tiny{\rm{Krein}}}_n, (.|.))$,
endowed with the indefinite inner product~\eqref{kreinip}.
In Section~\ref{secfockeff}, we introduced the {\em{effective Fock space}}~$(\Fock_n, (.|.))$
as the subspace obtained by anti-symmetrizing in the particles and symmetrizing in the lower indices~$b_l$.
The restriction of the inner product~$(.|.)$ to~$\Fock_n$ was positive definite, so that~$(\Fock_n, (.|.))$ is
a Hilbert space. Finally, in Section~\ref{secfield} we built up an indefinite inner product space
by acting with the creation operators~$(\Psi^\alpha_{[l,b]})^\dagger$ to the vacuum state~$|0\ket$.
For clarity, we now denote this fermionic Fock space by~$(\Fock^\text{\tiny{\rm{Krein}}}, \bra .|. \ket)$.
By~\eqref{Psi1} and~\eqref{Psi2}, we could identify~$\Fock^\text{\tiny{\rm{Krein}}}_n$ with
a subspace of~$\Fock^\text{\tiny{\rm{Krein}}}$, obtained by creating one fermion for each index~$l=1,\ldots, n$.
To summarize, we have the inclusions
\[ \Fock_n \subset \Fock^\text{\tiny{\rm{Krein}}}_n \subset \Fock^\text{\tiny{\rm{Krein}}} \:. \]
We denote the corresponding orthogonal projection operators by~$\Pi$
(the projection from~$\Fock^\text{\tiny{\rm{Krein}}}_n$ to~$\Fock_n$ was already introduced in~\eqref{proj}).

\subsubsection{\bf{The Effective Hamiltonian}}
It is a major advantage of the Fock space formalism that the
continual projection method~\eqref{Uproj} can be written in a compact form, as we now explain.
First, the projection operator~\eqref{proj} can be expressed by
\begin{align*}
\Pi &=  \frac{1}{n!}\sum_{\sigma \in S_n}  (-1)^{\sign(\sigma)} \prod_{l=1}^n \bigg(
\frac{1}{2} \sum_{b,b'=\pm1} \int_{\R^3} \hat{\Psi}_{[\sigma(l),b]}^\dagger(t_0,\vec{x})\,
\gamma^0 \Bd_{[k]}(t,\vec{x})  \, \hat{\Psi}_{[l,b']}(t_0,\vec{x})\: d^3x \bigg) \\
&\;:\; \Fock^\text{\tiny{\rm{Krein}}}_n \rightarrow \Fock_n \:.
\end{align*}
Obviously, this operator anti-symmetrizes the $n$ particle lines and symmetrizes
the lower indices~$b$, and thus maps~$\Fock^\text{\tiny{\rm{Krein}}}_n$ to the effective Fock space~$\Fock_n$.
Moreover, a direct computation using the anti-commutation relations~\eqref{ac} 
together with~\eqref{Tdef} shows that~$\Pi$ is indeed a projection operator.
Next, we introduce the field operators
\begin{align}
\hat{\Psi}^\alpha(x) &= \frac{1}{2} \sum_{b=\pm 1} \sum_{l=1}^n \hat{\Psi}^\alpha_{[l,b]}(x) 
&&\hspace*{-2.85cm}:\: \Fock_n \rightarrow \Fock^\text{\tiny{\rm{Krein}}} \label{Psi} \\
\intertext{as well as their adjoints}
\hat{\Psi}^\alpha(x)^\dagger &= \frac{1}{2} \:\Pi \sum_{b=\pm 1} \sum_{l=1}^n
\hat{\Psi}^\alpha_{[l,b]}(x)^\dagger &&\hspace*{-2.85cm}:\: \;\Fock^\text{\tiny{\rm{Krein}}} \rightarrow \Fock_n \:. \label{Psid}
\end{align}
Here~$\Pi$ is the projection operator from~$\Fock^\text{\tiny{\rm{Krein}}}$ to~$\Fock_n$. Its appearance can be understood
as follows: In~\eqref{Psi} we restrict the operator~$\hat{\Psi}^\alpha(x)$
to~$\Fock_n$. Therefore, its adjoint maps to~$\Fock_n$. This adjoint can be computed by taking
the adjoint of~$\hat{\Psi}^\alpha$ in the larger space~$\Fock^\text{\tiny{\rm{Krein}}}$ and projecting its image to~$\Fock_n$.

\begin{Lemma} \label{lemmaeff}
For all particles indices~$l \in \{1, \ldots, n\}$, lower indices~$b,b' \in \{\pm 1\}$
and spinor indices~$\alpha, \beta \in \{1, \ldots, 4\}$,
\[ \Pi \:\hat{\Psi}_{[l,b]}^\alpha(x)^\dagger\, \hat{\Psi}_{[l,b']}^\beta(y) \: \Pi
= \frac{1}{n} \,\hat{\Psi}^\alpha(x)^\dagger\, \hat{\Psi}^\beta(y)\: \Pi \:. \]
\end{Lemma}
\Proof Since~$\Pi$ anti-symmetrizes the particles and symmetrizes in the lower indices,
it is obvious by symmetry that
\[ \Pi \:\hat{\Psi}_{[l,b]}^\alpha(x)^\dagger\, \hat{\Psi}_{[l,b']}^\beta(y)\: \Pi =
\frac{1}{4n} \sum_{c,c'=\pm1} \sum_{k=1}^n
\Pi \:\hat{\Psi}_{[k,c]}^\alpha(x)^\dagger\, \hat{\Psi}_{[k,c']}^\beta(y) \: \Pi \:. \]
Next, we need to keep in mind that the operator~$\Pi$ gives zero unless for each lower index~$l$,
exactly one fermionic state is occupied. Therefore, by adding zeros we get
\[ \Pi \:\hat{\Psi}_{[l,b]}^\alpha(x)^\dagger\, \hat{\Psi}_{[l,b']}^\beta(y)\: \Pi =
\frac{1}{4n} \sum_{c,c'=\pm1} \sum_{k,k'=1}^n
\Pi \:\hat{\Psi}_{[k,c]}^\alpha(x)^\dagger\, \hat{\Psi}_{[k',c']}^\beta(y) \: \Pi \:. \]
The result now follows from~\eqref{Psi} and~\eqref{Psid}.
\QED

Using this Lemma, we can introduce and compute an effective Hamiltonian by
\beq \label{Heffcomp}
\begin{split}
\Heff(t) &:= \Pi \,\Hint\, \Pi \\
&\;= \frac{2}{n} \sum_{k=1}^n \int_{\R^3} \hat{\Psi}^\dagger(t,\vec{x})\,
\gamma^0 \Bd_{[k]}(t,\vec{x})  \, \hat{\Psi}(t,\vec{x})\: d^3x \\
&\quad+ \frac{\lambda}{n} \sum_{l=1}^n \int_{\R^3} 
\hat{\Psi}^\dagger(t,\vec{x})\, \gamma^0 \Bu_{[l]}(t,\vec{x}) \, \hat{\Psi}(t,\vec{x})\: d^3x \:.
\end{split}
\eeq
Combining the bosonic field operators to
\beq \label{Bdef}
\hat{\B}(x) = \frac{1}{\sqrt{2 |\lambda|}} \;\frac{1}{n}
\sum_{k=1}^n \big( 2\, \Bd_{[k]}(x) + \lambda\, \Bu_{[l]}(x) \big) \:,
\eeq
the effective Hamiltonian can be written in the short form
\beq \label{Heff}
\Heff(t) = \sqrt{2 |\lambda|} \int_{\R^3} \hat{\Psi}^\dagger(t,\vec{x})\, \gamma^0 \hat{\B}(t,\vec{x}) \,
\hat{\Psi}(t,\vec{x})\: d^3x \:.
\eeq
Using~\eqref{Bfield}, we obtain
\begin{align*}
[\hat{\B}(x), \hat{\B}(y)] &= \frac{\epsilon(\lambda)}{n^2} \sum_{k,l=1}^n
\Big( [\Bd_{[k]}(x), \Bu_{[l]}(y)] + [\Bu_{[l]}(x), \Bd_{[k]}(y)] \Big) \\
&= \frac{2 i}{n^2}\:\epsilon(\lambda) \sum_{k,l=1}^n \Theta(k-l+1) \\
&\qquad \times
\Big( \tau S_0^\vee(x,y) - (1-\tau) S_0^\wedge(x,y)
- \tau S_0^\wedge(y,x) + (1-\tau) S_0^\wedge(y,x) \Big) \\
&= \frac{2i}{n^2}\:\epsilon(\lambda) \sum_{k,l=1}^n \Theta(k-l+1)  
\left( S_0^\vee(x,y) - S_0^\wedge(x,y) \right) \\
&= 2 i\,\epsilon(\lambda) 
\big( S_0^\vee(x,y) - S_0^\wedge(x,y) \big) \:\bigg( \frac{1}{n^2} \sum_{p=0}^{n-1} p \bigg) ,
\end{align*}
showing that the field operators~$\B$ satisfy the commutation relations
\beq \label{Bc}
[\hat{B}(x), \hat{B}(y)] = -2 \pi \,\epsilon(\lambda)\,K_0(x,y) \:\frac{n-1}{n}
\eeq
with~$K_0$ according to~\eqref{K0def}.
It is remarkable that these commutation relations no longer depend on the
parameter~$\tau$ which describes our Green's function~\eqref{S0caus}.

\subsubsection{\bf{Treatment of the Fermion Loops}}
\begin{figure} %
\begin{picture}(0,0)%
\includegraphics{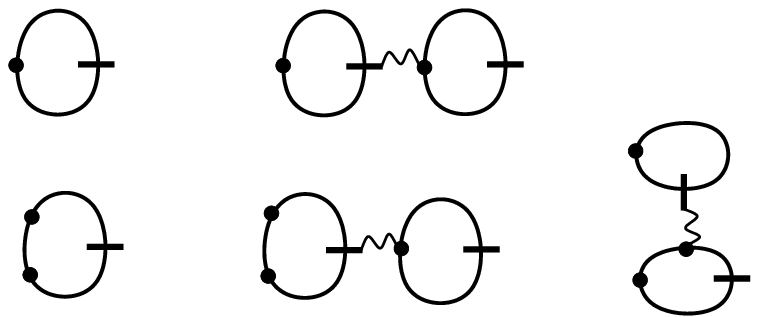}%
\end{picture}%
\setlength{\unitlength}{1533sp}%
\begingroup\makeatletter\ifx\SetFigFont\undefined%
\gdef\SetFigFont#1#2#3#4#5{%
  \reset@font\fontsize{#1}{#2pt}%
  \fontfamily{#3}\fontseries{#4}\fontshape{#5}%
  \selectfont}%
\fi\endgroup%
\begin{picture}(13760,3835)(-4846,-5139)
\put(514,-2176){\makebox(0,0)[lb]{\smash{{\SetFigFont{11}{13.2}{\rmdefault}{\mddefault}{\updefault}$x$}}}}
\put(-4026,-2166){\makebox(0,0)[lb]{\smash{{\SetFigFont{11}{13.2}{\rmdefault}{\mddefault}{\updefault}$L_1(x,y) =$}}}}
\put(5584,-2176){\makebox(0,0)[lb]{\smash{{\SetFigFont{11}{13.2}{\rmdefault}{\mddefault}{\updefault}$x \quad + \quad \cdots$}}}}
\put(1269,-2176){\makebox(0,0)[lb]{\smash{{\SetFigFont{11}{13.2}{\rmdefault}{\mddefault}{\updefault}$+$}}}}
\put(629,-4431){\makebox(0,0)[lb]{\smash{{\SetFigFont{11}{13.2}{\rmdefault}{\mddefault}{\updefault}$x$}}}}
\put(8404,-4806){\makebox(0,0)[lb]{\smash{{\SetFigFont{11}{13.2}{\rmdefault}{\mddefault}{\updefault}$x$}}}}
\put(1254,-4431){\makebox(0,0)[lb]{\smash{{\SetFigFont{11}{13.2}{\rmdefault}{\mddefault}{\updefault}$+$}}}}
\put(5274,-4436){\makebox(0,0)[lb]{\smash{{\SetFigFont{11}{13.2}{\rmdefault}{\mddefault}{\updefault}$x$}}}}
\put(8899,-4431){\makebox(0,0)[lb]{\smash{{\SetFigFont{11}{13.2}{\rmdefault}{\mddefault}{\updefault}$+ \quad \cdots$}}}}
\put(-4831,-4458){\makebox(0,0)[lb]{\smash{{\SetFigFont{11}{13.2}{\rmdefault}{\mddefault}{\updefault}$L_2(x,y_1, y_2) =$}}}}
\put(-1196,-2171){\makebox(0,0)[lb]{\smash{{\SetFigFont{11}{13.2}{\rmdefault}{\mddefault}{\updefault}$y$}}}}
\put(2109,-2171){\makebox(0,0)[lb]{\smash{{\SetFigFont{11}{13.2}{\rmdefault}{\mddefault}{\updefault}$y$}}}}
\put(6314,-3221){\makebox(0,0)[lb]{\smash{{\SetFigFont{11}{13.2}{\rmdefault}{\mddefault}{\updefault}$y_1$}}}}
\put(6374,-4791){\makebox(0,0)[lb]{\smash{{\SetFigFont{11}{13.2}{\rmdefault}{\mddefault}{\updefault}$y_2$}}}}
\put(5909,-4431){\makebox(0,0)[lb]{\smash{{\SetFigFont{11}{13.2}{\rmdefault}{\mddefault}{\updefault}$+$}}}}
\put(1804,-3946){\makebox(0,0)[lb]{\smash{{\SetFigFont{11}{13.2}{\rmdefault}{\mddefault}{\updefault}$y_1$}}}}
\put(1774,-4761){\makebox(0,0)[lb]{\smash{{\SetFigFont{11}{13.2}{\rmdefault}{\mddefault}{\updefault}$y_2$}}}}
\put(-1161,-4766){\makebox(0,0)[lb]{\smash{{\SetFigFont{11}{13.2}{\rmdefault}{\mddefault}{\updefault}$y_2$}}}}
\put(-1156,-3991){\makebox(0,0)[lb]{\smash{{\SetFigFont{11}{13.2}{\rmdefault}{\mddefault}{\updefault}$y_1$}}}}
\end{picture}%
\caption{The kernels~$L_\ell$ describing the fermion loops.}
\label{figloops}
\end{figure}

In order to treat the fermion loops, we apply Lemma~\ref{lemmaeff} to the time-evolution
operator in Proposition~\ref{prploop} and use computations similar to~\eqref{Heffcomp}--\eqref{Heff}.
For convenience, we now work with different kernels~$L_\ell$,
which may be composed of several connected loops (see Figure~\ref{figloops}).

\begin{Thm} \label{prploop2}
Suppose that the Fock state~$\Psi$ describes the initial data at time~$t_0$
of an effective Fock state (similar as explained in Section~\ref{secinitial} for a Fock-Krein state).
Moreover, assume that the system is background-synchronized (see Section~\ref{secbackground}).
Then in the liming case of an instantaneous recombination (see Section~\ref{secfockeff}),
the effective Fock state at a later time~$t$ is the time-ordered expectation value
\beq \label{fockex}
\Psi^{\alpha_1 \cdots \alpha_n}_{b1 \cdots b_n}(t; z_1, \ldots, z_n)
= \bra 0 \,|\, \hat{\Psi}^{\alpha_1}_{[1,b_1]}(z_1) \cdots \hat{\Psi}^{\alpha_n}_{[n,b_n]}(z_n)
\,|\, U^\text{\tiny{\rm{eff}}}_{t, t_0} ,\Psi \ket \:,
\eeq
with
\begin{align*}
U^\text{\tiny{\rm{eff}}}_{t, t_0} =\,& \Texp \bigg( -i \int_{t_0}^t \Heff(t) \\
&+ \sum_{\ell=1}^n |\lambda|^{\frac{\ell+1}{2}} \Big( \sum_{p=1}^n \frac{\ell \,p^\ell}{n^{\ell+1}} \Big)
 \int d^4x \int d^4y_1 \cdots \int d^4y_\ell \\
& \hspace*{3cm} \times \Bu(x) \:L_\ell(x,y_1, \ldots, y_\ell)\:\Bd(y_1) \cdots \Bd(y_\ell) \bigg)
\end{align*}
and~$\Heff$ according to~\eqref{Heff}. Here the fermionic field operators are defined by~\eqref{Psi}
and~\eqref{Psid}. The bosonic field operators satisfy the commutation relations~\eqref{Bc} and
\begin{align}
[\Bd(x), \Bu(y)]  &= [\Bd(x), \Bd(y)] = [\Bu(x), \Bu(y)] = 0 \label{cc0} \\
[\Bd(x), \hat{\B}(y)] &= i \Big( \tau \,S_0^\wedge(x,y) - (1-\tau) \,S_0^\vee(x,y) \Big) \,\frac{n-1}{n} \label{cc4} \\
[\Bu(x), \hat{\B}(y)] &= i \Big( (1-\tau) \,S_0^\wedge(x,y) - \tau \,S_0^\vee(x,y) \Big) \,\frac{n-1}{n} \label{cc5}
\end{align}
(and all other commutators vanish).
\end{Thm}
\Proof We introduce the field operators
\[ \Bd(x) = \frac{\sqrt{2}}{n\, \sqrt{|\lambda|}} \sum_{k=1}^n \Bd_{[k]}(x) \:,\qquad
\Bu(x) = \frac{\sqrt{|\lambda|}}{n \,\sqrt{2}} \sum_{k=1}^n \Bu_{[k]}(x) \:. \]
According to~\eqref{Bfield} and~\eqref{Bdef}, they satisfy the commutation relations~\eqref{cc4},
\eqref{cc5} and
\begin{align*}
[\Bd(x), \Bu(y)] &= i
\Big( \tau \,S_0^\wedge(x,y) - (1-\tau) \,S_0^\vee(x,y) \Big) \,\frac{n-1}{n} \\
[\Bd(x), \Bd(y)] &= 0 = [\Bu(x), \Bu(y)] \:.
\end{align*}
When treating the bosonic loops, the combinatorics of the
contractions of the field operators~$\Bu$ and~$\Bd$ becomes somewhat complicated.
In order to simplify the situation, we redefine the kernels such that
no kernels are connected to each other by bosonic lines.
To this end, we consider the diagrams of Proposition~\ref{prploop}. We remove the~$n$
particle lines as well as all the bosonic lines connected to these particle lines.
The kernels~$L_\ell$ are then defined as the connected components of the
resulting diagrams.
With this definition, the bosonic lines connected to the kernels~$L_\ell$ 
all end or begin at the particle lines. After projecting to the effective Fock space,
this means that the field operators~$\Bu$ and~$\Bd$ in~$U^\text{\tiny{\rm{eff}}}_{t, t_0}$ should
all be contracted with the operators~$\B$. This explains the commutation relations~\eqref{cc0}.
\QED

\subsubsection{\bf{Restriction to a Small Subsystem}}
So far, the fermionic field operators~$\hat{\Psi}(x)$ are defined only on the space~$\Fock_n$
involving~$n$ particles (see~\eqref{Psi} and~\eqref{Psid}). In particular, it is impossible to
take powers of these operators or to write down anti-commutation relations.
This restrictive framework describes the physical dynamics completely.
Nevertheless, it is not sufficient for the applications, because in most physical situations
one considers a small subsystem of the whole universe. Then the number of particles in the subsystems
is typically much smaller than~$n$, and we would like to formulate an effective Hamiltonian on
this smaller Fock space. This can be accomplished by the following mathematical construction:
First, a straightforward calculation shows that the image of the operators~$\hat{\Psi}(x)$
in~\eqref{Psi} is a positive definite subspace of~$\Fock^\text{\tiny{Krein}}$. Forming its completion,
we obtain a Hilbert space denoted by~$(\Fock_{n-1}, \bra .|. \ket)$.
Proceeding iteratively, we obtain Hilbert spaces~$\Fock_n$, $\Fock_{n-1}$, \ldots, $\Fock_0$ and
annihilation operators
\[ \hat{\Psi}^\alpha(x) := \frac{1}{2} \sum_{b=\pm 1} \sum_{l=1}^n \hat{\Psi}^\alpha_{[l,b]}(x) 
\::\: \Fock_p \rightarrow \Fock_{p-1} \:. \]
The space~$\Fock_p$ can be regarded as the fermionic Fock space for~$p$ particles.
Clearly, the space~$\Fock_0$ is spanned by the vacuum state.
Evaluating the scalar product~$\bra .|. \ket$ on the spaces~$\Fock_p$
has the disadvantage that the components involving different values of the indices~$l$
are orthogonal. This property was arranged in order to obtain a ``separate dynamics''
of the~$n$ particle lines. However, after the recombination, this is not quite
what we want, because we would better like to ``ignore'' the values of the indices~$l$.
The latter can be achieved simply by rescaling the scalar products on the spaces~$\Fock_p$ by
a combinatorial factor,
\beq \label{rescale}
\bra .|. \ket_{\Fock_p} := \begin{pmatrix} n \\ p \end{pmatrix} \bra .|. \ket \:.
\eeq
Now we introduce the creation operators~$\hat{\Psi}^\dagger$ by taking the adjoints
with respect to the new scalar products,
\[ \hat{\Psi}^\alpha(x)^\dagger \::\: \Fock_{p-1} \rightarrow \Fock_p \:. \]
A direct computation shows that these field operators satisfy the usual canonical anti-commutation relations
\beq \label{acomm}
\begin{split}
\{ \hat{\Psi}^\alpha(x), \hat{\Psi}^\beta(y)^\dagger \} &= 2 \pi\,
\big( k_m(x,y) \,\gamma^0 \big)^\alpha_\beta \\
\{ \hat{\Psi}^\alpha(x), \hat{\Psi}^\beta(y) \} &= 0 =
\{ \hat{\Psi}^\alpha(x)^\dagger, \hat{\Psi}^\beta(y)^\dagger \} \:.
\end{split}
\eeq
For clarity, we point out that the rescaling~\eqref{rescale} is unproblematic from the physical point of
view because the physics is described completely by the space~$\Fock_n$, whereas the
spaces~$\Fock_p$ for~$p<n$ are only a mathematical device for an effective description of
subsystems.

Now we can restrict attention to a subsystem whose particle number is much smaller than~$n$.
When doing so, we can no longer restrict attention to the fermionic part by
taking the expectation value~\eqref{fockex}. Instead, we need to take the bosons into account
and describe the system by a vector in the tensor product of the fermionic and bosonic Fock spaces.
This different description can be understood in view of the measurement process as follows:
In the formulation in~$\Fock_n$, the measurement device is considered as part of the total system.
Thus even if photons are exchanged between the experimental sample and the measurement device,
no photons leave or enter the total system, making it possible to take the expectation value~\eqref{fockex}.
However, if we restrict attention to a subsystem, the measurement device will in general not be
part of this subsystem. Therefore, we must allow for the possibility that the subsystem exchanges
photons with its environment, making it impossible to take an expectation value as in~\eqref{fockex}.

Taking the limit~$n \rightarrow \infty$, the result of Theorem~\ref{prploop2} simplifies as follows.
\begin{Thm} \label{thmeffective}
Consider a background-synchronized system (see Section~\ref{secbackground})
in the liming case of an instantaneous recombination (see Section~\ref{secfockeff}).
Moreover, we assume that the number of fermions in the physical system under considerations
is much smaller than the total number~$n$ of particles.
Then the time evolution is described by the unitary operator on the effective Fock space
\begin{align}
&U^\text{\tiny{\rm{eff}}}_{t, t_0} = \Texp \bigg( -i \int_{t_0}^t \sqrt{2 |\lambda|}
\int_{\R^3} \hat{\Psi}^\dagger(t,\vec{x})\, \gamma^0 \hat{\B}(t,\vec{x}) \, \hat{\Psi}(t,\vec{x})\: d^3x
\label{U1} \\
&+ \sum_{\ell=1}^n |\lambda|^{\frac{\ell+1}{2}}
\int d^4x \int d^4y_1 \cdots \int d^4y_\ell  \;\Bu(x) \:L_\ell(x,y_1, \ldots, y_\ell)\:\Bd(y_1) \cdots \Bd(y_\ell) \bigg) \,. \label{U2}
\end{align}
The field operators satisfy the canonical anti-commutation relations~\eqref{acomm}
as well as the commutation relations
\begin{align}
[\hat{\B}(x), \hat{\B}(y)] &= -2 \pi \,\epsilon(\lambda)\,K_0(x,y) \label{c1} \\
[\Bd(x), \Bu(y)] &= [\Bd(x), \Bd(y)] = [\Bu(x), \Bu(y)] = 0 \label{c2} \\
[\Bd(x), \hat{\B}(y)] &= i \Big( \tau \,S_0^\wedge(x,y) - (1-\tau) \,S_0^\vee(x,y) \Big) \label{c4} \\
[\Bu(x), \hat{\B}(y)] &= i \Big( (1-\tau) \,S_0^\wedge(x,y) - \tau \,S_0^\vee(x,y) \Big) \label{c5}
\end{align}
(and all other commutators vanish).
Here the parameter~$\tau$ describes the Green's function of the bosonic field~\eqref{S0caus}.
\end{Thm} \noindent
The effective Hamiltonian in~\eqref{U1} together with the
(anti-)commutation relations \eqref{acomm} and~\eqref{c1}
reproduces precisely quantum field theory on the tree level. 
Furthermore, the bosonic loop diagrams are described equivalently.
However, as explained after Proposition~\ref{prploop},
the description of the loop diagrams is different from standard quantum field theory.

\section{Interpretation and Outlook} \label{secinterpret}
\subsection{Comparison to the Standard Formulation of Quantum Field Theory}
In Section~\ref{secfock2} we rewrote the effective dynamics in the Fock space formalism.
The resulting dynamics in Theorem~\ref{thmeffective}
has striking similarity with standard perturbative quantum field theory.
In particular, we get complete agreement on the tree level as
well as for all bosonic loop diagrams.

The only difference concerns the description of the fermion loops.
Namely, in our formalism the fermion loops are described by integral kernels~$L_\ell$,
which are coupled to the bosonic field via the field operators~$\Bu$ and~$\Bd$.
The appearance of these two different field operators is reminiscent of the
fact that we are working with classical bosonic fields, which are generated by
a Dirac current (as described by~$\Bu$) and then couple to the fermions (as
described by~$\Bd$). The real parameter~$\tau$ in the commutation
relations~\eqref{c2}, \eqref{c4} and~\eqref{c5} corresponds to the freedom
in choosing the Green's functions of the classical fields.
A major difference to the standard formulation of quantum field theory is
that the divergences of the fermionic loop diagrams do not occur.
Instead, the integral kernels~$L_\ell$ are all finite. This can be understood
by the fact that the divergent part of the usual diagrams drop out of the
Euler-Lagrange equations corresponding to the causal action principle
(for a more detailed explanation see the review article~\cite{srev}).

We also point out that the bosonic loop diagrams appear only in the
limiting case of an instantaneous recombination.
Namely, in the dynamics of Proposition~\ref{prploop} without instantaneous recombination,
the commutation relations~\eqref{Bfield} imply that the bosonic lines move ``from the left to
the right,'' making it impossible to form loops. As a consequence,
the dynamics of the \ansybs\ is ultraviolet finite (to every order in perturbation theory).
Thus in our formulation, on could avoid the ultraviolet divergences of quantum field theory 
simply by replacing the instantaneous recombination by a more
appropriate limiting case. We come back to this point in Section~\ref{secopen}.

\subsection{Wave-Particle Dualism and Collapse} \label{seccollapse}
The fermionic projector approach is based on the physical concept that the
wave function should be considered as the basic physical concept, whereas the
``particle character'' should merely be a consequence of the interaction as described
by the causal action principle. For details we refer the reader to the exposition in~\cite{dice2010}.
Here we only remark that with our concept of microscopic mixing of wave functions,
it becomes possible to interpret the wave function as the basic physical object without
encountering the inconsistencies noted by Schr\"odinger~\cite{schroedinger2}.
This interpretation is also in agreement with Barut's ideas~\cite{barut}.

We now explain how our ``microscopic mixing of wave-functions'' relates to the
``microscopic mixing of decoherent space-time regions'' as introduced in~\cite{entangle}.
We begin with a system involving microscopic mixing of the wave functions as
introduced in Section~\ref{secmicmix}. Then in each subsystem~$M_\as$
we have a classical interaction described by a bosonic field~$\B_\as$ (see~\eqref{Badef}).
In particular, the wave functions in the subsystem~$M_\as$ (including all the sea states)
satisfy the Dirac equation with the potential~$\B_\as$ (see~\eqref{UPab}).
This bosonic field will be different in each subsystem.
As a consequence, as time evolves, the wave functions in the different subsystems will
get ``out of phase''. In other words, they become decoherent, just as explained in detail
in~\cite{joos}. In our context, we need to take into account that also the sea states become
decoherent. This leads to {\em{decoherent space-time regions}} as analyzed in~\cite{entangle}.
Once the subsystems have become decoherent, they no longer interact with each other.
Considering the observer or measurement device as being part of the physical system,
this amounts to restricting attention to a small number of still coherent subsystems.

We finally remark that in~\cite{dice2010} a mechanism was proposed that
should reduce the number of decoherent subsystems.
The reader who is willing to accept this mechanism can understand the above situation
alternatively as follows: The causal action principle penalizes a too large number of decoherent
subsystems. Hence if the number of decoherent subsystems gets too large, a ``{\em{collapse}}''
reduces the number of decoherent subsystems. This ``collapse'' amounts to removing
many subsystems from space-time and to rescaling the remaining subsystems.

\subsection{Open Problems} \label{secopen}
We finally mention a few open problems which hint towards possible directions of future research.
Generally speaking, the main task is to work out the differences to standard quantum field theory
in detail with the aim of getting experimental predictions.
\begin{itemize}
\item[(1)] In Section~\ref{secfock2} we obtained agreement to
the standard formulation of quantum field theory only up to the description of the fermionic loop diagrams.
In~\cite[\S8.2]{sector}, it was shown that the one-loop correction to the photon propagator gives
agreement with the standard Uehling potential.
However, it is an important open question whether the fermionic projector approach
also reproduces all the higher loop corrections of standard quantum field theory.
In this context, one should also take into account the corrections caused by
the microlocal transformation (see~\cite[\S7.10]{sector} and~\cite[\S4.4]{lepton}).
Some of these issues will be analyzed in~\cite{norm}.

Another difference to the standard formulation of quantum field theory is that
we need to take into account the ``polarization'' of the mixing states in~$\I_\as$ by the particles.
We expect that the resulting effect is very small due to the anti-symmetrization
(similar as worked out in~\cite{pickl} for a Fermi gas), but the details still need to be investigated.
\end{itemize}
Next, one should keep in mind that the connection to standard quantum field theory
was obtained in Section~\ref{secfock2} only as limiting case under additional assumptions.
Therefore, it is a major task to question these assumptions and to compute potential corrections:
\begin{itemize}
\item[(2)] The assumption that the system is {\em{background synchronized}}
implies that the stochastic bosonic background field is so weak that it does not give rise to
observable corrections, but on the other hand its interaction time should be so large that it
synchronizes the~$n$ particle lines of the \ansyb\ (see Section~\ref{secbackground}).
In order to question the weakness of the background field, one should
specify the covariance (for example by~\eqref{Cansatz} and~\eqref{Covscale}), compute
Feynman diagrams which involve the stochastic background field and analyze the resulting
effect on observations.
In order to question the synchronization by the background field, one needs to
work out the constraints coming from the condition that all~$n$ particle lines must
be synchronized by either the stochastic background field or the
outgoing and incoming bosonic lines (as shown in Figure~\ref{figmixing}).
\item[(3)] In order to question the assumption of {\em{instantaneous recombination}},
one needs to specify the recombination time and work out corrections due to the fact
that the recombination time is finite. An interesting point is that the bosonic loops
appear only in the limit of instantaneous recombination, so that working with a finite
recombination time should remove all ultraviolet divergences. 
Ultimately, the dynamics of \ansybs\ should be understood quantitatively by minimizing
the causal action principle.
\end{itemize}

\appendix
\section{Estimating Fluctuations on the Tensor Product} \label{appyoung}
We now analyze the representations of~$\U(n)$ on the $p$-fold tensor product.
Our goal is to give a proof of Proposition~\ref{prpcombi}. Moreover, we will
comment on the case~$p>n$.
The reduction of the tensor product~\eqref{reduce2} is carried out in detail in~\cite[Section~5]{sternberg}.
The irreducible representations are labelled by Young diagrams with~$p$ entries, i.e.\
by numbers~$\lambda = (\lambda_1, \ldots, \lambda_r)$ with~$\lambda_i \geq \lambda_{i+1}$
and~$\sum_{i=1}^r \lambda_i = p$. As usual, we denote the Young diagrams
by drawing an array of boxes with~$r$ rows, with~$\lambda_1$ boxes in the first row, $\lambda_2$
boxes in the second row, etc.\ (for basics on Young diagrams cf.~\cite[Section~2.8]{sternberg}).
For example,
\begin{align*}
&\begin{tabular}{|c|c|c|}
\hline $\:\:$ & $\:\:$ & $\:\:$ \\
\hline
\end{tabular} \\[-.4em]
\lambda = (3,2,1,1) =\; &\begin{tabular}{|c|c|}
$\:\:$ & $\:\:$ \\
\hline
\end{tabular}\\[-.4em]
&\begin{tabular}{|c|}
$\:\:$ \\
\hline $\:\:$ \\
\hline
\end{tabular}  \quad . 
\end{align*}
We denote the irreducible representation of~$\U(n)$ corresponding to a Young diagram~$\lambda$
by~$U_\lambda$. Every direct summand in~\eqref{reduce2} carries one of the irreducible
representations~$U_\lambda$. We denote the multiplicity with which each representation~$U_\lambda$
appears by~$n_\lambda$. As shown in~\cite[Section~5.4]{sternberg}, these multiplicities
coincide with the dimensions of the corresponding representations of the symmetric group.
More specifically,
\beq \label{nlam}
n_\lambda = \frac{p!}{\prod {\text{(all hook lengths in~$\lambda$)}}} \:,
\eeq
where the {\em{hook length}} of any position in a Young diagram is defined
as the sum of positions to its right plus the number of positions
below it plus one. For example, writing the hook lengths into the above Young diagram, we obtain
\begin{align*}
&\begin{tabular}{|c|c|c|}
\hline 6 & 3 & 1 \\
\hline
\end{tabular} \\[-.4em]
&\begin{tabular}{|c|c|}
4 & 1 \\
\hline
\end{tabular} \\[-.4em]
&\begin{tabular}{|c|}
2 \\
\hline 
\end{tabular} \\[-.4em]
&\begin{tabular}{|c|}
1 \\
\hline
\end{tabular} \quad .
\end{align*}
The dimensions of the irreducible representations $\dim U_\lambda$
is also worked out in~\cite[Section~5.4]{sternberg}. First, only those representations
occur for which the number of rows of~$\lambda$ is at most~$n$.
We denote the set of these Young diagrams by~$\Lambda_n(p)$,
\beq \label{Lamdef}
\Lambda_n(p) = \{ \text{Young diagrams with~$p$ boxes and at most~$n$ rows} \} \:.
\eeq
For any~$\lambda \in \Lambda_n(p)$, we write the number~$n+j-i$
into the box in the~$i^\text{th}$ row and the~$j^\text{th}$ column, for example
\beq \label{nprodex}
\begin{split}
&\begin{tabular}{|c|c|c|}
\hline $\;\;\;n\;\:\:\,$ & $n+1$ & $n+2$ \\
\hline
\end{tabular} \\[-.4em]
&\begin{tabular}{|c|c|}
$n-1$ & $\;\;\;n\;\:\:\,$ \\
\hline
\end{tabular} \\[-.4em]
&\begin{tabular}{|c|}
$n-2$ \\
\hline 
\end{tabular} \\[-.4em]
&\begin{tabular}{|c|}
$n-3$ \\
\hline
\end{tabular}  \quad .
\end{split}
\eeq
Expressed in terms of these so-called {\em{$n$-entries}}, we have
\beq \label{dimUl}
\dim U_\lambda = \frac{\prod \text{(all $n$-entries)}}{\prod {\text{(all hook lengths in~$\lambda$)}}} \:.
\eeq
Combining this formula with~\eqref{nlam}, we obtain
\beq \label{cres}
\sum_{k=1}^L \frac{1}{\dim I_k} = \sum_{\lambda \in \Lambda_n(p)} \frac{n_\lambda}{\dim U_\lambda}
= \sum_{\lambda \in \Lambda_n(p)} \frac{p!}{\prod \text{(all $n$-entries)}}
\eeq
with~$\Lambda_n(p)$ according to~\eqref{Lamdef}.

This formula tells us about the contributions by the different irreducible representations
to the fluctuations on the tensor product (cf.~\eqref{Iksum}).
Let us discuss the summands on the very right of~\eqref{cres}. 
We first note that the totally antisymmetric representation corresponds to the Young diagram
\begin{align*}
&\begin{tabular}{|c|}
\hline 1 \\
\hline
\end{tabular} \\[-.2em]
&\begin{tabular}{c}
$\,\vdots$ \\[0.4em]
\hline 
\end{tabular} \\[-.4em]
&\begin{tabular}{|c|}
n \\
\hline
\end{tabular} \quad
\end{align*}
in the case~$p=n$. In this case, the product of all $n$-entries in~\eqref{nprodex} equals~$p!$,
so that the corresponding summand in~\eqref{cres} equals one.
For all other Young diagrams in the case~$p=n$, the product of all $n$-entries will be
larger than~$p!$, so that the corresponding summand in~\eqref{cres} is smaller than one.
As we shall quantify below, even the sum over all irreducible representations except for
the totally antisymmetric representation tends to zero as~$n \rightarrow \infty$.
In the case~$p<n$, the totally antisymmetric representation does not exist.
In this case, we will show that the sum over all irreducible representations
tends to zero as~$n \rightarrow \infty$, uniformly in~$p$.
The remaining case~$p>n$ is more subtle for two reason. First, in the cases~$p=2n, 3n, \ldots$,
there are representations involving the tensor product of several totally antisymmetric representations.
The second, more serious problem is the
appearance of so-called {\em{exceptional Young diagrams}} where the first column has~$n$ entries,
and the remaining number of boxes~$p-n$ is small. For example, in the case~$p=n+1$, there is the
exceptional Young diagram
\begin{align*}
&\begin{tabular}{|c|c|c|}
\hline 1 & $\:\:$ \\
\hline
\end{tabular} \\[-.4em]
&\begin{tabular}{|c|c|}
2 \\
\hline
\end{tabular} \\[-.2em]
&\begin{tabular}{c}
$\,\vdots$ \\[0.4em]
\hline 
\end{tabular} \\[-.4em]
&\begin{tabular}{|c|}
n \\
\hline
\end{tabular} \quad .
\end{align*}
A numerical study shows that these exceptional Young diagrams give a significant contribution
to~\eqref{cres}, which does not tend to zero if~$n$ tends to infinity and~$p-n$ is kept fixed.
On the other hand, exceptional diagrams involve an anti-symmetrization
in~$n$ out of~$p$ basis vectors, and therefore it seems that in the limit~$n \rightarrow \infty$,
the exceptional representations should be ``just as good'' as the totally antisymmetric representation.
Unfortunately, we do not know how to make the statement ``just as good'' mathematically precise.
For this reason, we decided not to treat this case here.
Nevertheless, the above consideration suggests that the main conclusion of
Proposition~\ref{prpcombi}, namely the justification of the restriction to \ansybs, can be made
in the case~$p>n$ as well.

The remainder of this appendix is devoted to the proof of Proposition~\ref{prpcombi}.
We assume throughout that~$p \leq n$.
In the next lemma we give an estimate for the contribution of all irreducible representations
except for the totally antisymmetric representation.
\begin{Lemma} There is a numerical constant~$c$ such that
\[ \sum_{\lambda \in \Lambda_{\min(p-1,n)}(p)} \frac{p!}{\prod \text{\rm{(all $n$-entries)}}}
\leq \frac{c\,p!}{n!} \:\sum_{a=1}^{p-1} \frac{(n-a)!}{(n+2-p/2)^{p-a}} \:
\exp \left( \pi\: \sqrt{\frac{2 \,(p-a)}{3}} \:\right) . \]
\end{Lemma}
\Proof It is useful to introduce the abbreviation
\beq \label{abbrev}
P_r(m,n) = \sum_{\lambda \in \Lambda_{\min(r,n)}(m)} \frac{1}{\prod \text{(all $n$-entries)}} \:.
\eeq
We first derive a simple estimate for~$P_r(m,n)$.
The Young diagrams in~$\Lambda_{\min(r,n)}(m)$ have at most~$\min(m,n,r)$
rows. Therefore, the $n$-entries are bounded from below by~$n-\min(m,n,r)+1$
(see~\eqref{nprodex}). As a consequence,
\beq \label{Pes2}
P_r(m,n) \leq \frac{\# \Lambda_{\min(r,n)}(m)}{(n-\min(m,n,r)+1)^m} \:.
\eeq
Clearly, the number of Young diagrams is bounded from above by the number
of partitions of the set~$\{1, \ldots, m\}$. Using the asymptotic formula for the number
of partitions by Hardy and Ramanujan~\cite[Chapter~5]{andrews}, there is a numerical constant~$c$ such that
\[ \# \Lambda_{\min(r,n)}(m) \leq \# \Lambda_m(m) \leq \frac{c}{m}\: \exp \left(\pi\: \sqrt{\frac{2 m}{3}} \right) . \]
Using this inequality in~\eqref{Pes2}, we obtain the estimate
\beq \label{Pes1}
P_r(m,n) \leq \frac{c}{m}\: \frac{1}{(n-\min(m,n,r)+1)^m}\:
e^{\pi\: \sqrt{\frac{2 m}{3}}}\: .
\eeq

In order to improve this estimate, we expand~\eqref{abbrev} in the first column.
Denoting the number of boxes in the first column by~$a$, we we obtain the iteration formula
\[ P_r(p,n) = \sum_{a=1}^{\min(r,p-1)} \frac{(n-a)!}{n!} \:P_a(p-a, n+1) \]
(here the factor~$(n-a)!/n!$ describes the first column, and~$P_a(p-a, n+1)$
describes all the other columns).
Estimating the terms~$P_a(p-a, n+1)$ with the help of~\eqref{Pes1}, we obtain
\[ P_r(p,n) \leq \sum_{a=1}^{\min(r,p-1)} \frac{(n-a)!}{n!} \:\frac{c}{(p-a)}\:
\frac{e^{\pi\: \sqrt{\frac{2 \,(p-a)}{3}}}}{(n+1-\min(p-a,n+1,a)+1)^{p-a}} \:. \]
Finally, we simplify this estimate by using the inequalities
\[ \min(p-a,n+1,a) \leq \frac{p}{2} \qquad \text{and} \qquad \frac{1}{p-a} \leq 1 \]
and choose~$r=p-1$.
\QED

Next, we estimate the factorials with the Stirling formula (see for example~\cite[eq.~(5.11.3)]{DLMF})
\[ \frac{1}{c} \:\sqrt{n}\: \left( \frac{n}{e} \right)^n \leq n! \leq c \:\sqrt{n}\: \left( \frac{n}{e} \right)^n \:, \]
where~$c$ is again a numerical constant. We thus obtain the estimate
\begin{align*}
G&:= \sum_{\lambda \in \Lambda_{\min(p-1,n)}(p)} \frac{p!}{\prod \text{\rm{(all $n$-entries)}}}
\leq c' \:\sqrt{\frac{p}{n}}\; \sum_{a=1}^{p-1} \left( n+2 - \frac{p}{2} \right)^{a-p} \\
&\quad \times
\exp \bigg( \pi\: \sqrt{\frac{2 \,(p-a)}{3}} - (p-a) + p \log p - n \log n + \Big( n-a+\frac{1}{2} \Big) \,\log(n-a) \bigg)
\end{align*}
with a new numerical constant~$c'$.
In order to analyze the dependence on~$p$, it is convenient to introduce the
new summation variable~$\xi = p-a$. Then
\begin{align*}
G&\leq c' \:\sqrt{\frac{p}{n}}\; \sum_{\xi=1}^{p-1} \left( n+2 - \frac{p}{2} \right)^{-\xi} \\
&\times
\exp \bigg( \pi\: \sqrt{\frac{2 \xi}{3}} - \xi + p \log p - n \log n + \Big( n-p+\xi + \frac{1}{2} \Big) \log(n-p+\xi) \bigg)
\end{align*}

We want to show that the right side of this inequality is monotone increasing in~$p$.
First, an elementary estimates gives
\begin{align*}
G&\leq c' \:\sum_{\xi=1}^{p-1} \left( \frac{n}{2}+2 \right)^{-\xi} \\
&\times
\exp \bigg( \pi\: \sqrt{\frac{2 \xi}{3}} - \xi + p \log p - n \log n + \Big( n-p+\xi + \frac{1}{2} \Big) \,\log
\Big(n-p+\xi+\frac{1}{2} \Big) \bigg) .
\end{align*}
Now the $p$-dependence of the exponent can be removed with the following estimate.
\begin{Lemma} For all~$p<n$ and~$1 \leq \xi \leq p-1$,
\[ p \log p + \Big( n-p+\xi + \frac{1}{2} \Big) \,\log
\Big(n-p+\xi+\frac{1}{2} \Big)
\leq \Big(\xi+\frac{1}{2} \Big) \,\log \Big( \xi + \frac{1}{2} \Big) +  n \log n \:. \]
\end{Lemma}
\Proof For any parameters~$0<a \leq b$, we consider the function
\[ f(x) = (a-x) \log(a-x) + (b+x) \log (b+x) \qquad \text{for} \qquad 0 < x < a\:. \]
Computing its derivative,
\[ f'(x) = -\log(a-x) + \log(b+x)  \geq 0 \:, \]
one sees that~$f$ is monotone increasing and thus
\[ a \log a + b \log b \leq (a-x) \log(a-x) + (b+x) \log (b+x) \:, \]
valid for all~$0<x<a \leq b$.

In the case~$p \leq n-p+\xi + \frac{1}{2}$, we apply this inequality
choosing~$a=p$, $b=n-p+\xi + \frac{1}{2}$ and~$x=p-\xi-\frac{1}{2} \geq \frac{1}{2}$.
Likewise, in the case~$p > n-p+\xi + \frac{1}{2}$, we
choose~$a=n-p+\xi + \frac{1}{2}$, $b=p$ and~$x=n-p \geq 1$.
This gives the result.
\QED
Applying this lemma and adding summands for~$\xi=p,\ldots, n-1$, we conclude that
\beq \label{Gsum}
G \leq c' \:\sum_{\xi=1}^{n-1} h(\xi, n) \:,
\eeq
where~$h$ is the function
\beq \label{hdef}
h(\xi, n) := \left( \frac{n}{2}+2 \right)^{-\xi} 
\: \exp \bigg( \pi\: \sqrt{\frac{2 \xi}{3}} - \xi + \Big( \xi + \frac{1}{2} \Big) \log \Big(\xi+\frac{1}{2} \Big) \bigg) .
\eeq

\begin{figure}
\begin{center}
\includegraphics[width=9cm]{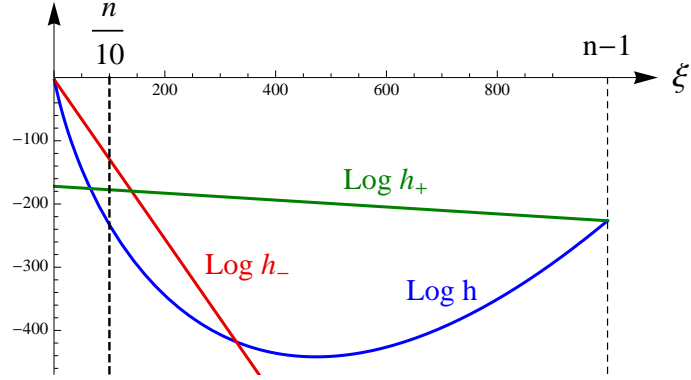}
\caption{The functions~$h$ and~$h_\pm$ in the example~$n=1000$ and~$\xi_1 = n/10$.}
\label{figyoung}
\end{center}
\end{figure}

In Figure~\ref{figyoung}, the function~$\log h$ is shown for a typical large value of~$n$.
One sees that~$h(\xi, n)$ is extremely small except for small values of~$\xi$.
This suggests that the main contribution to~\eqref{Gsum} should come from the first summands.
Next, it is obvious from~\eqref{hdef} that~$\lim_{n \rightarrow \infty} h(\xi,n)=0$ for any fixed~$\xi$.
This gives hope that~$G$ tends to zero as~$n \rightarrow \infty$.
In order to prove it, we proceed as follows. We choose an intermediate value~$\xi_1>1$.
On the interval~$[1, \xi_1]$, we use that the square root function is concave
and thus lies below the tangent at~$\xi=1$,
\[ \sqrt{\xi} \leq s_-(\xi) := 1 + \frac{1}{2}\, (\xi-1) \:. \]
The function~$f(\xi) := (\xi+1/2) \log (\xi+1/2)$, on the other hand, is convex and thus lies below
the secant through the points~$\xi=1$, and~$\xi=\xi_1$,
\[ f(\xi) \leq f_-(\xi) := \frac{1}{\xi_1-1} \Big( (\xi-1)\: f(\xi_1) + (\xi_1-\xi)\: f(1) \Big) 
\qquad \text{for all~$\xi \in [1, \xi_1]\:.$} \]
Hence
\[ h(\xi, n) \leq h_-(\xi, n) :=
\left( \frac{n}{2}+2 \right)^{-\xi} \: \exp \bigg( \pi\: \sqrt{\frac{2}{3}}\, s_-(\xi) - \xi + f_-(\xi) \bigg)
\quad \text{for all~$\xi \in [1, \xi_1]\:.$} \]
For~$\xi$ in the remaining interval~$[\xi_1, n-1]$, we use similarly the estimates
\begin{align*}
\sqrt{\xi} &\leq \sqrt{n-1} \\
f(\xi) &\leq f_+(\xi) := \frac{1}{n-1-\xi_1} \Big( (\xi-\xi_1)\: f(n-1) + (n-1-\xi)\: f(\xi_1) \Big) \\
h(\xi, n) &\leq h_+(\xi, n) :=
\left( \frac{n}{2}+2 \right)^{-\xi} \: \exp \bigg( \pi\: \sqrt{\frac{2 \,(n-1)}{3}} - \xi + f_+(\xi) \bigg) \:.
\end{align*}
Typical plots of the functions~$h_\pm$ are shown in Figure~\ref{figyoung}.
The $\xi$-sums of~$h_\pm$ can be estimated by integrals, which can be computed in closed
form. Choosing~$\xi_1=[(n-1)/10]$, one can expand the resulting expressions in terms of~$n$.
This shows that
\[ \sum_{\xi=1}^{\xi_1} h_-(\xi,n) + \sum_{\xi=\xi_1}^{n-1} h_+(\xi,n) = \O \Big( \frac{1}{n} \Big) . \]
This concludes the proof of Proposition~\ref{prpcombi}.

\Thanks{\em{Acknowledgments:}}
I would like to thank Theodor Br\"ocker, David Cherney, Dirk-Andr{\'e} Deckert,
Domeni\-co Giulini, Johannes Kleiner, Peter Pickl, Hermann Schulz-Baldes, Alexander Strohmaier
and the referees for helpful discussions or comments on the manuscript. I am very grateful to
 J\"urgen Tolksdorf for intensive discussions on the subject.
I would like to thank the Max Planck Institute for Mathematics in the Sciences, Leipzig,
for its hospitality.

\providecommand{\bysame}{\leavevmode\hbox to3em{\hrulefill}\thinspace}
\providecommand{\MR}{\relax\ifhmode\unskip\space\fi MR }
\providecommand{\MRhref}[2]{%
  \href{http://www.ams.org/mathscinet-getitem?mr=#1}{#2}
}
\providecommand{\href}[2]{#2}

\end{document}